\documentclass[a4paper,12pt]{article}
\usepackage{amscd,amssymb,amsmath,amsthm}
\usepackage{rotating}
\usepackage{hyperref}
\usepackage{graphicx,caption,dsfont}
\usepackage{mathtools}

\usepackage{subcaption}
\usepackage{amsfonts,enumerate}
\usepackage{fullpage}
\usepackage[numbers,sort & compress]{natbib}
\usepackage{soul}
\usepackage{xfrac}
 \usepackage{xcolor}
\usepackage{epstopdf}
\usepackage{booktabs}
\numberwithin{equation}{section}
\usepackage{xcolor}

\newtheorem{theorem}{Theorem}[section]
\newtheorem{example}{Example}[section]

\newtheorem{definition}{Definition}[section]

\newtheorem{note}{ Note}[section]

\newtheoremstyle{case}{}{}{}{}{}{:}{ }{}
\theoremstyle{case}

\allowdisplaybreaks

\makeatletter
\def\and{%
\end{tabular}%
\begin{tabular}[t]{c}}%
\def\@fnsymbol#1{\ensuremath{\ifcase#1\or 1\or 2\or 3\or
		4\or 5\or 6\or 7\or 8\or 9\else\@ctrerr\fi}}
\makeatother
\vspace{-8ex}
\date{}
\begin{document}
	\baselineskip=0.70cm
\baselineskip=0.70cm
\begin{center}\large{\textbf{{{Nonlinear two-component system} of  time-fractional PDEs in $(2+1)$-dimensions: Invariant subspace method combined with variable transformation}}}
\end{center}
\vspace{-0.5cm}
\begin{center}
	P. Prakash$^{1*}$,  K.S. Priyendhu$^{1}$,  M. Lakshmanan$^{2}$\\
	%Department of Mathematics, Amrita School of Engineering, Coimbatore, Amrita Vishwa
	%Vidyapeetham, India
	$^{1}$Department of Mathematics,\\
	Amrita School of Engineering, Coimbatore-641112,\\
	Amrita Vishwa Vidyapeetham, INDIA.\\
	$^{2}$Department of Nonlinear Dynamics,\\
	Bharathidasan University,\\
	Tiruchirappalli-620 024, INDIA.\\
	E-mail ID's: vishnuindia89@gmail.com \&\ p$_{-}$prakash@cb.amrita.edu (P Prakash)\\
	\hspace{1.6cm} : priyendhusasikumar@gmail.com (K.S. Priyendhu)\\
	\hspace{1.6cm} : lakshman.cnld@gmail.com (M. Lakshmanan)\\
	$*$-Corresponding Author
\end{center}
%%%% Abstract text to be placed here %%%%%%%%%%%%
\begin{abstract}
	In this article, we develop a systematic approach of the invariant subspace method combined with variable transformation to find the generalized separable exact solutions of the {nonlinear two-component system} of time-fractional PDEs (TFPDEs) in $(2+1)$-dimensions for the first time. Also, we explicitly explain how to construct various kinds of finite-dimensional invariant linear product spaces for the given system using the invariant subspace method combined with variable transformation. Additionally, we present how to use the obtained invariant linear product spaces to derive the generalized separable exact solutions of the discussed system. We also note that the discussed method will help to reduce the {nonlinear two-component system} of TFPDEs in $(2+1)$-dimensions into the {nonlinear two-component system} of TFPDEs in $(1+1)$-dimensions, which again reduces to a system of time-fractional ODEs through the obtained invariant linear product spaces. More specifically,  the significance and efficacy of the systematic investigation of the discussed method have been investigated through the initial and boundary value problems of the generalized {nonlinear two-component system} of time-fractional reaction-diffusion equations (TFRDEs) in $(2+1)$-dimensions for finding the generalized separable exact solutions, which can be expressed  in terms of the exponential, trigonometric, polynomial, Euler-Gamma and Mittag-Leffler functions.
  Also, 2D and 3D graphical representations of some of the obtained solutions are presented for different values of fractional orders.
\end{abstract}
 \section{Introduction}\label{Intro}
%%%% Insert A head here
The theory of fractional-order (arbitrary-order) integrals  and derivatives is a rapidly emerging branch of mathematics,  commonly known as  fractional calculus \cite{Podlubny1999,Diethelm2010,Tarasov2011,Tarasov2013a,Tarasov2013b,povstenko2013}. Recent studies have shown many interesting and efficient applications of fractional calculus in various disciplines of science and engineering. The theory of  arbitrary-order  integrals and derivatives has gained immensely from the contributions of many eminent mathematicians like Riemann, Liouville, Caputo, Weyl, Riesz, Letnikov,  Gr\"unwald and many others, which helped to build a strong theoretical foundation and  provided a wide range of applications to various real-world phenomena \cite{Podlubny1999,Diethelm2010,Tarasov2011,Tarasov2013a,Tarasov2013b,povstenko2013}.  These mathematicians have   proposed specific  significant definitions of arbitrary-order integrals and derivatives to generalize classical integer-order differential operators  and analyzed the salient features of these fractional-order operators \cite{Podlubny1999,Diethelm2010,Tarasov2011,Tarasov2013a,Tarasov2013b}. We note that the presence of the power-law singular kernel in the definitions of the Riemann-Liouville fractional operator as well as the Caputo fractional derivative have resulted in specific distinguished properties such as the ability to capture long-term memory of the considered physical process, to describe its nonlocal behaviour and  the  fractal property \cite{Tarasov2011,Tarasov2013a,Tarasov2013b}.
In this work, we will use one of the prominent definitions of the arbitrary-order  derivative as proposed by Caputo,  which is commonly known as the Caputo fractional derivative
and defined in \cite{Podlubny1999} as follows:
	\begin{eqnarray}
	\begin{aligned}\label{caputo}
		\dfrac{\partial^\alpha y(t)}{\partial t^\alpha}=
		\left\{
		\begin{array}{ll}
			\dfrac{1}{\Gamma(k-\alpha)}\int\limits_{0}^{t}\dfrac{d ^{k}y(s)}{d s^{k} }(t-s)^{k-\alpha-1}ds, & k-1<\alpha<k,\ k\in\mathbb{N}; \\
			\dfrac{d^k y}{d t^k}, & \alpha=k\in\mathbb{N}.
		\end{array}
		\right.
	\end{aligned}
\end{eqnarray}
 The main advantage of the  Caputo fractional derivative is that it can  be effectively used as an efficient tool to describe many physical processes that are primarily nonlocal concerning space and time. Also, while dealing with Caputo TFPDEs, one can easily interpret the initial conditions  as similar to  integer-order differential equations.
 \subsection{Motivation}\label{Intro.1}
%%%% Insert B head here
In the literature, we can observe that one of the main motivations to study fractional differential equations (FDEs), specifically FPDEs, emerged due to its ability to model many real-world physical processes more accurately and precisely  \cite{Podlubny1999,Diethelm2010,Tarasov2011,Tarasov2013a,Tarasov2013b,povstenko2013}.
One of the famous applications of FDEs can be observed in the study of anomalous diffusion processes that occur in various transport problems.  {   
	  Now, let us consider the  two-component system of time-fractional diffusion-wave equations \cite{povstenko2013} of the form
\begin{equation}\label{diffusionequation}
	\dfrac{\partial ^{\alpha_s} u_s}{\partial t^{\alpha_s}}=C_s\dfrac{\partial ^2 u_s}{\partial x^2},\alpha_s\in(0,2],s=1,2,
\end{equation}
where $C_s$ is the coefficient of diffusion, $u_s=u_s(x,t),x\in\mathbb{R},t>0,$ and $ \dfrac{\partial ^{\alpha_s} }{\partial t^{\alpha_s}}(\cdot)$ is the  $\alpha_s$-th order  Caputo  fractional derivative \eqref{caputo}. From a mathematical point of view, we can observe that the above system  \eqref{diffusionequation} represents both parabolic (if $\alpha_s=1$) and hyperbolic  (if $\alpha_s=2$)  types of PDEs \cite{prakash2022a} simultaneously.
Additionally, we would like to mention  that the above {two-component system of time-fractional diffusion-wave equations} \eqref{diffusionequation} represents different types of physical processes depending upon the different values of  orders $\alpha_s$ of the Caputo fractional derivative. When  the parameters $\alpha_s,s = 1, 2,$ take  non-integer values as  $	\alpha_s\in(0,1)$ and $\alpha_s\in(1,2),$ then the given system   \eqref{diffusionequation}  represents the sub-diffusion process \cite{bakkyaraj2015,prakash2022a,hanygad2022,pov,povstenko2013} and the  super-diffusion process \cite{bakkyaraj2015,prakash2022a,hanygad2022,pov,povstenko2013}, respectively.
On the other hand,  for $ \alpha_s=1$ and $\alpha_s=2,$  the given system \eqref{diffusionequation}  coincides with the standard (normal) diffusion and the classical wave systems, respectively.} In the literature, these types of time-fractional diffusion equations are widely applied in the theories of neutron migration, viscous flow, electrochemical problems,  flow through porous media, and many other transport phenomena.
 In general, the system of the Caputo TFPDEs has been used to study many significant diffusion processes or heat conduction in complex physical phenomena that consist of more than one particle substance or species, and it is used  even in higher dimensions. Recently, Zhang and Li   \cite{zhang2017} have formulated a  two-component system of TFPDEs in $(2+1)$-dimensions   to investigate the co-occurrence of moisture diffusion and  heat conduction in hygrothermoelastic medium  using the  following conservation laws:
\begin{eqnarray} \label{conservation-law-zhang}
	\begin{aligned}
		(i)\,&\text{Conservation law of mass: }&\bigtriangledown\cdot \mathbf{Q}_1=-\dfrac{\rho_1}{\nu_1}\dfrac{\partial \mathcal{M}}{\partial t}-\dfrac{\partial \mathcal{C}_2}{\partial t}, \qquad\\
		(ii)\,&\text{Conservation law of energy: }&\bigtriangledown\cdot\mathbf{Q}_2={\rho_1}{\nu_2}\dfrac{\partial \mathcal{M}}{\partial t}-{\rho_1}{\rho_2}\dfrac{\partial \mathcal{C}_2}{\partial t} ,
			\end{aligned}
\end{eqnarray}
where $\mathbf{Q}_1$ is the heat flux vector  with temperature $\mathcal{C}_1=\mathcal{C}_1(x_1,x_2,t),$  $\mathbf{Q}_2$ is the matter flux vector of the water vapour with concentration $\mathcal{C}_2=\mathcal{C}_2(x_1,x_2,t),$  $\mathcal{M}=a_0-a_1\mathcal{C}_1+a_2\mathcal{C}_2$ represent the quantity of moisture absorbed per unit mass,  $a_i\in\mathbb{R},i=0,1,2,$ are material constants, and $\rho_1,\rho_2,\nu_1,$ and $\nu_2$ denote the density of the material, specific heat at constant pressure, volume fraction of voids and the heat released per unit mass of moisture, respectively.
They  have considered the fluxes $\mathcal{Q}_s,s=1,2,$ which follow the time-fractional Fourier's and Fick's laws. That is, for $s=1,2,$ we have \begin{eqnarray}\label{fluxes-zhang}
\mathbf{Q}_s=\left\{\begin{array}{ll}
-\kappa_s\, {\mathcal{D}}^{\alpha_s}_{t}\bigtriangledown\mathcal{C}_s(x_1,x_2,t),&0<\alpha_s\leq1,\\
-\kappa_s\,{\mathcal{I}}^{\alpha_s-1}_{t}\bigtriangledown\mathcal{C}_s(x_1,x_2,t),&1<\alpha_s\leq2,
\end{array}\right.
\end{eqnarray}
where ${\mathcal{D}}^{\alpha_s}(\cdot) $ and ${\mathcal{I}}^{\alpha_s-1}(\cdot)$ denote the Riemann-Liouville fractional derivative of order $\alpha_s$ and  $(\alpha_s-1)$-th order  Riemann-Liouville fractional integral, respectively, which are defined in \cite{Podlubny1999} as follows:
	$$\begin{aligned}
		&{ D^{\alpha_s}_t}C(x_1,x_2,t)= \dfrac{1}{\Gamma(1-\alpha_s)}\dfrac{\partial }{	\partial t}\left( {\int_{0}^{t}(t-y)^{-\alpha_s}C(x_1,x_2,y)dy}\right) ,0<\alpha_s\leq1, \\
		&{I^{\alpha_s-1}_t}C(x_1,x_2,t)=\dfrac{1}{\Gamma(\alpha_s-1)}{\int_{0}^{t}(t-y)^{\alpha_s-2}C(x_1,x_2,y)dy},1<\alpha_s\leq2.
\end{aligned}$$
Using the above equations with  \eqref{conservation-law-zhang} and \eqref{fluxes-zhang}, they derived  the following   two-component system of TFPDEs in $(2+1)$-dimensions:
\begin{eqnarray}\label{zhang-model}
	\begin{aligned}
		\dfrac{\partial^{\alpha_s}}{\partial t^{\alpha_s}}(c_{s1}\mathcal{C}_1+c_{s2}\mathcal{C}_2)=\kappa_s\bigtriangledown^2 \mathcal{C}_s,\alpha_s\in(0,2],s=1,2,	\end{aligned}
\end{eqnarray}  where $c_{11}=\rho_1(\rho_2+\nu_2a_2),c_{12}=-\rho_1\rho_2a_1,c_{21}=-\dfrac{\rho_1a_1}{\nu_1},$ and $c_{22}=\dfrac{\nu_1+\rho_1a_2}{\nu_1}.$
We note that when we substitute $u_s=c_{s1}\mathcal{C}_1+c_{s2}\mathcal{C}_2$ in the above system \eqref{zhang-model}, it gets reduced into the  two-component  system of time-fractional diffusion equations in $(2+1)$-dimensions as follows:
\begin{eqnarray}\label{reduced-zhang-model}
	\begin{aligned}
		\dfrac{\partial^{\alpha_s} u_s}{\partial t^{\alpha_s}}=\bigtriangledown^2 (k_{s1}u_1+k_{s2}u_2),\alpha_s\in(0,2],s=1,2,	\end{aligned}
\end{eqnarray}
where $\dfrac{\partial^{\alpha_s} }{\partial t^{\alpha_s}}(\cdot)$ represents the $\alpha_s$-order Caputo fractional derivative \eqref{caputo} for $s=1,2,$ $k_{11}= \dfrac{c_{22}}{c_{11}c_{22}-c_{12}c_{21}},$ $k_{12}=-\dfrac{c_{12}}{c_{11}c_{22}-c_{12}c_{21}},$ $ k_{21}= \dfrac{c_{21}}{c_{12}c_{21}-c_{22}c_{11}},$  and $k_{22}=-\dfrac{c_{11}}{c_{12}c_{21}-c_{22}c_{11}}.$
%We observe that different analytical and numerical methods were developed to solve the  two-component  system of time-fractional diffusion equations in $(2+1)$-dimensions, and its generalizations arise naturally in order to analyze the behaviour of the corresponding physical phenomena that occur in various disciplines of science and engineering, such as the population dynamics, chemical reactions, pattern formations and so on.
\\
We know that the study of exact solutions for FDEs is an essential requirement due to their  crucial role in describing the quantitative-qualitative properties and the long-term asymptotic behaviour of the physical system under consideration.
Note that the Riemann-Liouville and the Caputo fractional derivatives fail to obey the standard  forms of the Leibniz rule, the semigroup property and  the chain rule, which can be called the unusual (non-standard) properties. Thus, finding the exact solutions for FDEs that overcome these unusual properties is a very challenging and difficult task, which manifests when one tries to solve  nonlinear TFPDEs.
  In the literature,  many numerical and analytical methods have been discussed to solve the FDEs, including  the Adomian decomposition method \cite{gejji2005},
the   homotopy perturbation method \cite{jafari2007}, the  new hybrid method \cite{ma2020}, {the Lie symmetry method \cite{bakkyaraj2015,sahadevan2017b,prakash2017,nass2019,prakash2021,sethu2021,bakkyaraj2020,paper1,paper2,paper3,paper4},}  separation of the  variable method \cite{wu2018,rui2020,rui2022a,rui2022b,xu2023,uma2},
and  the invariant subspace method \cite{uma1,gazizov2013,harris2013,sahadevan2015,sahadevan2016, sahadevan2017a,choudary2017,harris2017,prakash2020a,rui2018,prakash2020b,choudary2019a,priyendhu2023,prakash2022a,choudary2019b,prakash2022b,gara2021,prakash2023,chu2022,kader2021,prakash2024}.
However, no well-defined analytic methods are available to solve  {nonlinear  systems} of FPDEs systematically,  mainly when one looks for the exact solutions of  systems of FPDEs in higher dimensions.
  To the best of our knowledge, the invariant subspace method combined with variable transformation has not been developed so far to solve a system of {nonlinear two-component system} of TFPDEs in $(2+1)$-dimensions.

It is well-known that the  invariant subspace method,  initially proposed by Galaktionov and Svirshchevski \cite{Galaktionov2007}, is a powerful analytic tool to solve nonlinear evolution PDEs of integer order.  Ma \cite{ma2012a} and many others \cite{ma2012b,ma2012c,ye2014,qu2009,song2013,zhu2016,prakash2019}  investigated various significant extensions and theoretical improvements of this method to solve the general class of nonlinear scalar and  systems of integer-order PDEs in the literature.  Later, the invariant subspace method was extended  to solve FPDEs initially by Gazizov and  Kasatkin \cite{gazizov2013}.
It should be noted that the invariant subspace method is known as the generalized separation of the variable method, which can be efficiently applied to get generalized separable exact solutions for the various classes of  FPDEs \cite{gazizov2013,uma1,uma2,harris2013,sahadevan2015,sahadevan2016, sahadevan2017a,choudary2017,harris2017,prakash2020a,rui2018,prakash2020b,choudary2019a,priyendhu2023,prakash2022a,choudary2019b,prakash2022b,gara2021,prakash2023,chu2022,prakash2024}.
{
Additionally, we can see that many investigations were made to study the explicit solutions for nonlinear integer-order evolution PDEs using different techniques such as invariant conditions, group-invariance, scaling, and similarity reductions \cite{Galaktionov2007}. Note that the self-similar solutions, specifically the soliton-type solutions for the given nonlinear system of integer-order PDEs, help one to investigate the singularity patterns and asymptotic behavior of the considered physical system. Ma and others have investigated the  blow-up solutions, soliton solutions, and multiple wave solutions for many complex nonlinear systems of PDEs such as the generalized coupled multiple components nonlinear Schrödinger and the modified Korteweg-de Vries  type equations \cite{ma2010phys,ma2020opt,ma2023chinjphys,ma2023romjphys,ma2023romrepphys,ma2023theormathphys,Galaktionov2007}.  
It is interesting to note that invariant subspace method can be used to obtain soliton solutions for integer-order PDEs. For example, consider the well-known integrable integer-order Korteweg-de Vries   equation \cite{Galaktionov2007} of the form 
\begin{eqnarray}
	\dfrac{\partial u}{\partial t}=-6u	\dfrac{\partial u}{\partial x}-	\dfrac{\partial^3 u}{\partial x^3},
\end{eqnarray}
 which under the substitution $u=2\dfrac{\partial^2 (\text{ln}|v|)}{\partial x^2}$ reduces into the following form
\begin{eqnarray}\label{redkdv}
	v\dfrac{\partial^2 v}{\partial x\partial t }-	\dfrac{\partial v}{\partial t}	\dfrac{\partial v}{\partial x}+v	\dfrac{\partial^4 u}{\partial x^4}-4	\dfrac{\partial v}{\partial x}	\dfrac{\partial^3 v}{\partial x^3}+3\left( 	\dfrac{\partial^2 u}{\partial x^2}\right)^2 =0.
\end{eqnarray}
In \cite{Galaktionov2007}, it is observed that the reduced equation \eqref{redkdv}
admits the linear space  $$ \mathit{W}_1=\mathcal{L}\{1,e^{ax}\},a\in\mathbb{R}.$$
Thus,   the 1-soliton solution with respect to the linear space $\mathit{W}_1$ obtained in \cite{Galaktionov2007} as:
\begin{eqnarray}
	v(x,t)=K(t)+bK(t)e^{ax-{a^3}t},
\end{eqnarray}
where $K(t)\neq0$ is an arbitrary smooth function and $b\in\mathbb{R}.$
However, there exist only a few investigations were available to study the soliton-type solutions for nonlinear fractional PDEs.
In \cite{rui2022comments}, Rui commented that  for time-fractional PDEs with time derivatives involving the Caputo and the Riemann-Liouville fractional-order derivatives it may not be possible to find the soliton solutions using existing methods in the literature.   In future, we can expect to develop different methods to find soliton-type self-similar solutions for complex nonlinear systems of fractional PDEs with the Riemann-Liouville and the Caputo fractional-order derivatives.}
\subsection{Novelty}\label{Intro.2}
In this work, our main aim is to develop the invariant subspace method combined with variable transformation in a systematic and algorithmic way to solve the {nonlinear two-component system} of TFPDEs in $ (2+1) $-dimensions for the first time. We know that,  in general, the following particular forms of the exact solutions $\mathbf{U}=(u_1,u_2)$ with $u_s=u_s(x_1,x_2,t),s=1,2,$  can be expected for the integer-order {nonlinear two-component system} of PDEs in $ (2+1) $-dimensions:
\begin{itemize}
\item[(1)] Travelling-wave solution:
\begin{equation}
	\label{tws}
	u_s(x_1,x_2,t)=F_s(a_1x_1+a_2x_2+a_0t) ,s=1,2.
\end{equation}
\item[(2)]Functional separable solution:
 \begin{equation}
	\label{fss}
	u_s(x_1,x_2,t)=F_s(v),v=\sum\limits^{2}_{i=1}\sum\limits^{n_{s,i}}_{r_i=1}\varPhi^s_{r_1r_2}(t)\varPsi^s_{1r_1}(x_1)\varPsi^s_{2r_2}(x_2), n_{s,i}\in\mathbb{N},i,s=1,2.
\end{equation}
\item[(3)] Generalized separable solutions:
 \begin{eqnarray}
	\label{gssa}
%\begin{aligned}
&	(i)\,\, u_s(x_1,x_2,t)=\sum\limits^{2}_{i=1}\sum\limits^{n_{s,i}}_{r_i=1}\varPhi^s_{r_1r_2}(t)\varPsi^s_{1r_1}(x_1)\varPsi^s_{2r_2}(x_2), n_{s,i}\in\mathbb{N},i,s=1,2,	\&
\\\
	&
		\label{gssb}	(ii)\,\, u_s(x_1,x_2,t)=\sum\limits^{n_s}_{r=1}\varPhi^s_r(t)\varPsi^s_{r}(a_1x_1+a_2x_2), a_i\in\mathbb{R},n_s\in\mathbb{N},i,s=1,2,\quad
\end{eqnarray}
\end{itemize}
where the functions $\varPsi^s_{ir_i}(x_i),i=1,2,$ and $\varPhi^s_{r_1r_2}(t)$ that depend on the considered system must be determined. We observe that the above forms of the exact solutions \eqref{tws} and \eqref{fss} cannot be obtained  for the {nonlinear two-component system} of PDEs of fractional-order in $ (2+1) $-dimensions due to the unusual properties of the Caputo fractional derivative. However, we wish to point out that the particular form of the generalized separable exact solution  given in \eqref{gssa} for  the {nonlinear two-component system} of TFPDEs in $ (2+1) $-dimensions has been studied using the invariant subspace method \cite{choudary2019a}.
Thus, the main aim of this work is  to give the systematic procedure of the invariant subspace method combined with variable transformation for finding the generalized separable exact solution as given in  \eqref{gssb} for the  {nonlinear two-component system} of TFPDEs in $ (2+1) $-dimensions.

%%%% 	Original RD system
Another aim of this work is to effectively apply the invariant subspace method combined with variable transformation to find generalized separable exact solutions of the above-given type \eqref{gssb} for  the  following generalized  {nonlinear two-component system} of TFRDEs in $(2+1)$-dimensions having the form
\begin{eqnarray}
\begin{aligned}{\label{2+1rd}}
	\dfrac{\partial^{\alpha_s} u_s}{\partial t^{\alpha_s}}=
	&	\dfrac{\partial }{\partial x_1}\left( f_{s1}(u_1,u_2)\frac{\partial u_1}{\partial x_1}+f_{s2}(u_1,u_2)\frac{\partial u_2}{\partial x_1}\right)
 \\
 &+\dfrac{\partial }{\partial x_2}\left( g_{s1}(u_1,u_2)\frac{\partial u_1}{\partial x_2}
 +g_{s2}(u_1,u_2)\frac{\partial u_2}{\partial x_2}\right)
	+h_s(u_1,u_2),
	\alpha_s\in(0,2],s=1,2,
	\end{aligned}
\end{eqnarray}
{along with } the initial and the Dirichlet boundary conditions as follows:
\begin{eqnarray}
&	\begin{array}{ll}
		\label{2+1:ic}
\begin{array}{rll}\text{Initial Conditions: }(a)&	u_s(x_1,x_2,0)=\delta_{s1}(x_1,x_2) \text{ if }\alpha_s\in(0,1],s=1,2,
	 \\ (b) &  u_s(x_1,x_2,0)=\delta_{s1}(x_1,x_2), \&
	 	\\
	 &\dfrac{\partial u_s}{\partial t} \big{|}_{t=0}=\delta_{s2}(x_1,x_2) \text{ if } \alpha_s\in(1,2],s=1,2,
 \end{array}\end{array}
\\
 &\begin{array}{rll}
		\label{2+1:bc}
\text{Boundary Conditions: }	(i)&\,	u_s(x_1,x_2,t)\Big{|}_{\kappa_1x_1+\kappa_2x_2=0}=\tau_{s1}(t),s=1,2,\label{bc}
\\
		(ii)&\,		u_s(x_1,x_2,t)\Big{|}_{\kappa_1x_1+\kappa_2x_2=r}=\tau_{s2}(t),s=1,2,\qquad\qquad\qquad
	\end{array}
\end{eqnarray}
where $f_{si}(u_1,u_2),g_{si}(u_1,u_2)$ and $h_s(u_1,u_2),s,i=1,2,$ represent the  coefficients of diffusion and reaction kinetics, respectively, of the corresponding physical system, $u_s=u_s(x_1,x_2,t),t>0,$ $x_i,\kappa_i,s,i=1,2, r\in\mathbb{R},$ and $\dfrac{\partial^{\alpha_s} }{\partial t^{\alpha_s}}(\cdot),s=1,2,$ denotes the $ \alpha_s$-th order fractional derivative \eqref{caputo} with respect to $t$ in the Caputo sense.
The given {nonlinear two-component system} of TFRDEs \eqref{2+1rd} includes various significant physical processes in both classical  and fractional orders, such as the Lotka-Volterra system \cite{cherniha2017}, prey-predatory models \cite{cherniha2017,murray2002},  Belousov-Zhabotinskii system \cite{cherniha2017},  activator-inhibitor systems\cite{cherniha2017,murray2002}, and the  system of time-fractional Nernst-Planck equations \cite{yang2019}. Note that such systems play a crucial role in  different fields of science and engineering,   such as in chemical kinetics, oceanic waves, climatic variations, models of visco-elastic fluids, porous medium diffusion,  and many more. Additionally,  we note that the above-mentioned system \eqref{reduced-zhang-model} is a particular case of the considered system \eqref{2+1rd}.
%, for which Zhang and Li  \cite{zhang2017} has used numerical Laplace transform technique to study the solution.
Recently, Choudary et al. \cite{choudary2019a} have investigated the exact solutions of the form \eqref{gssa} for the particular class  of the above-system of TFRDEs \eqref{2+1rd} through the invariant subspace method. To the best of our knowledge, no  one has developed an invariant subspace method combined with variable transformation to solve the generalized {nonlinear two-component system} of TFPDEs  in $(2+1)$-dimensions. More specifically,
 by using  the invariant subspace method combined with variable transformation, we aim to find  exact solutions of the form \eqref{gssb} for the given generalized {nonlinear two-component system} of TFRDEs \eqref{2+1rd} along with the initial and the non-zero Dirichlet boundary conditions for the first time.
\subsection{Structure of the article}\label{Intro.3}
%%%%%%%%%%%%%%%%%%%%%%%%%%%%%%%%%%%%%%%%%%%%%%%%%%%%%%%%%%%%%%%%%%%%%%%%%%%%%%%%%%%%%%%%%%%%%%%%%%%%%%%%%%%%%%%%%%%%%%%%%%%%
This paper is structured as follows:
In section \ref{theory}, we provide the systematic description of how to develop  the invariant subspace method combined with variable transformation for a  {nonlinear two-component system} of TFPDEs in $ (2+1) $-dimensions. Additionally, we explain how to construct finite-dimensional invariant linear product space for the {nonlinear two-component system} of TFPDEs in $ (2+1) $-dimensions and, thus, to find its exact solution. Next, we will show the applicability and efficacy of the developed method through the {nonlinear two-component system} of TFRDEs  \eqref{2+1rd} in $ (2+1) $-dimensions in section \ref{TFRDEs}. Also, we present a  systematic study to find the different types of invariant linear product spaces for the system \eqref{2+1rd}, and explain how to obtain generalized separable exact solutions of the initial and boundary value problems associated with  \eqref{2+1rd} using the invariant linear product spaces. Finally, in section \ref{Discussions-conclusions}, we provide the importance of the discussed system of TFRDEs with their various physical phenomena and significance. Additionally, we briefly present the concluding remarks of the developed method and the obtained results of our work.
%
%%%%%%%%%%%%%%%%%%%%%%%%%%%%%%%%%%%%%%%%%%%%%%%%%%%%%%%%%%%%%%%%%%%%%%%%%%%%%%%%%%%%%%%%%%%%%%%%%%%%%%%%%%%%%%%%%%%%%%%%%%
\section{{Nonlinear two-component system} of TFPDEs in $ (2+1) $-dimensions: Invariant subspace method combined with variable transformation}\label{theory}
This  section provides a detailed systematic investigation of obtaining the invariant linear product spaces for the following   {nonlinear two-component system} of TFPDEs in $ (2+1) $-dimensions having the form
\begin{eqnarray}
	\begin{aligned}
		\label{coupledsystem}
		&	\left(\dfrac{\partial^{\alpha_1}u_1}{\partial t^{\alpha_1}}, \dfrac{\partial^{\alpha_2}u_2}{\partial t^{\alpha_2}} \right) =\mathbf{\hat{H}}\big(\mathbf{U}\big) \equiv \Big({{H}_1(u_1,u_2)},{{H}_2(u_1,u_2)}\Big),\alpha_s>0,
	\end{aligned}
\end{eqnarray}
{where} $ \dfrac{\partial^{\alpha_s} }{\partial t^{\alpha_s}} (\cdot)$ is the $ \alpha_s$-th order  Caputo  fractional derivative given in \eqref{caputo},  $\mathbf{U}=(u_1,u_2),$  $ u_s=u_s(x_1,x_2,t),s=1,2,$ and $ {{H}_s(u_1,u_2)},s=1,2,$ is the $m_s$-th order sufficiently smooth  partial differential operator, that is
$$	\small
{{H}_s(u_1,u_2)}
=
{{H}_s}\left(x_1,x_2,u_1,u_2,
\dfrac{\partial u_1}{\partial x_{1}},
\dfrac{\partial u_2}{\partial x_{1}},
\dfrac{\partial u_1}{\partial x_{2}},
\dfrac{\partial u_2}{\partial x_{2}},
\dfrac{\partial^2 u_1}{\partial x_{1}\partial x_{2}},
\dfrac{\partial^2 u_2}{\partial x_{1}\partial x_{2}}, \dots, \dfrac{\partial^{m_s} u_1}{\partial x_{1}^i\partial x_{2}^j},\dfrac{\partial^{m_s} u_2}{\partial x_{1}^i\partial x_{2}^j}\right),$$
with $i+j=m_s,i,j=0,1,\dots,m_s,m_s\in\mathbb{N},s=1,2.$
\\
Now, we consider  the following particular form of  the exact solution  for the given {nonlinear two-component system}  \eqref{coupledsystem} as
\begin{eqnarray}
	\begin{aligned}\label{transformation}
		&u_s(x_1,x_2,t)=v_s(z,t), z=\kappa_1x_1+\kappa_2x_2,s=1,2,
	\end{aligned}
\end{eqnarray}
where  $\kappa_1,\kappa_2\in\mathbb{R}.$
Substituting   the above particular form \eqref{transformation} into the given {nonlinear two-component system} \eqref{coupledsystem}, we obtain the   reduced  {nonlinear two-component system} of TFPDEs in $ (1+1) $-dimensions as follows:
\begin{eqnarray}
	\begin{aligned}
		\label{transformedcoupledsystem}
		&	\left(\dfrac{\partial^{\alpha_1}v_1}{\partial t^{\alpha_1}}, \dfrac{\partial^{\alpha_2}v_2}{\partial t^{\alpha_2}} \right) =\mathbf{\hat{F}}\big(\mathbf{V}\big) \equiv \Big({{F}_1(v_1,v_2)},{{F}_2(v_1,v_2)}\Big), \alpha_s>0,
	\end{aligned}
\end{eqnarray}
{where}  $ {{F}_s(v_1,v_2)},s=1,2,$  is the reduced $m_s$-th order sufficiently smooth ordinary  differential operator, which reads  as follows:
$$	
{{F}_s(v_1,v_2)}
=
{{F}_s}\left(z,v_1,v_2, \dfrac{\partial  v_1}{\partial z}, \dfrac{\partial  v_2}{\partial z},\dots, \dfrac{\partial ^{m_s} v_1}{\partial z^{m_s}},\dfrac{\partial ^{m_s} v_2}{\partial z^{m_s}}\right).
$$
%and $ \dfrac{\partial^{\alpha_s} }{\partial t^{\alpha_s}}(\cdot) $ is the $ \alpha_s$-th order Caputo fractional derivative  for $ s=1,2.$

Next, applying the invariant subspace method to the reduced {nonlinear two-component system} \eqref{transformedcoupledsystem}, we systematically explain how to obtain the  generalized separable exact solution for the given  {nonlinear two-component system} \eqref{coupledsystem}
in the following form
\begin{equation}
	\label{solutionform}
	\mathbf{U}(x_1,x_2,t):=\mathbf{V}(z,t)
	=\left(
	\sum\limits_{i=1}^{n_1}\phi_{1,i}(t)\varphi_{1,i}(z),\sum\limits_{i=1}^{n_2}\phi_{2,i}(t)\varphi_{2,i}(z)
	\right) ,z=\kappa_1x_1+\kappa_2x_2,
\end{equation}
where $\mathbf{V}(z,t)=(v_1(z,t),v_2(z,t))$ is the vector-valued function,  and  $\phi_{s,i}(t),\varphi_{s,i}(z),$ $i=1,2,\dots,n_s,n_s\in\mathbb{N},$ are the  unknowns functions, which need to be found using the invariant subspace method. The following steps summarize the details of the working methodology followed in this work.
\begin{itemize}
	\item[Step-1:] First, we substitute the particular form \eqref{transformation} into
	 the given  {nonlinear two-component system} of TFPDEs  \eqref{coupledsystem} in $ (2+1) $-dimensions and reduce it into a    {nonlinear two-component system} of TFPDEs in $ (1+1) $-dimensions as given in  \eqref{transformedcoupledsystem}.
	\item[Step-2:]
Next, we find the invariant linear product spaces for the  reduced {nonlinear two-component system} of TFPDEs \eqref{transformedcoupledsystem} in $ (1+1) $-dimensions using the appropriate invariant conditions.
	\item[Step-3:]
The obtained invariant linear product spaces will allow us to convert the reduced {nonlinear two-component system} of TFPDEs \eqref{transformedcoupledsystem} in $(1+1)$-dimensions into a system of time-fractional ODEs.
	\item[Step-4:] Finally, we can solve the obtained system of time-fractional ODEs in step-3 using the well-known analytical methods for obtaining the generalized separable exact solution for the given system \eqref{coupledsystem} as expected in the form of \eqref{solutionform}.
\end{itemize}
Now, we provide a systematic study of  each of the steps given above. So, let us first explain how to construct the  invariant linear product spaces $	\mathbf{W}_{n} \,(n<\infty,n\in\mathbb{N})$  for the  reduced {nonlinear two-component system}  \eqref{transformedcoupledsystem} in the following subsection.
\subsection{Invariant linear product spaces $	\mathbf{W}_{n}$ for the given system \eqref{transformedcoupledsystem}:}
Let us introduce the  $n_s$-th order homogeneous linear ODEs as follows:
\begin{eqnarray}
	\begin{aligned}
		\label{LHODEs}
		D_{s,n_s}[y_s(z)]= y^{(n_s)}_s(z)+\xi_{s,n_s-1}(z)y^{(n_s-1)}_s(z) +\dots
		+\xi_{s,1}(z)y^{(1)}_s(z)
		+\xi_{s,0}(z)y_s(z)=0,
	\end{aligned}
\end{eqnarray}
where $y^{(i)}_s(z)=\dfrac{d^{i} y_s(z)}{ d z^{i}},i=1,2,\dots,n_s,$ and  the functions $\xi_{s,j}(z),j=0,1,2,\dots,n_s-1,n_s\in\mathbb{N},$ $s=1,2,$  are arbitrary continuous functions of $ z.$
From this, we can  define the solution space of the above homogeneous linear ODEs $ D_{s,n_s}[y_s(z)]=0,s=1,2,$ as  follows:
\begin{eqnarray}
	\label{component linear spaces}
	\begin{aligned}
		\mathit{W}_{s,n_s}=&\mathcal{L}(\mathcal{B}_{s})
			%\\	=		&\text{Span}\left\lbrace		\varphi_{s,1}(z),\varphi_{s,2}(z),\dots,		\varphi_{s,n_s}(z),\right\rbrace		\\
		=
		\left\{\psi_s(z)=\sum\limits_{i=1}^{n_s}k_{s,i}\varphi_{s,i}(z)\, \big{|} k_{s,i}\in\mathbb{R}, D_{s,n_s}[\psi_s(z)] =0\right\}
		,
	\end{aligned}
\end{eqnarray}
where  the set
$\mathcal{B}_{s}= \left\lbrace
\varphi_{s,i}(z){|} i=1,2,\dots,n_s\right\rbrace
$ is the basis for the solution space $\mathit{W}_{s,n_s},$  which consists of $n_s$-linearly independent solutions of the given system of homogeneous linear ODEs \eqref{LHODEs} for $s=1,2,$ and $\mathcal{L}(\cdot)$ represents the linear span of the functions in the set  $\mathcal{B}_{s},s=1,2,$ over $\mathbb{R}.$

Now, we define the $n$-dimensional linear product space $	\mathbf{W}_{n}$   as follows:
\begin{eqnarray}\label{invariantproductspaces}
	\begin{aligned}
		\mathbf{W}_{n} =&\mathit{W}_{1,n_1}\times\mathit{W}_{2,n_2}
			=\left\{\psi(z)=(\psi_1(z),\psi_2(z))\big{|}\psi_s(z)\in\mathit{W}_{s,n_s},s=1,2 \right\},
	\end{aligned}
\end{eqnarray}
where the component linear spaces $\mathit{W}_{s,n_s},s=1,2,$ are given in \eqref{component linear spaces}.
The  following theorem presents the relation between the dimension of a linear product space and the dimensions of its component linear spaces.
\begin{theorem}\cite{Axler2014}
	\label{productspacedimensiontheorem}
Suppose that the
 Cartesian product space 	\begin{eqnarray}\begin{aligned}
		\label{catresianproduct}
		\tilde{\mathbf{W}}_n&
		= \tilde{\mathit{W}}_{1,n_1}\times
		\tilde{\mathit{W}}_{2,n_2}
		=\{\mathbf{w}=(w_1,w_2)\big{|} \, w_s\in \tilde{\mathit{W}}_{s,n_s} ,s=1,2 \}
	\end{aligned}
\end{eqnarray} is defined under the   component-wise addition  `{$ +$}' and scalar multiplication  `{$\cdot$}'
as:
	%	$\mathbf{v}=(v_1,v_2),\mathbf{w}=(w_1,w_2)\in \tilde{\mathbf{W}}_n,$ and $ K\in\mathbb{R}, $ we have
	\begin{eqnarray*}
		\begin{aligned}
			&	\mathbf{v}+\mathbf{w}=(v_1 + w_1,v_2 + w_2),\forall\,\mathbf{v}=(v_1,v_2),\mathbf{w}=(w_1,w_2)\in \tilde{\mathbf{W}}_n,\\
			&	K \cdot 	\mathbf{v} =(K\cdot v_1,K\cdot v_2),\forall\,K\in\mathbb{R}, \,\forall\,\mathbf{v}=(v_1,v_2)\in \tilde{\mathbf{W}}_n.
		\end{aligned}
	\end{eqnarray*}
	Then,  the dimension of the Cartesian product space $\tilde{\mathbf{W}}_n$ is   $n=n_1+n_2,$ where $n_1$ and $n_2$ are the dimensions of the linear spaces $\tilde{\mathit{W}}_{1,n_1}$ and $\tilde{\mathit{W}}_{2,n_2},$ respectively.
\end{theorem}

\begin{definition}\label{invariant conditions}
	The $n$-dimensional linear product space $	\mathbf{W}_{n}$ given in \eqref{invariantproductspaces} is called the invariant linear product space of the   nonlinear vector differential operator $\mathbf{\hat{F}}\big(\mathbf{V}\big) $  given in \eqref{transformedcoupledsystem} if either  one of the following conditions holds:
	\begin{itemize}
		\item[(a)]
		$\mathbf{\hat{F}}\big(\mathbf{W}_{n}\big)\subseteq\mathbf{W}_{n}$ that is, for every $\mathbf{V}=(v_1,v_2)\in \mathbf{W}_{n},$ we have $\mathbf{\hat{F}}(\mathbf{V})\in \mathbf{W}_{n}.$
		\item[(b)] For every $v_s\in\mathit{W}_{s,n_s}, $ $\mathit{{F}}_s\big(\mathbf{V}\big)=\mathit{{F}}_s\big((v_1,v_2)\big)\in\mathit{W}_{s,n_s},s=1,2.$ Equivalently, $\mathit{{F}}_s\big(\mathbf{W}_{n}\big)\subseteq\mathit{W}_{s,n_s},s=1,2.$
		\item[(c)]  For all $ \mathbf{V}=(v_1,v_2)$, it is true that
		$D_{s,n_s}\left[ \mathit{{F}}_s\big(\mathbf{V}\big)\right] =0$ whenever $D_{s,n_s}[v_s]=0$ for each $s=1,2.$
	\end{itemize}
\end{definition}
Thus,  using  Theorem \ref{productspacedimensiontheorem}, we can find the dimension of the invariant linear product space $\mathbf{W}_{n},$  that is, dim$(\mathbf{W}_{n})=$ dim$(\mathit{W}_{1,n_1})+$dim$(\mathit{W}_{1,n_1})=n(=n_1+n_2).$
Additionally, we would like to mention  that  the possible maximal dimension of the linear space to be invariant under the given nonlinear differential operator $\mathbf{\hat{F}}\big(\mathbf{V}\big)$ plays a crucial role in estimating and completing the classification of  invariant linear spaces for a nonlinear  differential operator under consideration.
 \begin{definition}
 	The vector differential operator $\mathbf{\hat{F}}\big(\mathbf{V}\big) = \Big({{F}_1(v_1,v_2)},{{F}_2(v_1,v_2)}\Big)$ given in \eqref{transformedcoupledsystem}
 	is said to be nonlinear and really coupled if it satisfies the following
 		$$ \begin{aligned}
 		(i)&
 		\dfrac{\partial^2 {{F}_s} }{\partial v_{p}^{(i)}(z) \, \partial v_{q}^{(j)}(z)}\neq0, i,j=0,1,\dots,m_s,s,p,q
 		=1,2,
 		 		\text{{ and}}&\\
 			(ii)& 	\sum\limits_{i=0}^{m_1}\left( \dfrac{\partial \mathit{{F}}_1 }{\partial\, v_2^{(i)}(z)}\right)^2\neq 0,
 			\text{ and }
 				\sum\limits_{i=0}^{m_2}\left( \dfrac{\partial \mathit{{F}}_2 }{\partial\, v_1^{(i)}(z)}\right)^2\neq 0, \text{ respectively, }
 			 		\end{aligned}
  	$$
  		where $v_{p}^{(0)}(z)=v_{p}(z)$ and $ v_p^{(i)}(z)=\dfrac{\partial^{i} v_p}{\partial z^{i}},i=1,2,\dots,m_s,p,s=1,2.$	
 \end{definition}
\begin{definition}
	The order of the differential operators ${{F}_s(v_1,v_2)},s=1,2,$ is said to be $m_s$ if it satisfies the following:
	$$\begin{aligned} 	&\left( \dfrac{\partial \mathit{{F}}_s }{\partial\, v_1^{(m_s)}(z)}\right)^2+\left( \dfrac{\partial \mathit{{F}}_s }{\partial\, v_2^{(m_s)}(z)}\right)^2\neq 0, v_i^{(m_s)}(z)=\dfrac{\partial^{m_s} v_i}{\partial z^{m_s}},i,s=1,2.
		\end{aligned}
	$$
\end{definition}
Next, the  following theorem presents the maximal dimensions  possible for  linear product space $	\mathbf{W}_{n}$ to be invariant under  the nonlinear vector  differential operator $\mathbf{\hat{F}}\big(\mathbf{V}\big) .$
\begin{theorem}\cite{qu2009,song2013,sahadevan2017a}
	Let the vector  differential operator $\mathbf{\hat{F}}\big(\mathbf{V}\big)= (\mathit{{F}}_1(v_1,v_2),\mathit{{F}}_2(v_1,v_2)) $ given in \eqref{transformedcoupledsystem} be  really coupled and nonlinear.
Also, let the orders of differential operators  $\mathit{{F}}_s(v_1,v_2)$ be $m_s,s=1,2.$
Moreover,  if the  vector  differential operator $\mathbf{\hat{F}}\big(\mathbf{V}\big)$ admits the linear product space $	\mathbf{W}_{n}=\mathit{W}_{1,n_1}\times\mathit{W}_{2,n_2}$ given in \eqref{invariantproductspaces},	 then there holds
the bounds	$0\leq n_1-n_2\leq m_2, \text{and } n_1\leq2(m_1+m_2)+1, $ where $n_1,n_2\in\mathbb{N}.$
%{	where $n_s$  is the dimension of component space $\mathit{W}_{s,n_s}$ of the product space $	\mathbf{W}_{n}=\mathit{W}_{1,n_1}\times\mathit{W}_{2,n_2}$ for each $s=1,2.$}
	\label{maximalthm1}
\end{theorem}
\subsection{Construction  of the generalized separable exact solutions \eqref{solutionform} for the given {nonlinear two-component system} \eqref{coupledsystem}}
This subsection provides a detailed study of how to obtain the particular form of the   generalized separable exact solutions   \eqref{solutionform} for the {nonlinear two-component system} \eqref{coupledsystem}  using the above-discussed invariant linear product spaces $	\mathbf{W}_{n}$ given in \eqref{invariantproductspaces}.

Let the nonlinear vector  differential operator $\mathbf{\hat{F}}\big(\mathbf{V}\big) $   admit the invariant linear product space $	\mathbf{W}_{n}$  given  in  \eqref{invariantproductspaces}.
Then, using the definition \ref{invariant conditions},  we obtain   $\mathit{{F}}_s(v_1,v_2)\in \mathit{W}_{s,n_s},$ for every $v_s\in \mathit{W}_{s,n_s},s=1,2,$
which can be deduce from the fact that  the vector differential operator  $\mathbf{\hat{F}}\big(\mathbf{V}\big) $ admits the following invariant conditions:
\begin{equation}\label{transformedinvariantcriteria}
	D_{s,n_s}\left[ \mathit{{F}}_s\big(\mathbf{V}\big)\right] =0,s=1,2,\text{  whenever } D_{s,n_s}[v_s]=0 ,s=1,2.
\end{equation}
%This can be rewritten as follows:
The above invariant conditions \eqref{transformedinvariantcriteria} read as follows:
\begin{eqnarray}\label{transformedinvariantcondition}
	\begin{aligned}
		& \mathit{{F}}_s\left( \sum\limits_{i=1}^{n_1}k_{1,i}\ \varphi_{1,i}(z),\sum\limits_{i=1}^{n_2}k_{2,i}\ \varphi_{2,i}(z)\right)  =\sum\limits_{i=1}^{n_s} \mathbf{K}_{s,i}(k_{1,1},k_{2,1},\dots,k_{1,n_1},k_{2,n_2})\varphi_{s,i}(z) \in \mathit{W}_{s,n_s},
		\\
	&	\text{ whenever  }  	v_s=\sum\limits_{i=1}^{n_s}k_{s,i}\ \varphi_{s,i}(z)\in\mathit{W}_{s,n_s},k_{s,i}\in\mathbb{R},s=1,2,  	
	\end{aligned}
\end{eqnarray}
where the quantities $\mathbf{K}_{s,i}(\cdot)$ are the expansion coefficients corresponds to the basis $\mathcal{B}_{s},$ for each $s=1,2.$

Next,  the following theorem shows that the  nonlinear vector  partial  differential operator $\mathbf{\hat{H}}\big(\mathbf{U}\big) $ given in \eqref{coupledsystem} and the corresponding reduced  nonlinear  vector ordinary differential operator $\mathbf{\hat{F}}\big(\mathbf{V}\big) $ given in \eqref{transformedcoupledsystem} admit the same invariant linear product space $	\mathbf{W}_{n}$  given  in  \eqref{invariantproductspaces}  along with $u_s(x_1,x_2)=v_s(z),z=\kappa_1x_1+\kappa_2x_2,s=1,2.$
\begin{theorem}
	\label{theorem:operators}
	Let
	$	\mathbf{W}_{n} =\left\{\psi(z)=(\psi_1(z),\psi_2(z))\big{|}\psi_s(z)\in\mathit{W}_{s,n_s},s=1,2 \right\}$ be a linear product space
	with the component linear spaces $\mathit{W}_{s,n_s},s=1,2,$ as given in \eqref{component linear spaces}. Then, 	the	 nonlinear  vector partial differential operator
	$\mathbf{\hat{H}}\big(\mathbf{U}\big) $
given in	\eqref{coupledsystem}
% along with $\mathbf{U}=(u_1,u_2)$ having the particular form \eqref{transformation}
	admits the invariant  linear product space  $	\mathbf{W}_{n}$ if and only if  $	\mathbf{W}_{n}$ is  an invariant linear product space for the reduced  nonlinear vector  ordinary  differential operator $\mathbf{\hat{F}}\big(\mathbf{V}\big) $
	given in \eqref{transformedcoupledsystem} along with $u_s(x_1,x_2)=v_s(z),z=\kappa_1x_1+\kappa_2x_2.$
\end{theorem}
\begin{proof}
	Assume that  $u_s(x_1,x_2)=v_s(z),z=\kappa_1x_1+\kappa_2x_2,$ $s=1,2,$ and $\mathbf{\hat{H}}\big(\mathbf{U}\big) $ admits the invariant linear product space $ \mathbf{W}_n. $
		Then, for every $\mathbf{U}=(u_1,u_2)\in 	\mathbf{W}_{n},$ we have $\mathbf{\hat{H}}\big(\mathbf{U}\big)\in 	\mathbf{W}_{n}. $
Thus, we obtain
	\begin{eqnarray}
		\begin{aligned}
			\label{eq:H_s-u_s}
			&	{{H}_s(u_1,u_2)}=  \sum\limits_{i=1}^{n_s} \mathbf{K}_{s,i}(k_{1,1},k_{2,1},\dots,k_{1,n_1},k_{2,n_2})\varphi_{s,i}(\kappa_1x_1+\kappa_2x_2)\in \mathit{W}_{s,n_s},s=1,2,
				\end{aligned}
	\end{eqnarray}
whenever $	u_s=\sum\limits_{i=1}^{n_s}k_{s,i}\  \varphi_{s,i}(\kappa_1x_1+\kappa_2x_2)\in \mathit{W}_{s,n_s},s=1,2
,$	where the coefficients $\mathbf{K}_{s,i}(\cdot),s=1,2,\dots,n_s,s=1,2 $ are obtained from the expansion with respect to the basis $\mathcal{B}_s.$
Since the vector ordinary  differential operator $\mathbf{\hat{F}}\big(\mathbf{V}\big) $
	given in \eqref{transformedcoupledsystem} is obtained for the original operator 	$\mathbf{\hat{H}}\big(\mathbf{U}\big) $ under the substitution $u_s(x_1,x_2)=v_s(z),z= \kappa_1x_1+\kappa_2x_2,s=1,2,$
then by	applying the differential consequences $ \dfrac{\partial^{i}u_s(x_1,x_2)}{\partial x_j^{i}} = \kappa_j^i\left( \dfrac{\partial^{i}v_s(z)}{\partial z^{i}}\right) ,s,j=1,2,i\in\mathbb{N},$  we obtain
	\begin{eqnarray}
		\begin{aligned}
			& \mathit{{F}}_s\left(\sum\limits_{i=1}^{n_1}k_{1,i}\ \varphi_{1,i}(z),\sum\limits_{i=1}^{n_2}k_{2,i}\ \varphi_{2,i}(z)\right) =\sum\limits_{i=1}^{n_s} \mathbf{K}_{s,i}(k_{1,1}, \dots,k_{1,n_1},k_{2,n_2})\varphi_{s,i}(z)\in \mathit{W}_{s,n_s},
	\end{aligned}
\label{thm:invarianceconditiontransformed}
	\end{eqnarray}
	{whenever} $	v_s(z)=\sum\limits_{i=1}^{n_s}k_{s,i}\ \varphi_{s,i}(z)\in \mathit{W}_{s,n_s},z= \kappa_1x_1+\kappa_2x_2,k_{s,i}\in\mathbb{R},s=1,2. $ 	
{Hence, under the transformation  $u_s(x_1,x_2)=v_s(z),z=  \kappa_1x_1+\kappa_2x_2,s=1,2,$ we can conclude that $\mathbf{\hat{F}}\big(\mathbf{V}\big)$  admits the linear product space $\mathbf{W}_{n}$ whenever $\mathbf{W}_{n}$ is an invariant linear product space of the partial differential operator $\mathbf{\hat{H}}\big(\mathbf{U}\big). $ }
	\\
Similarly, we can prove the converse part, which completes the proof.
\end{proof}
Now, we show that for the given systems \eqref{coupledsystem} and \eqref{transformedcoupledsystem}, there exists a generalized separable exact solution of the form \eqref{solutionform} whenever the corresponding reduced system  \eqref{transformedcoupledsystem} admits the linear product space $ \mathbf{W}_n . $
%%%%%%%%%%%%%%%%%%%%%%%%%%%%%%%%%%%%%%%%%%%%%%%%%%%%%%%%%%%%%%%%%%%%%%%%%%%%%%%%%%%%%%%%%%%%%%%%%%%%%%%%%%%%%%%%
\begin{theorem}\label{thm:existenceofsolution}
	Let the linear product space
	$ \mathbf{W}_n  $  given in \eqref{invariantproductspaces}  be an invariant linear product space  for the  nonlinear  vector ordinary differential operator $\mathbf{\hat{F}}\big(\mathbf{V}\big)=
	\big(\mathit{{F}}_1(v_1,v_2), \mathit{{F}}_1(v_1,v_2)\big) $
	given in \eqref{transformedcoupledsystem}.
	Then,
	\begin{equation} \label{thm:GSesoftransformedcoupledsystem}
		v_s(z,t)= \sum\limits_{i=1}^{n_s}k_{s,i}(t)\varphi_{s,i}(z),s=1,2,
	\end{equation}
	is a generalized separable exact solution  for the following  {nonlinear two-component system} of TFPDEs \eqref{transformedcoupledsystem} in $ (1+1) $-dimensions
	\begin{eqnarray}
		\begin{aligned}
			\label{thm:transformedcoupledsystem}
			&	\left(\dfrac{\partial^{\alpha_1}v_1}{\partial t^{\alpha_1}}, \dfrac{\partial^{\alpha_2}v_2}{\partial t^{\alpha_2}} \right) =	\big(\mathit{{F}}_1(v_1,v_2), \mathit{{F}}_1(v_1,v_2)\big) ,{\alpha_1},{\alpha_2}>0,
		\end{aligned}
	\end{eqnarray}
	if and only if  the   {nonlinear two-component system} of TFPDEs \eqref{coupledsystem} in $ (2+1) $-dimensions
	\begin{eqnarray}
		\begin{aligned}
			\label{thm:coupledsystemeqn}
			&	\left(\dfrac{\partial^{\alpha_1}u_1}{\partial t^{\alpha_1}}, \dfrac{\partial^{\alpha_2}u_2}{\partial t^{\alpha_2}} \right) = \Big({{H}_1(u_1,u_2)},{{H}_2(u_1,u_2)}\Big),{\alpha_1},{\alpha_2}>0
		\end{aligned}
	\end{eqnarray}
	has the generalized separable exact solution as  follows:
	\begin{eqnarray}
		\begin{aligned}\label{thm:GSesofcoupledsystem}
			u_s(x_1,x_2,t)&=v_s(z,t)
			= \sum\limits_{i=1}^{n_s}k_{s,i}(t)\varphi_{s,i}(z), z=\kappa_1x_1+\kappa_2x_2,s=1,2.
		\end{aligned}
	\end{eqnarray}
Moreover,  the coefficients $  k_{s,i}(t),i=1,2,\dots,n_s, s=1,2 $ satisfy the following system of fractional ODEs
	\begin{equation}\label{thm:systemoffractionalODEs}
		\dfrac{d ^{\alpha_s} k_{s,i}(t) }{d t^{\alpha_s}}=\mathbf{K}_{s,i}(k_{1,1}(t),k_{2,1}(t),\dots,k_{1,n_1}(t),k_{2,n_2}(t)),i=1,2,\dots,n_s,s=1,2.
	\end{equation}
\end{theorem}
\begin{proof}
	Assume that the given   {nonlinear two-component system} of TFPDEs \eqref{thm:transformedcoupledsystem} in $ (1+1) $-dimensions admits a generalized separable exact solution as given in \eqref{thm:GSesoftransformedcoupledsystem}. Now, we consider  the  particular form
  $u_s=u_s(x_1,x_2,t),s=1,2,$  as follows:
	\begin{equation}
		\label{thm:solution form}	u_s(x_1,x_2,t):=v_s(z,t)=\sum\limits_{i=1}^{n_s}k_{s,i}(t)\varphi_{s,i}(z),z= \kappa_1x_1+\kappa_2x_2,s=1,2.
	\end{equation}
We substitute \eqref{thm:solution form} into the   {nonlinear two-component system} of TFPDEs \eqref{thm:coupledsystemeqn} in $ (2+1) $-dimensions.
	Then, applying the linearity property of fractional derivative in the Caputo sense,
	we obtain
	\begin{eqnarray}
		\begin{aligned}
			\label{lhscoupledsystemeqn}
			\dfrac{\partial^{\alpha_s}u_s}{\partial t^{\alpha_s}}=&	\dfrac{\partial^{\alpha_s}}{\partial t^{\alpha_s}}\left( \sum\limits_{i=1}^{n_s}k_{s,i}(t)\varphi_{s,i}(\kappa_1x_1+\kappa_2x_2)\right)  \\=			&	\sum\limits_{i=1}^{n_1} \dfrac{d^{\alpha_s}k_{s,i}(t) }{d  t^{\alpha_s}}\varphi_{s,i}(\kappa_1x_1+\kappa_2x_2),s=1,2.		\end{aligned}
	\end{eqnarray}
	Since $\mathbf{W}_n $ is an invariant linear product space of the given  reduced {nonlinear two-component system} of TFPDEs  \eqref{thm:transformedcoupledsystem} in $ (1+1) $-dimensions, then  using Theorem \ref{theorem:operators}, we obtain that the given   {nonlinear two-component system} of TFPDEs  \eqref{thm:coupledsystemeqn} in $ (2+1) $-dimensions also admits the same linear product space $\mathbf{W}_n $. Thus,
	using the equation \eqref{eq:H_s-u_s}, we obtain
	\begin{eqnarray}\label{thm: H_s-u_s}
		{{H}_s(u_1,u_2)}=  \sum\limits_{i=1}^{n_s} \mathbf{K}_{s,i}(k_{1,1}(t),k_{2,1}(t),\dots,k_{1,n_1}(t),k_{2,n_2}(t))\varphi_{s,i}(\kappa_1x_1+\kappa_2x_2),
	\end{eqnarray}
where $s=1,2.$
Substituting \eqref{lhscoupledsystemeqn} and \eqref{thm: H_s-u_s} into the given {nonlinear two-component system} \eqref{thm:coupledsystemeqn}, and equating the coefficients of the functions $\varphi_{s,i}(\kappa_1x_1+\kappa_2x_2),i=1,2,\dots,n_s,s=1,2,$  we deduce the required condition  \eqref{thm:systemoffractionalODEs}.
	\\	
	Conversely, suppose that the given {nonlinear two-component system} \eqref{thm:coupledsystemeqn} has the generalized separable exact  solution of the form \eqref{thm:GSesofcoupledsystem} and  $\mathbf{W}_n $ is an invariant linear product space of the given{nonlinear two-component system} \eqref{thm:coupledsystemeqn}.
	Then, using  Theorem \ref{theorem:operators}, we find that  $\big(\mathit{{F}}_1(v_1,v_2), \mathit{{F}}_1(v_1,v_2)\big)$ also admits the same invariant linear product space $\mathbf{W}_n. $  	
	Thus, from the equation \eqref{thm:invarianceconditiontransformed}, the component differential operators  $\mathit{{F}}_s(v_1,v_2),s=1,2,$  given in \eqref{thm:transformedcoupledsystem} satisfy the following conditions:
	\begin{eqnarray}\label{thm: F_s-u_s}
		\mathit{{F}}_s(v_1,v_2)=  \sum\limits_{i=1}^{n_s} \mathbf{K}_{s,i}(k_{1,1}(t),k_{2,1}(t),\dots,k_{1,n_1}(t),k_{2,n_2}(t))\phi_{s,i}(z).
	\end{eqnarray}
Using the linearity property of the Caputo fractional derivative,  we obtain
	\begin{eqnarray}
		\begin{aligned}
			\label{lhstransformedcoupledsystemeqn}
			\left( 	\dfrac{\partial^{\alpha_1}v_1}{\partial t^{\alpha_1}},	\dfrac{\partial^{\alpha_2}v_2}{\partial t^{\alpha_2}}\right) =&	
			\left(
			\dfrac{\partial^{\alpha_1}}{\partial t^{\alpha_1}}
			\left( \sum\limits_{i=1}^{n_1}k_{1,i}(t)\varphi_{1,i}(z)
			\right),
			\dfrac{\partial^{\alpha_2}}{\partial t^{\alpha_2}} \left(
			\sum\limits_{i=1}^{n_2}k_{2,i}(t) \varphi_{2,i}(z)
			\right)
			\right)
			\\
			=	&
			\left(
			\sum\limits_{i=1}^{n_1} \dfrac{d^{\alpha_1}k_{1,i}(t) }{d t^{\alpha_1}} \varphi_{1,i}(z),
			\sum\limits_{i=1}^{n_2} \dfrac{d^{\alpha_2}k_{2,i}(t) }{d  t^{\alpha_2}} \varphi_{2,i}(z)
			\right)
		\end{aligned}
	\end{eqnarray}
{when  $v_s=\sum\limits_{i=1}^{n_s}k_{s,i}(t)\varphi_{s,i}(z),z=\kappa_1x_1+\kappa_2x_2.$}
Now, we substitute the equations \eqref{thm:invarianceconditiontransformed} and \eqref{lhstransformedcoupledsystemeqn} into \eqref{thm:transformedcoupledsystem}. Then, by equating the coefficients of the functions $\varphi_{s,i}(z),i=1,2,\dots,n_s,s=1,2,$ to zero, we obtain the required system of fractional ODEs  \eqref{thm:systemoffractionalODEs}.
	Hence, we conclude  that if $u_s(x_1,x_2,t)=v_s(z,t),z= \kappa_1x_1+\kappa_2x_2,s=1,2,$ then the given  reduced {nonlinear two-component system} of TFPDEs \eqref{thm:transformedcoupledsystem} in $ (1+1) $-dimensions admits the generalized separable exact solutions  \eqref{thm:GSesoftransformedcoupledsystem} if and only if  the {nonlinear two-component system} of TFPDEs \eqref{thm:coupledsystemeqn} in $ (2+1) $-dimensions admits the generalized separable  exact solutions   \eqref{thm:GSesofcoupledsystem}.
\end{proof}
The following section presents the applicability of the above-developed invariant subspace method for determining   various types of finite-dimensional invariant linear product spaces $\mathbf{W}_n$ for the     {nonlinear two-component system} of TFRDEs \eqref{2+1rd} in $(2+1)$-dimensions. Additionally,  we give a detailed systematic procedure to find the  generalized separable exact solutions for the given {nonlinear two-component system} of TFRDEs \eqref{2+1rd} using the determined  invariant linear product spaces $\mathbf{W}_n$.
%%%%%%%%%%%%%%%%%%%%%%%%%%%%%%%%%%%%%%%%%%%%%%%%%%%%
%
%
%%%%%%%%%%%%%%%%%%%%%%%%%%%%%%%%%%%%%%%%%%%%%%%%%%%%
%
%
%%%%%%%%%%%%%%%%%%%%%%%%%%%%%%%%%%%%%%%%%%%%%%%%%%%
\section{Invariant linear product spaces and generalized separable exact solutions for the given system \eqref{2+1rd}}\label{TFRDEs}
The nonlinear vector partial differential  operator for the given {nonlinear two-component system} of TFRDEs \eqref{2+1rd} in $(2+1)$-dimensions  reads as follows:
\begin{eqnarray}
	\begin{aligned}
		{\label{2+1rdop}}
		\mathbf{A}(u_1,u_2)=&
		\left(	\mathit{A}_1(u_1,u_2),	\mathit{A}_2(u_1,u_2)
		\right) ,\text{ where } \\
		\mathit{A}_s(u_1,u_2)=&	\sum\limits_{i=1}^2\dfrac{\partial }{\partial x_1}\left( f_{si}(u_1,u_2)\frac{\partial u_i}{\partial x_1}\right)
		 +\sum\limits_{i=1}^2\dfrac{\partial }{\partial x_2}\left( g_{si}(u_1,u_2)\frac{\partial u_i}{\partial x_2}\right)
		+h_s(u_1,u_2),s=1,2.
	\end{aligned}
\end{eqnarray}
Now, let us reduce the  given {nonlinear two-component system} of TFRDEs \eqref{2+1rd}  in $(2+1)$-dimensions into the {nonlinear two-component system} of TFRDEs in $(1+1)$-dimensions  using the particular form
\begin{eqnarray}\label{particularform-rd}
	u_s(x_1,x_2,t)=v_s(z,t),
	z=\kappa_1x_1+\kappa_2x_2, s=1,2.
\end{eqnarray}
Thus, by substituting the above  particular form \eqref{particularform-rd} into \eqref{2+1rd}, we obtain  the  reduced {nonlinear two-component system} of TFRDEs in $(1+1)$-dimensions as follows:
%\begin{flushleft}Reduced DR system\end{flushleft}
\begin{eqnarray}
	\begin{aligned}{\label{transformedrd}}
		&\left( 	\dfrac{\partial^{\alpha_1} v_1}{\partial t^{\alpha_1}},\dfrac{\partial^{\alpha_2} v_2}{\partial t^{\alpha_2}}\right) =\mathbf{\hat{A}}(v_1,v_2)\equiv (\mathit{{A}}_1(v_1,v_2),\mathit{{A}}_2(v_1,v_2)),\alpha_s\in(0,2], \text{ with }	\\
		&	\mathit{{A}}_s(v_1,v_2)=\sum\limits_{i=1}^2\dfrac{\partial }{\partial z}\left[(\kappa_1^2 f_{si}(v_1,v_2)+\kappa_2^2g_{si}(v_1,v_2))\left(\frac{\partial v_i}{\partial z} \right)\right]
		%+(\kappa_1^2 f_{s2}(v_1,v_2)+\kappa_2^2g_{s2}(v_1,v_2)\left(\frac{\partial v_2}{\partial z} \right)
		+h_s(v_1,v_2),s=1,2.
	\end{aligned}
\end{eqnarray}
Now,  for the reduced nonlinear vector  differential operator $\mathbf{\hat{A}}(\mathbf{V})$ given in \eqref{transformedrd}, we aim to construct invariant linear product spaces
\begin{eqnarray}
	\begin{aligned}\label{rd:invariantprodtspace2+1}
		\mathbf{W}_{n} =&\mathit{W}_{1,n_1}\times\mathit{W}_{1,n_1}
		\\
		=&\left\{\psi(\kappa_1x_1+\kappa_2x_2)=(\psi_1,\psi_2)\big{|}\psi_s=\psi_s(\kappa_1x_1+\kappa_2x_2)\in\mathit{W}_{s,n_s},\kappa_i\in\mathbb{R},i,s=1,2 \right\},
	\end{aligned}
\end{eqnarray}	
where  $\mathit{W}_{s,n_s}$ are $n_s$-dimensional component linear spaces.
\subsection{Estimation of invariant linear product spaces $\mathbf{W}_n$ for the given {nonlinear two-component system} \eqref{2+1rd}}
Let us first consider the $n_s$-dimensional linear space $\mathit{W}_{s,n_s},s=1,2,$ as follows:
 $$
\begin{aligned}
	\mathit{W}_{s,n_s} =&		
	 \mathcal{L}\left\lbrace
	\varphi_{s,1}(z),\varphi_{s,2}(z),\dots,
	\varphi_{s,n_s}(z) \right\rbrace
\\&	=
	\left\{\psi_s(z)=\sum\limits_{i=1}^{n_s}k_{s,i}\varphi_{s,i}(z)\, \big{|} k_{s,i}\in\mathbb{R}, D_{s,n_s}[\psi_s(z)] =0\right\}
	,
\end{aligned}
$$
  where
   \begin{eqnarray}	\begin{aligned}		\label{rd:LHODEs}	D_{s,n_s}[y_s(z)]:= y^{(n_s)}_s(z)+\lambda_{s\,n_s-1}y^{(n_s-1)}_s(z) +\dots		+\lambda_{s1}y^{(1)}_s(z)		+\lambda_{s0}y_s(z)=0,s=1,2,
	\end{aligned}
\end{eqnarray}
$	\lambda_{sj}\in\mathbb{R},j=0,1,\dots,n_s-1,$ along with the fundamental set of solutions as $ \{\varphi_{s,i}(z){|} i=1,2,\dots,n_s\}
.$
Note that the linear space $\mathit{W}_{s,n_s},s=1,2,$
 is defined as the solution space for the given system of linear homogeneous ODEs  \eqref{rd:LHODEs}.
Now, we consider  the invariant linear product space $\mathbf{W}_{n}$  of the given  reduced {nonlinear two-component system} of TFRDEs  \eqref{transformedrd} in $(1+1)$-dimensions  as follows:
\begin{eqnarray}
	\begin{aligned}\label{rd:invariantprodtspace1+1}
		\mathbf{W}_{n} =&\mathit{W}_{1,n_1}\times\mathit{W}_{2,n_2}
		=\left\{\psi(z)=(\psi_1,\psi_2)\big{|}\psi_{s,i}(z)\in\mathit{W}_{s,n_s},s=1,2 \right\},
	\end{aligned}
\end{eqnarray}
with dim$(\mathbf{W}_{n})=n=n_1+n_2,$  by using Theorem \ref{productspacedimensiontheorem}.
{Thus, from the given  invariant conditions \eqref{transformedinvariantcriteria}, we observe that the operator $\mathbf{\hat{A}}(\mathbf{V})=(\mathit{{A}}_1(v_1,v_2),\mathit{{A}}_2(v_1,v_2))$  admits the  linear product space   $\mathbf{W}_{n}$ given in \eqref{rd:invariantprodtspace1+1} if $\mathbf{\hat{A}}(\mathbf{V})\subseteq \mathbf{W}_{n}$ or $\mathit{{A}}_s(v_1,v_2)\in \mathit{W}_{s,n_s},s=1,2,$  for every $(v_1,v_2)\in\mathbf{W}_n,$ which can be obtained as }
\begin{eqnarray}
	\begin{aligned}
		\label{rd:invariant conditions}
		&		D_{s,n_s}[\mathit{{A}}_s] \equiv \dfrac{d^{n_s} \mathit{{A}}_s(v_1,v_2)}{ d z^{n_s}}+\lambda_{s\,n_s-1}\dfrac{d^{(n_s-1)} \mathit{{A}}_s}{ d z^{(n_s-1)}} +\dots
		%+\lambda_{s,1}(z)Y^{(1)}_s(z)
		+\lambda_{s0}\mathit{{A}}_s=0,
		\\
		&	\text{whenever } D_{s,n_s}[v_s]\equiv \dfrac{d^{n_s} v_s}{ d z^{n_s}}+\lambda_{s\,n_s-1}\dfrac{d^{(n_s-1)} v_s}{ d z^{(n_s-1)}} +\dots
		%+\lambda_{s,1}(z)Y^{(1)}_s(z)
		+\lambda_{s0}\ v_s=0,s=1,2,
	\end{aligned}
\end{eqnarray}
where $\lambda_{sj}\in\mathbb{R},j=0,1,\dots,n_s-1,$ and $\mathit{{A}}_s=\mathit{{A}}_s(v_1,v_2)$ given in \eqref{transformedrd}.% for $v_s=v_s(z,t),s=1,2.$

{Moreover, using Theorem \ref{maximalthm1}, we obtain that the possible dimensions of the invariant linear product space $\mathbf{W}_{n}$ is $n=n_1+n_2, $ where   $n_1\leq 9,$ and $0<n_1-n_2\leq2.$
Hence, the possible values of $(n_1,n_2)$ are $$(1,1), (2,1),(2,2),(3,1),(3,2),(3,3),\dots,(8,6),(8,7),(8,8),(9,7),(9,8),(9,9).$$}
Now, let us consider the case $n_1=n_2=2.$  For this case, we aim to present how to  construct the invariant linear product space $\mathbf{W}_4$ for the given {nonlinear two-component system} \eqref{transformedrd}. Thus, the nonlinear vector differential operator $\mathbf{\hat{A}}(\mathbf{V})=(\mathit{{A}}_1(v_1,v_2),\mathit{{A}}_2(v_1,v_2))$ given  in \eqref{transformedrd} admits the invariant linear product space $\mathbf{W}_{4}=	\mathit{W}_{1,2}\times 	\mathit{W}_{2,2}=		
	\mathcal{L}\left\lbrace
	\varphi_{1,1}(z),\varphi_{1,2}(z) \right\rbrace\times	\mathcal{L}\left\lbrace
\varphi_{2,1}(z),\varphi_{2,2}(z) \right\rbrace
	,
$
$s=1,2,$
if the following invariant conditions hold:
\begin{eqnarray}
	%	\begin{aligned}
		\label{rd:invariant conditionn=4(a)}
		& \dfrac{d^{2} v_s}{ d z^{2}}=-\lambda_{s1}\dfrac{d v_s}{ d z}
		%+\lambda_{s,1}(z)Y^{(1)}_s(z)
		-\lambda_{s0}\,v_s	,s=1,2,
		\text{ implies that }	&	 \dfrac{d^{2} \mathit{{A}}_s}{ d z^{2}}+\lambda_{s1}\dfrac{d \mathit{{A}}_s}{ d z}
		%+\lambda_{s,1}(z)Y^{(1)}_s(z)
		+\lambda_{s0}\mathit{{A}}_s=0,
		\label{rd:invariant conditionn=4(b)}
		%		\end{aligned}	
\end{eqnarray}
where  $\mathit{{A}}_s=\mathit{{A}}_s(v_1,v_2),v_s=v_s(z,t),s=1,2.$
\\
Thus, we substitute  $\mathit{{A}}_s(v_1,v_2),s=1,2,$  given in \eqref{transformedrd} into the given invariant conditions \eqref{rd:invariant conditionn=4(b)}, and then equate the coefficients of $	\left(\dfrac{\partial v_1}{\partial z}\right) ^4,$ $	\left( \dfrac{\partial v_2}{\partial z}\right) \left( \dfrac{\partial v_1}{\partial z}\right) ^3,$ $ \dots, \left( \dfrac{\partial v_1}{\partial z}\right) ,	\left( \dfrac{\partial v_2}{\partial z}\right) $    equal to 0.
Hence, we obtain an over-determined system of equations with unknown arbitrary functions $f_{si}(v_1,v_2),g_{si}(v_1,v_2),$ and $h_s(v_1,v_2),$ and unknown arbitrary constants $\lambda_{{1j}}, s,i=1,2,$ and $j=0,1,$  as follows:
{  
	%\subsection{oDS1} 
	%The system of equations  obtained from the given equation \eqref{rd:invariant condition(a)} is as follows:
	\begin{eqnarray}
	\small	\begin{aligned}\label{rd:ODS(a)}
		&	
3 \dfrac{\partial^{3}\Psi_1}{\partial  v_1\partial  v_2^2}	
+ \dfrac{\partial^{3}\Phi_1}{\partial  v_2^3}=0,		\lambda_{{20}}(2\lambda_{{11}}+\lambda_{{21}})v_2
\dfrac{\partial\Phi_1}{\partial  v_2}+7\lambda_{{11}}\lambda_{{10}}\dfrac{\partial\Phi_1}{\partial  v_1} 
+
\lambda_{{20}}(3\lambda_{{21}}
+\lambda_{{11}})v_2\dfrac{\partial	\Psi_1}{\partial  v_1}=0,	
	\\&	\dfrac{\partial^{2} h_2}{\partial  v_1^2}	-3\lambda_{{20}}v_2\left( \dfrac{\partial^{2}\Psi_2}{\partial  v_1^2} + \dfrac{\partial^{2}\Phi_2}{\partial  v_1\partial  v_2}\right) 
-6\lambda_{{10}}v_1\dfrac{\partial^{2}\Phi_2}{\partial  v_1^2}
+ (7\lambda^2_{{11}}-3\lambda_{{11}}\lambda_{{21}}-4\lambda_{{10}}
+\lambda_{{20}})\dfrac{\partial\Phi_2}{\partial  v_1}
=0,	
\end{aligned}
\end{eqnarray}	\begin{eqnarray}
\small \begin{aligned}\label{rd:ODS(b)}	&
		\lambda_{{10}}(\lambda_{{11}}
		+2\lambda_{{21}})v_1\dfrac{\partial  \Psi_2}{\partial  v_1} 
		+7\lambda_{{21}}\lambda_{{20}}v_2\dfrac{\partial  \Psi_2}{\partial  v_2} 
		+
		\lambda_{{10}}(3\lambda_{{11}}
		+2\lambda_{{21}})v_1
		\dfrac{\partial  \Phi_2}{\partial  v_2} 
		=0,
		(\lambda_{{21}}-6\lambda_{{11}})\dfrac{\partial^{2}\Phi_2}{\partial  v_1^2} 
		=0,
		\\
		&
			(2\lambda_{{11}}+3\lambda_{{21}})	\dfrac{\partial^{2}\Psi_1}{\partial  v_1^2}	
			+
			(7\lambda_{{11}}+3\lambda_{{21}})	\dfrac{\partial^{2}\Phi_1}{\partial  v_1\partial  v_2}	
			=0,
			%%%%%%%%%%%%%%%%%%%%%%%%%%%%%%%%%%%%%%%%%%%%%%%%%%%%%%%%%%%%%%%%%%%%%%%%%%%%%%%%%%%%%%
			(2\lambda_{{11}}+3\lambda_{{21}})	\dfrac{\partial^{2}\Phi_1}{\partial  v_2^2}	
			+(\lambda_{{11}}
			+9\lambda_{{21}})	\dfrac{\partial^{2}\Psi_1}{\partial  v_1\partial  v_2}	
			=0,
			\\&
		%%%%%%%%%%%%%%%%%%%%%%%%%%%%%%%%%%%%%%%%%%%%%%%%%%%%%%%%%%%%%%%%%%%%%%%%
			(3\lambda^2_{{21}}+\lambda_{{11}}\lambda_{{21}}-3\lambda_{{20}})
		\dfrac{\partial\Psi_1}{\partial  v_1}
		-3\lambda_{{10}}v_1\dfrac{\partial^2\Psi_1}{\partial  v_1^2}-9
			\lambda_{{20}}v_2\dfrac{\partial^{2}
				\Psi_1}{\partial  v_1\partial  v_2}
			+
			[(\lambda_{{11}}+\lambda_{{21}})^2+2\lambda_{{10}}-\lambda_{{20}}]\dfrac{\partial\Phi_1}{\partial  v_2}
			\\&
		2\dfrac{\partial^{2}h_1}{\partial  v_1\partial  v_2}	-3\lambda_{{20}}v_2\dfrac{\partial^2\Phi_1}{\partial  v_2^2}-9(
			\lambda_{{10}}v_1+		\lambda_{{20}}v_2)\dfrac{\partial^{2}\Phi_1}{\partial  v_1\partial  v_2}
			=0,
				(\lambda_{{11}}-6\lambda_{{21}})\dfrac{\partial^2\Psi_1}{\partial  v_2^2}	=0,	\lambda_{{11}} \dfrac{\partial^{2}\Phi_1}{\partial v_1^2}	 		
				=0,
		\\	&
			[\lambda_{{21}}(\lambda_{{21}}\lambda_{{11}}-\lambda_{{21}}^2-\lambda_{{10}}
			+2\lambda_{{20}})-\lambda_{{11}}\lambda_{{20}}]\Psi_1 
			+
			\lambda_{{20}}\lambda_{{11}}	(10\lambda_{{21}}-3\lambda_{{11}})v_2	\dfrac{\partial\Psi_1}{\partial  v_2}
			+3\lambda_{{10}}\lambda_{{11}}v_1\dfrac{\partial\Psi_1}{\partial  v_1}
			\\
			&
			+\lambda_{{10}}	(\lambda_{{11}}+3\lambda_{{21}})v_1	\dfrac{\partial\Phi_1}{\partial  v_2}	+\lambda_{{11}}\dfrac{\partial h_1}{\partial  v_2}-\lambda_{{21}}\dfrac{\partial h_2}{\partial  v_2}=0, 	\dfrac{\partial^{3}\Psi_1}{\partial v_2^3}	=0,
	3 \dfrac{\partial^{3}\Phi_1}{\partial  v_1^2\partial  v_2}	
+ \dfrac{\partial^{3}\Psi_1}{\partial  v_1^3}		
=0,	\\&	\dfrac{\partial^{3}\Psi_2}{\partial  v_2^3}  
=0,		\dfrac{\partial^2 h_2}{\partial  v_2^2}
			%%%%%%%%%%%%%%%%%%%%%%%%%%%%%%%%%%%%%%%%%%%%%%%%%%%%%%%%%%%%%%%%%%%%%%%%%%%%%%%%%%
			-3\lambda_{{10}}v_1\left( \dfrac{\partial ^2 \Phi_2}{\partial  v_2^2}+\dfrac{\partial^2  \Psi_2}{\partial  v_1\partial  v_2}\right) 
			+
			(4\lambda^2_{{21}}-3\lambda_{{20}})\dfrac{\partial \Psi_2 }{\partial  v_2}-6\lambda_{{20}}v_2\dfrac{\partial^2  \Psi_2}{\partial  v_2^2}
			=0,
			\\&		\lambda_{{10}}[3\lambda_{{20}}v_2\dfrac{\partial\Phi_2 }{\partial  v_2}- \dfrac{\partial h_2 }{\partial  v_1}
	+
			3\lambda_{{10}}v_1 \dfrac{\partial \Phi_2}{\partial  v_1}
			+
			(\lambda_{{21}}\lambda_{{11}}
			-\lambda^2_{{11}}
			-\lambda_{{20}})\Phi_2]	v_1
			+3(\lambda^2_{{20}}v_2\dfrac{\partial \Psi_2}{\partial  v_2 } 	\\&	+\lambda_{{20}}\lambda_{{10}}v_1\dfrac{\partial \Psi_2}{\partial  v_1}-\dfrac{\partial h_2  }{\partial  v_2})v_2
			+\lambda_{{20}} 
			h_2=0,
				%%%%%%%%%%%%%%%%%%%%%%%%%%%%%%%%%%%%%%%%%%%%%%%%%%%%%%%%%%%%%%%%%%%%%%%%%%%
					%%%%%%%%%%%%%%%%%%%%%%%%%%%%%%%%%%%%%%%%%%%%%%%%%%%%%%%%%%%%%%%%%%%%%%%%%%%%%%%%%%%%%%%%
			(\lambda_{{11}}\lambda_{{21}}-3\lambda_{{10}}
			+3\lambda^2_{{11}})\dfrac{\partial\Phi_2}{\partial  v_2}+2\dfrac{\partial^{2} h_2}{\partial  v_1\partial  v_2}
			\\&-3\lambda_{{20}}v_2\left( \dfrac{\partial^2\Phi_2}{\partial  v_2^2} +3\dfrac{\partial^{2}\Psi_2}{\partial  v_1\partial  v_2}\right) -3\lambda_{{10}}v_1\left( \dfrac{\partial^{2}\Psi_2}{\partial  v_1^2}+3
			\dfrac{\partial^{2}\Phi_2}{\partial  v_1\partial  v_2}\right) 	
			+[(\lambda_{{21}}+\lambda_{{11}})^2
			-\lambda_{{10}}
		\\
		&	-2\lambda_{{20}}]
			\dfrac{\partial\Psi_2}{\partial  v_1}
			=0, (3\lambda_{{11}}+2\lambda_{{21}})\dfrac{\partial^{3}\Phi_2}{ \partial  v_2^3}
			+(3\lambda_{{11}}
			+7\lambda_{{21}})\dfrac{\partial^{2}\Psi_2}{\partial  v_1\partial  v_2}  
			=0, \dfrac{\partial^{3}\Phi_1}{\partial v_1^3} 		
			=0,	
		\lambda_{{10}}[3v_1(\lambda_{{20}}v_2\dfrac{\partial\Phi_1}{\partial  v_2} 	\\
		& +\lambda_{{10}}v_1\dfrac{\partial\Phi_1}{\partial  v_1})+ h_1-v_1\dfrac{\partial h_1}{\partial  v_1} 
			]
			%%%%%%%%%%%%%%%%%%%%%%%%%%%%%%%%%%%%%%%%%%%%%%%%%%%%%%%%%%%%%%%%%%%%%%%%%%%%%%%%%%%%%%%
			%%%%%%%%%%%%%%%%%%%%%%%%%%%%%%%%%%%%%%%%%%%%%%%%%%%%%%%%%%%%%%	
			+
			\lambda_{{20}} [3	(\lambda_{{10}}v_1 \dfrac{\partial\Psi_1}{\partial  v_1}
			+\lambda_{{20}}v_2	\dfrac{\partial\Psi_1}{\partial  v_2})
			-\dfrac{\partial h_1}{\partial  v_2}
			+(\lambda_{{21}}\lambda_{{11}}
			-\lambda_{{21}}^2-\lambda_{{10}}
		\\&	+\lambda_{{20}})
			\Psi_1]v_2=0,	(9\lambda_{{11}}+\lambda_{{21}}) \dfrac{\partial^{2}\Phi_2}{\partial  v_1\partial  v_2} 
			+(3\lambda_{{11}}
			+2\lambda_{{21}})\dfrac{\partial^{2}\Psi_2}{\partial  v_1^2} =0,
			\dfrac{\partial^{3}\Phi_1}{\partial  v_1\partial  v_2^2}	
					+	 \dfrac{\partial^{3}\Psi_1}{\partial  v_1^2\partial  v_2}	
			=0,			
			%%%%%%%%%%%%%%%%%%%%%%%%%%%%%%%%%%%%%%%%%%%%%%%%%%%%%%%%%%%%%%%%%%%%%%%%%%%%%%%%%%%%%%%
		\\	&	\lambda_{{10}}(10\lambda_{{11}}-3\lambda_{{21}})v_1 \dfrac{\partial\Phi_2 }{\partial  v_1}
			+3\lambda_{{11}}\lambda_{{20}}v_2
			\dfrac{\partial\Phi_2 }{\partial  v_2}
			+[\lambda_{{11}}(\lambda_{{11}}\lambda_{{21}}
			-\lambda_{{20}}+2\lambda_{{10}}
		-\lambda_{{11}}^2)-\lambda_{{21}}\lambda_{{10}}] \Phi_2	\\
		&+\lambda_{{21}}\dfrac{\partial h_2}{\partial  v_1}
				+
		\lambda_{{20}}	(3\lambda_{{11}}+\lambda_{{21}})v_2
			\dfrac{\partial \Psi_2}{\partial  v_1}
					-
			\lambda_{{11}} \dfrac{\partial h_1}{\partial  v_1}=0,
					5\lambda_{{21}}  
			\dfrac{\partial^2 \Psi_2}{\partial  v_2^2}
			=0, 	
			\dfrac{\partial^{3}\Psi_2}{\partial  v_1^3}   +
			3\dfrac{\partial^{3}\Phi_2}{\partial  v_1^2\partial  v_2} 
			=0,
			%%%%%%%%%%%%%%%%%%%%%%%%%%%%%%%%%%%%%%%%%%%%%%%%%%%%%%%%%%%%%%%%%%%%
			\\
		&	\dfrac{\partial^{3}\Phi_2}{\partial  v_1\partial  v_2^2} 
			+
			\dfrac{\partial^{3}\Psi_2}{\partial  v_1^2\partial  v_2} 
			=0,
		(4\lambda_{{11}}^2-3\lambda_{{10}})\dfrac{\partial\Phi_1}{\partial  v_1}		-6\lambda_{{10}}v_{{1}} 
\dfrac{\partial^{2} \Phi_1}{\partial  v_1^2}	 -3\lambda_{{20}}v_{{2}}
\dfrac{\partial^{2}\Phi_1}{\partial  v_1\partial  v_2}
-3\lambda_{{20}}v_{{2}}
\dfrac{\partial^{2}\Psi_1}{\partial  v_1^2} 
\\& +\dfrac{\partial^{2}h_1}{\partial  v_1^2}
=0,	
	\dfrac{\partial^{3}\Phi_2}{\partial  v_2^3}  
+3\dfrac{\partial^{3}\Psi_2}{\partial  v_1\partial  v_2^2}  =0,
\dfrac{\partial^{2}h_1}{\partial  v_2^2}
-3		\lambda_{{10}}v_1  
 \left( 	\dfrac{\partial^{2}\Phi_1}{\partial v_2^2} +
\dfrac{\partial^{2}\Psi_1}{\partial  v_1\partial  v_2} \right) 
\\
&+
(7\lambda^2_{{21}}-3\lambda_{{21}}\lambda_{{11}}
+
\lambda_{{10}}
-4\lambda_{{20}})\dfrac{\partial\Psi_1}{\partial  v_2}
-6\lambda_{{20}}v_2\dfrac{\partial^2\Psi_1}{\partial  v_2^2}	=0,\dfrac{\partial^{3}\Phi_2}{\partial  v_1^3} =0,	
		\end{aligned}\label{rd:ODS(c)}
	\end{eqnarray}
	where $\Phi_s=\kappa_1^2 f_{s1}(v_1,v_2)+\kappa_2^2g_{s1}(v_1,v_2), \Psi_s=\kappa_1^2 f_{s2}(v_1,v_2)+\kappa_2^2g_{s2}(v_1,v_2),$ and $ h_s=h_s(v_1,v_2),s=1,2.$  
}
{Solving the obtained over-determined system, we can get various types of   four-dimensional invariant linear product spaces $\mathbf{W}_4$ and the corresponding  nonlinear vector differential operators $\mathbf{\hat{A}}(\mathbf{V})= (\mathit{{A}}_1(v_1,v_2),\mathit{{A}}_2(v_1,v_2)).$
 However,  we observe that  the obtained over-determined system of equations  \eqref{rd:ODS(a)}-\eqref{rd:ODS(b)}
  is not solvable, in general.
 Thus, we consider the parametric restrictions for $\lambda_{{1i}},i=0,1.$
So, let us first consider  the case $\lambda_{{10}}=0.$ For this case, the various types of obtained four-dimensional invariant linear product spaces $\mathbf{W}_4$  with their corresponding coefficient functions  $\Phi_s, \Psi_s,$ and $h_s,s=1,2,$  of $\mathit{{A}}_s(v_1,v_2),s=1,2,$ given in  \eqref{transformedrd} are listed in Tables \ref{table1}. %where  $\Phi_s=\kappa_1^2 f_{s1}(v_1,v_2)+\kappa_2^2g_{s1}(v_1,v_2), \Psi_s=\kappa_1^2 f_{s2}(v_1,v_2)+\kappa_2^2g_{s2}(v_1,v_2),$ and $ h_s=h_s(v_1,v_2),s=1,2.$
  Also, we have listed other different  four-dimensional invariant linear product spaces $\mathbf{W}_4$  with their corresponding coefficient functions  $\Phi_s, \Psi_s,$ and $h_s,s=1,2,$  of $\mathit{{A}}_s(v_1,v_2),s=1,2,$ for the case $\lambda_{{11}}=0$  in Tables \ref{table3}.  Similarly, we can find the other four-dimensional invariant linear product spaces   $\mathbf{W}_4$ for the cases $\lambda_{{2j}}=0,j=0,1.$  Further,    we can also obtain 
  other different dimensional invariant linear product spaces $\mathbf{W}_n$  with their corresponding differential operators  $\mathbf{\hat{A}}(\mathbf{V})= (\mathit{{A}}_1(v_1,v_2),\mathit{{A}}_2(v_1,v_2))$ given in \eqref{transformedrd}  using   procedures similar to the above.}
 %\subsection{Tabulation $\lambda_{10}=0$}

\begin{sidewaystable}
		\tiny
\caption{Invariant linear product spaces $\mathbf{W}_4$ for the given {nonlinear two-component system}  \eqref{transformedrd}  when $\lambda_{10}=0$ in \eqref{rd:invariant conditionn=4(a)}.}
\label{table1}
\begin{tabular}{cll}
\toprule
Cases &{\qquad\qquad\qquad\qquad\qquad\qquad Coefficient functions of \eqref{transformedrd} } &  $\mathbf{W}_4=\mathit{W}_{1,2}\times 	\mathit{W}_{2,2}$
\\
\midrule
1.
%%% FIRST COMPONENT%%%%%
& $\left\{\begin{array}{ll}
	\Phi_{s}=\kappa_1^2[\gamma_{s0}+\gamma_{s1}v_1+(\gamma_{s2}+\gamma_{s3}v_1)v_2
	+  \frac{1}{2}(\gamma_{s4}v_2^2	
	+\gamma_{s5}v_1^2)
		- \frac{1}{2}(\frac{\eta_{s8}}{3}v_1^2
		+\eta_{s7}v_1v_2
+\eta_{s6}v_2^2) v_2]
\\
	\Psi_{s}=\kappa_1^2[\eta_{s0}+\eta_{s1}v_1+(\eta_{s3}v_1 +\eta_{s2}	)v_2
	+  \frac{1}{2}(\eta_{s4}v_2^2	+\eta_{s5}v_1^2)
+ \frac{1}{2}(\frac{\eta_{s8}}{3}v_1^2
+\eta_{s7}v_1v_2
+\eta_{s6}v_2^2)v_1 ]
\\
	%%%% SECOND COMPONENT%%%%%%
	h_{s}=\mu_{s0}+\mu_{s1}v_1
	+\mu_{s2}v_2 ,s=1,2
\end{array}\right.$
%%%% INVARIANT SPACE%%%%%%
&
$\mathbf{W}_4=\mathcal{L}\{1,z \}\times \mathcal{L}\{1,z\}$
\\
\\
2. %%% FIRST COMPONENT%%%%%
& $\left\{\begin{array}{ll}
	\Phi_{1}=\kappa_1^2(\gamma_{10}+\gamma_{11}v_1+  \frac{\gamma_{12}}{2}v_1^2)
	- \frac{1}{\lambda_{20}}(\frac{\mu_{15}}{8}v_2
	+\mu_{13})v_2
\\
	\Psi_{1}=\frac{1}{\lambda_{20}}[\frac{1}{4}(\mu_{15}v_1 +\mu_{14})v_2
	+(\mu_{13}v_1+\mu_{12})]
\\	h_{1}=\mu_{10}+v_1(\mu_{11}+\frac{\mu_{15}}{2}v_2^2)+(\mu_{12}+\mu_{13}v_1
		+\frac{\mu_{14}}{2}v_2)v_2
			\end{array}\right.
	%%%% SECOND COMPONENT%%%%%%
	\left\{\begin{array}{ll}
	\Phi_{2}=\kappa_1^2(\gamma_{20}+\gamma_{21}v_2)-\frac{\mu_{23}}{\lambda_{20}}v_1v_2
	\\
	\Psi_{2}=\kappa_1^2\eta_{20}+\frac{1}{\lambda_{20}}(\frac{\mu_{23}}{2}v_1^2+\frac{\mu_{22}}{2}v_2^2+\mu_{21}v_1)
	\\
	h_{2}=(\mu_{20}+\mu_{21}v_1+\mu_{22}v_2^2+\frac{\mu_{23}}{2}v_1^2)v_2
\end{array}\right.$
%%%% INVARIANT SPACE%%%%%%
&
$\begin{array}{ll}\mathbf{W}_4=
	\mathcal{L}\{1,z \}
\times \mathcal{L}\{\sin(\sqrt{\lambda_{20}}z),
\cos(\sqrt{\lambda_{20}}z)\}
\end{array}$
\\
\\
3.
%%% FIRST COMPONENT%%%%%
&  $\left\{\begin{array}{ll}
	\Phi_{1}=\kappa_1^2(\gamma_{10}+\gamma_{11}v_1+  \frac{\gamma_{12}}{2}v_1^2)
	- \frac{1}{\lambda_{20}}(\frac{\mu_{15}}{8}v_2
	+\mu_{13}) v_2
	,\\
	\Psi_{1}=\frac{1}{\lambda_{20}}[\frac{1}{4}(\mu_{15}v_1 +\mu_{14})v_2+\mu_{13}v_1
	+\mu_{12}]\\
	h_{1}=\mu_{10}+v_1(\mu_{11}+\frac{\mu_{15}}{2}v_2^2)+(\mu_{12}
	+\mu_{13}v_1
		+\frac{\mu_{14}}{2}v_2)v_2
			\end{array}\right.
	%%%% SECOND COMPONENT%%%%%%
\left\{\begin{array}{ll}
		\Psi_{2}=-\frac{1}{\lambda_{21}^2}
		(\mu_{23}v_1^2 +\mu_{22})
	\\
	%%%% SECOND COMPONENT%%%%%%
	\Phi_{2}=\frac{\mu_{23}}{\lambda_{21}^2}v_2+\kappa_1^2(\gamma_{20}+\gamma_{21}v_1
		+\frac{\gamma_{22}}{2}v_1^2)
\\
	h_{2}=\mu_{20}+\mu_{21}v_1+\mu_{22}v_2+\mu_{23}v_1v_2
\end{array}\right.$
%%%% INVARIANT SPACE%%%%%%
&
$\mathbf{W}_4=\mathcal{L}\{ 1,z\}\times \mathcal{L}\{1,e^{-\lambda_{21}z}\}$
\\
\\
4. %%% FIRST COMPONENT%%%%%
& $\left\{\begin{array}{ll}
	\Phi_{1}=\kappa_1^2(\gamma_{10}+\gamma_{11}v_1+  \frac{\gamma_{12}}{2}v_1^2)\qquad\qquad\qquad\quad\quad\quad
\,\,\,	\\
	\Psi_{1}=0,~	h_{1}=\mu_{10}+\mu_{11}v_1
		\end{array}\right.
%%%% SECOND COMPONENT%%%%%%
\left\{\begin{array}{ll}
\Phi_{2}=\frac{2\mu_{23}}{\lambda_{20}}v_1v_2+\kappa_1^2(\gamma_{20}+\gamma_{21}v_2)
\\
\Psi_{2}=\kappa_1^2\eta_{20}-\frac{1}{\lambda_{20}}(\frac{\mu_{23}}{2}v_1^2+\mu_{22}v_1)	
\\h_{2}=(\mu_{21}+\mu_{22}v_1+\frac{\mu_{23}}{2}v_1^2)v_2
\end{array}\right.$
%%%% INVARIANT SPACE%%%%%%
&
$\left\{\begin{array}{ll}\mathbf{W}_4=
	\mathcal{L}\{1,z \}
	\times \mathcal{L}\{e^{-\sqrt{\lambda_{20}}z},ze^{-\sqrt{\lambda_{20}}z}\}
	\\
	\mathbf{W}_4=	\mathcal{L}\{1,z \}
		\times \mathcal{L}\{e^{\sqrt{\lambda_{20}}z}, ze^{\sqrt{\lambda_{20}}z}\}
\end{array}\right.$
\\
\\
5.
%%% FIRST COMPONENT%%%%%
&  $\left\{\begin{array}{ll}
	\Phi_{1}=\kappa_1^2(\gamma_{10}+\gamma_{11}v_1+  \frac{\gamma_{12}}{2}v_1^2)\qquad\qquad\qquad\quad\quad\quad
	\,\,\,	\\
		\Psi_{1}=0,~~h_{1}=\mu_{10}+\mu_{11}v_1
	\end{array}\right.
%%%% SECOND COMPONENT%%%%%%
\left\{\begin{array}{ll}
	\Phi_{2}=\kappa_1^2(\gamma_{20}+\gamma_{21}v_2)
	\\
	\Psi_{2}=\kappa_1^2\eta_{20}
,~~
h_{2}=\mu_{20}v_2
\end{array}\right.$
%%%% INVARIANT SPACE%%%%%%
&
$\begin{array}{ll}
		\mathbf{W}_4=\mathcal{L}\{ 1,z\}
	\times \mathcal{L}\{e^{-(\frac{1}{2}\lambda_{21}+\sqrt{\lambda_{21}^2-4\lambda_{20}})z},
	\\~~~~~~~~~~
	 	e^{-(\frac{1}{2}\lambda_{21}+ \sqrt{\lambda_{21}^2-4\lambda_{20}})z}\}
\end{array}$
\\
\\
6.
%%% FIRST COMPONENT%%%%%
&  $\left\{\begin{array}{ll}
	\Phi_{1}= \kappa_1^2\gamma_{10}-\frac{\mu_{12}}{4\lambda_{11}^2}v_1\\
	
	\Psi_{1}=\kappa_1^2(\eta_{10}+\eta_{11}v_1+\eta_{12}v_2)\qquad\qquad\qquad\quad\quad\quad
	\,\,\,
	\\
	h_{1}=\mu_{10}+\mu_{11}v_1+\frac{\mu_{12}}{2}v_1^2
		\end{array}\right.
%%%% SECOND COMPONENT%%%%%%
\left\{\begin{array}{ll}
	\Phi_{2}=-\frac{1}{\lambda_{11}^2}(\mu_{21}+\mu_{23}v_2)
	\\
	\Psi_{2}=\frac{\mu_{23}}{\lambda_{11}^2}v_1+\kappa_1^2(\eta_{20}+\eta_{21}v_2+\frac{\eta_{22}}{2}v_2^2)
	\\
	h_{2}=\mu_{20}+\mu_{21}v_1+(\mu_{22}+\mu_{23}v_1)v_2
\end{array}\right.$
%%%% INVARIANT SPACE%%%%%%
&
$\begin{array}{ll}
	\mathbf{W}_4=\mathcal{L}\{ 1,e^{-\lambda_{11}z}\}
	\times \mathcal{L}\{ 1,z\}
\end{array}$
\\
\\
7.
%%% FIRST COMPONENT%%%%%
&  $\left\{\begin{array}{ll}
	\Phi_{1}= \kappa_1^2\gamma_{10}-\frac{\mu_{13}}{4\lambda_{11}^2}v_1
,~
\Psi_{1}=\frac{1}{4\lambda_{20}}(\mu_{14}v_2+4{\mu_{12}})~~~~~
\\
	h_{1}=\mu_{10}+\mu_{12}v_2
	+\frac{1}{2}(
	\mu_{13}v_1^2+\mu_{14}v_2^2)
	+\mu_{11}v_1
	\end{array}\right.
%%%% SECOND COMPONENT%%%%%%
\left\{\begin{array}{ll}
	\Phi_{2}=-\frac{\mu_{21}}{\lambda_{20}+\lambda_{11}^2}v_2	\\
	\Psi_{2}=\frac{\mu_{22}}{\lambda_{20}}v_2^2+\kappa_1^2 \eta_{20}+\frac{\mu_{21}}{\lambda_{20}+\lambda_{11}^2}v_1\\
	h_{2}=(\mu_{20}+\mu_{21}v_1+\mu_{22}v_2^2)v_2
\end{array}\right.$
%%%% INVARIANT SPACE%%%%%%
&
$\begin{array}{ll}
	\mathbf{W}_4=\mathcal{L}\{ 1,e^{-\lambda_{11} z}\}
	\times \mathcal{L}\{ \sin(\sqrt{\lambda_{20}}z),
	\cos(\sqrt{\lambda_{20}}z)\}
\end{array}$
\\
\\
	8.
%%% FIRST COMPONENT%%%%%
&  $\left\{\begin{array}{ll}
	\Phi_{1}= \frac{\mu_{13}}{\lambda_{21}(\lambda_{21}+\lambda_{11})}v_2-\frac{\mu_{14}}{4\lambda_{11}^2}v_1+\kappa_1^2\gamma_{10}\\
	\Psi_{1}=-\frac{1}{\lambda_{21}^2}({\mu_{13}}v_1+{\mu_{12}})
	\\
	h_{1}=\mu_{10}+(\mu_{11}+\frac{	\mu_{14}}{2}
	v_1)v_1+(\mu_{12}+\mu_{13}v_1)v_2~~~~~~~~~~
	\end{array}\right.
%%%% SECOND COMPONENT%%%%%%
\left\{\begin{array}{ll}
	\Phi_{2}=-\frac{1}{\lambda_{11}^2}({\mu_{23}}v_2+{\mu_{21}})
	\\
	\Psi_{2}=\frac{\mu_{23}}{\lambda_{11}(\lambda_{21}+\lambda_{11})}v_1-\frac{\mu_{24}}{4\lambda_{21}^2}v_2+\kappa_1^2\eta_{20}
	\\
	h_{2}=\mu_{20}+\mu_{21}v_1+(\mu_{22}+\mu_{23}v_1+\frac{\mu_{24}}{2}v_2)v_2
\end{array}\right.$
%%%% INVARIANT SPACE%%%%%%
&
$\begin{array}{ll}
	\mathbf{W}_4=\mathcal{L}\{ 1,e^{-\lambda_{11} z}\}
		\times \mathcal{L}\{ 1,e^{-\lambda_{21} z}\}
\end{array}$\\
%\bottomrule
%\end{tabular}
%\end{sidewaystable}

%\begin{sidewaystable}
%	\small
%\caption{Invariant linear product spaces $\mathbf{W}_4$ for the given {nonlinear two-component system}  \eqref{transformedrd}  when $\lambda_{10}=0$ in \eqref{rd:invariant conditionn=4(a)}.}
%\label{table2}
%\begin{tabular}{lll}
%\toprule
%Cases  &{\qquad\qquad Coefficient functions of \eqref{transformedrd} } & $\mathbf{W}_4=\mathit{W}_{1,2}\times 	\mathit{W}_{2,2}$
%\\
%\midrule

		9.
	%%% FIRST COMPONENT%%%%%
	&  $\left\{\begin{array}{ll}
		\Phi_{1}=  \frac{1}{4\beta^2}(2{\mu_{14}}v_2
		-{\mu_{13}}v_1)
		+\kappa_1^2\gamma_{10}
		\\
		\Psi_{1}=\frac{1}{4\beta^2}(4{\mu_{14}}v_1
		-{\mu_{15}}v_2
		-4{\mu_{12}})
		\\
		h_{1}=\mu_{10}+(\mu_{11}+\frac{\mu_{14}}{2}v_1)v_1+(\mu_{12}+\mu_{14}v_1
				+\frac{\mu_{15}}{2}v_2)v_2,~
				\end{array}\right.
		%%%% SECOND COMPONENT%%%%%%
	\left\{\begin{array}{ll}
		\Phi_{2}=\kappa_1^2(\gamma_{20}-\eta_{21}v_2)
	\\
		\Psi_{2}=\frac{\mu_{21}}{3\beta^2}v_2^2+\kappa_1^2 (\eta_{20}+\eta_{21}v_1)
		\\
		h_{2}=(\mu_{20}+\mu_{21}v_2^2)v_2
	\end{array}\right.$
	%%%% INVARIANT SPACE%%%%%%
	&
	$\left\{\begin{array}{ll}
		\mathbf{W}_4=\mathcal{L}\{ 1,e^{-\beta z}\}
			\times
		\mathcal{L}\{ e^{-\beta z},e^{\beta z}\},	\beta\in\mathbb{R}	
		\\
		\mathbf{W}_4=\mathcal{L}\{ 1,e^{\beta z}\},
	\times \mathcal{L}\{ e^{-\beta z},e^{\beta z}\},
	\beta\in\mathbb{R}	
	\end{array}\right.$
	
	\\
	\\
	10.
	%%% FIRST COMPONENT%%%%%
	&  $\left\{\begin{array}{ll}
		\Phi_{1}= \frac{1}{16\beta^2}(2{\mu_{15}}v_2^2
		-{\mu_{13}}v_1)+\kappa_1^2\gamma_{10}
		\\
		\Psi_{1}=\frac{1}{4\beta^2}[(2{\mu_{15}}v_1-{\mu_{14}})v_2-4{\mu_{12}}]
		\\
		h_{1}=\mu_{10}+(\mu_{11}+\mu_{15}v_2^2)v_1+\mu_{12}v_2
		+\frac{1}{2}(
		\mu_{13}v_1^2+\mu_{14}v_2^2)
		\end{array}\right.
%%%% SECOND COMPONENT%%%%%%
\left\{\begin{array}{ll}
		\Phi_{2}=-(\frac{\kappa_1^2\eta_{21}}{2}+\frac{\mu_{21}}{6\beta^2})v_2
		\\
		\Psi_{2}=-\frac{\mu_{22}}{3\beta^2}v_2^2+\kappa_1^2 (\eta_{20}+\eta_{21}v_1)
		\\
		h_{2}=(\mu_{20}+\mu_{21}v_1+\mu_{22}v_2^2)v_2
	\end{array}\right.$
	%%%% INVARIANT SPACE%%%%%%
	&
	$\begin{array}{ll}
		\mathbf{W}_4=\mathcal{L}\{ 1,e^{-2\beta z}\}
				\times \mathcal{L}\{ e^{-\beta z},e^{\beta z}\}	, \beta\in\mathbb{R}
	\end{array}$
	
	\\
	\bottomrule
\end{tabular}
\\
\\
	\end{sidewaystable}
%\subsection{Tabulation $\lambda_{11}=0$}
\begin{sidewaystable}\tiny
\caption{Invariant linear product spaces $\mathbf{W}_4$ for the given {nonlinear two-component system} \eqref{transformedrd}  when $\lambda_{11}=0$ in \eqref{rd:invariant conditionn=4(a)}.}
\label{table3}
\begin{tabular}{cll}
\toprule
Cases & {\qquad\qquad\qquad Coefficient functions of \eqref{transformedrd} } & $\mathbf{W}_4=\mathit{W}_{1,2}\times 	\mathit{W}_{2,2}$
\\
\midrule
1.
%%% FIRST COMPONENT%%%%%
& $\left\{\begin{array}{ll}
	\Phi_{1}=\kappa_1^2[\gamma_{10}-( \eta_{13}v_2
	+\frac{\eta_{12}}{2}v_1
	+\frac{\eta_{11}}{2})v_2
\\~~~	+ \frac{1}{3\lambda_{20}}(\mu_{14}v_1^2+\mu_{13}v_1v_2 +\frac{\mu_{12}}{2}v_2^2) ]
	\\
	\Psi_{1}=\kappa_1^2[
	\eta_{10}+(\eta_{11} +\frac{1}{2}\eta_{12}v_1
	+\eta_{13}v_2)v_1]	+\frac{\mu_{15}}{18\lambda_{20}}v_2^2
	\\
	h_{1}=[\mu_{10}	+(\mu_{11}+\frac{\mu_{15}}{6}v_2^2)v_2
	+\frac{\mu_{12}}{2}v_2^2+\mu_{13}v_1v_2
\\~~~~	+\mu_{14}v_1^2]v_1\\
	\end{array}\right.
%%%% SECOND COMPONENT%%%%%%
\left\{\begin{array}{ll}
	\Phi_{2}=\kappa_1^2[\gamma_{20}
		-( \eta_{23}v_2
	+\frac{\eta_{22}}{2}v_1
	+\frac{\eta_{21}}{2})v_2
\\~~~~+ \frac{1}{3\lambda_{20}}(\mu_{25}v_1^2
+\mu_{22}v_1v_2+\frac{\mu_{23}}{2}v_2^2) ]
\\
	\Psi_{2}=\kappa_1^2[
	\eta_{20}+(\eta_{21} +\frac{\eta_{22}}{2}v_1+\eta_{23}v_2)v_1]	+\frac{\mu_{21}}{18\lambda_{20}}v_2^2 	
	\\
	h_{2}=(\mu_{10}+\frac{\mu_{23}}{2}v_2^2
	+\mu_{22}v_1v_2+\mu_{25}v_1^2)v_1
\\~~~~	+(\mu_{21}+\frac{\mu_{24}}{6}v_2^2)v_2
\end{array}\right.$
%%%% INVARIANT SPACE%%%%%%
&
$\begin{array}{ll}\mathbf{W}_4=\mathcal{L}\{\sin(\sqrt{\lambda_{20}}z),	\cos(\sqrt{\lambda_{20}}z)\}
	\\
~~~~~~~~~\times \mathcal{L}\{\sin(\sqrt{\lambda_{20}}z),	  \cos(\sqrt{\lambda_{20}}z)\}
\end{array}$
\\
\\
2.
%%% FIRST COMPONENT%%%%%
&  $\left\{\begin{array}{ll}
	\Phi_{1}=
	\kappa_1^2(\gamma_{10}-2\eta_{10}v_2)-\frac{4}{3\lambda_{21}^2}( {\mu_{12}}v_1^2+{\mu_{11}}v_2)~~~~~~~
	\\
	\Psi_{1}=\kappa_1^2\eta_{10}
	,~~
	h_{1}=(\mu_{12}v_1^2+\mu_{11}v_2+\mu_{10})v_1
	\end{array}\right.
%%%% SECOND COMPONENT%%%%%%
\left\{\begin{array}{ll}
	\Phi_{2}=-\frac{1}{\lambda_{21}^2}({\mu_{25}v_2+\mu_{24}v_1+4\mu_{22}})
	\\
	\Psi_{2}=\kappa_1^2\eta_{20}-\frac{1}{4\lambda_{21}^2}({\mu_{25}}v_1^2
	+{\mu_{23}}v_2)
\\	h_{2}=\mu_{20}
+\frac{1}{2}[\mu_{23}v_2^2+(\mu_{24}
+\mu_{25}v_2)v_1^2]\\~~~~~
+\mu_{21}v_2+\mu_{22}v_1
\end{array}\right.$
%%%% INVARIANT SPACE%%%%%%
&
$\begin{array}{ll}
	\mathbf{W}_4=\mathcal{L}\{e^{-\frac{\lambda_{21} }{\sqrt{2}} z},
	e^{\frac{\lambda_{21} }{\sqrt{2}}z}\}
	\times\mathcal{L}\{1, e^{\lambda_{21} z}\}
\end{array}$
\\
\\
	3.
%%% FIRST COMPONENT%%%%%
&  $\left\{\begin{array}{ll}
	\Phi_{1}
	=
	\frac{1}{15\lambda_{20}}
	(	
	2{\mu_{14}}v_1^2-5\mu_{13}v_2^2
	+10{\mu_{12}}v_2
	)
+\kappa_1^2\gamma_{10}~~~
\\
	\Psi_{1}
	=\frac{1}{3\lambda_{20}}
	[
(	2\mu_{13}v_2
	-2{\mu_{12}})v_1
	+3{\mu_{11}}
	]
	
	\\
	h_{1}=(\mu_{10}+\mu_{12}v_2
	+\frac{\mu_{13}}{2}v_2^2
	+\mu_{14}v_1^2)v_1+\mu_{11}v_2
	\end{array}\right.
%%%% SECOND COMPONENT%%%%%%
\left\{\begin{array}{ll}
	\Phi_{2}
	=\frac{1}{4\lambda_{20}}
	[
	5(\mu_{23}v_1
		+6{\mu_{22}})v_2
	+18{\mu_{20}}
	]
	\\
	\Psi_{2}
	=\kappa_1^2\eta_{20}-\frac{1}{18\lambda_{20}}
	[
	(\mu_{23}v_1
	+12{\mu_{22}})v_1
	-6{\mu_{24}}v_2^2
	]
\\
	h_{2}=\mu_{20}v_1
+v_2(\mu_{21}+\mu_{22}v_1
+\frac{\mu_{23}}{2}v_1^2+\mu_{24}v_2^2 )
\end{array}\right.$
%%%% INVARIANT SPACE%%%%%%
&
$\begin{array}{ll}\mathbf{W}_4=\mathcal{L}\{\sin(\theta z),
		\cos(\theta z)\}
 \times \mathcal{L}\{\sin(\sqrt{\lambda_{20}}z),
 \\~~~~~~~~~~
\cos(\sqrt{\lambda_{20}}z)\},
\theta=\sqrt{\frac{5\lambda_{20}}{2}}
\end{array}$
\\
\\
4.
%%% FIRST COMPONENT%%%%%
&  $\left\{\begin{array}{ll}
	\Phi_{1}
	=
	\frac{1}{6\lambda_{10}}
	(
	2\mu_{13}v_1^2
	+3{\mu_{12}}v_2^2
	+2{\mu_{12}}v_2
	)
	+\kappa_1^2\gamma_{10}~~~~~~~
	\\
	\Psi_{1}
	=\kappa_1^2(\eta_{11}v_1+\eta_{10})-\frac{{\mu_{12}}}{3\lambda_{10}}
	v_1v_2
	\\
	h_{1}=v_1(\mu_{10}+\mu_{11}v_2+\frac{\mu_{13}}{2}v_2^2+\mu_{14}v_1^2)
\end{array}\right.
%%%% SECOND COMPONENT%%%%%%
\left\{\begin{array}{ll}
	\Phi_{2}
	=\frac{1}{4\lambda_{10}}
	[	(
	\mu_{24}v_2
	+\mu_{25})v_1
	+4{\mu_{23}}v_2
	+4\mu_{21}
	]
	\\
	\Psi_{2}
	=\kappa_1^2(\frac{\eta_{22}}{2}v_2^2+\eta_{21}v_2+\eta_{20})+\frac{1}{8\lambda_{10}}
	(
	8{\mu_{23}}v_1
	+{\mu_{24}}v_1^2
	)
	\\
	h_{2}=\mu_{20}+(\mu_{22}
	+\mu_{21}+\mu_{23}v_2)v_1
	+\frac{1}{2}(\mu_{24}v_2+\mu_{25})v_1^2
\end{array}\right.$
%%%% INVARIANT SPACE%%%%%%
&
$\begin{array}{ll}\mathbf{W}_4=&\mathcal{L}\{\sin(\sqrt{\lambda_{10}}z),
	\cos(\sqrt{\lambda_{10}}z)\}
	\\&
	\times \mathcal{L}\{1,z)\}
\end{array}$
\\

5.
%%% FIRST COMPONENT%%%%%
&  $\left.\begin{array}{ll}
	\Phi_{1}= \kappa_1^2\gamma_{10}-\frac{\mu_{11}}{13\beta^2}v_1^2
	,~\Psi_{1}=0,~	h_{1}=(\mu_{11}v_1^2+\mu_{10})v_1,	
	~	\Phi_{2}=\frac{3\mu_{20}}{13\beta^2}
	%%%% SECOND COMPONENT%%%%%%
	,~
	\Psi_{2}=\kappa_1^2\eta_{20},~h_{2}=\mu_{20}v_1+\mu_{21}v_2
\end{array}\right.$
%%%% INVARIANT SPACE%%%%%%
&
$\begin{array}{ll}
	\mathbf{W}_4=\mathcal{L}\{e^{-\sqrt{\frac{13}{3}}\beta z},
	e^{\sqrt{\frac{13}{3}}\beta z}\}
	\\\qquad\,	\times\mathcal{L}\{e^{{\sqrt{3}}\beta z}, e^{-\frac{\beta}{\sqrt{3}} z}\},\beta\in\mathbb{R}
\end{array}$
\\
\\
6. %%% FIRST COMPONENT%%%%%
& $\left\{\begin{array}{ll}
	\Phi_{1}=\kappa_1^2\gamma_{10}+\frac{1}{\lambda_{20}}( \frac{\mu_{12}}{10}v_2+  \frac{\mu_{13}}{8}v_2^2+\frac{\mu_{14}}{27})
	\\
	\Psi_{1}=\frac{1}{\lambda_{20}}[  ({3\mu_{15}}v_2-\frac{\mu_{13}}{5})v_2-\frac{\mu_{12}}{8}v_1+{\mu_{11}}]~~~~~~
	\\
	h_{1}=[\mu_{10}+(\mu_{12}+\frac{\mu_{13}}{2}v_2)v_2
	+\mu_{14}v_1^2]v_1
	+\mu_{15}v_2^2v_2
	\\~~~~+\mu_{11}v_2
	\end{array}\right.
%%%% SECOND COMPONENT%%%%%%
\left\{\begin{array}{ll}
	\Phi_{2}=\frac{1}{\lambda_{20}}[(
	\frac{\mu_{24}}{35}v_1+
	\frac{\mu_{22}}{8}
+
	\frac{\mu_{23}}{18}v_2)v_2+
	\frac{\mu_{20}}{9}]
	\\
	\Psi_{2}=\kappa_1^2\eta_{20}+\frac{1}{\lambda_{20}}[\frac{\mu_{25}}{3}v_2^2
		+(\frac{\mu_{24}}{7}v_1
	+\frac{\mu_{23}}{15}v_2
	+\frac{\mu_{22}}{8})v_1]
	\\
	h_{2}=(\mu_{20}+\mu_{22}v_2+\frac{\mu_{23}}{2}v_2^2
	+\mu_{24}v_1^2)v_1+(\mu_{22}+\mu_{25}v_2^2)v_2	
\end{array}\right.$
%%%% INVARIANT SPACE%%%%%%
&
$\begin{array}{ll}\mathbf{W}_4=\mathcal{L}\{\sin(3\sqrt{\lambda_{20}}z),
	\cos(3\sqrt{\lambda_{20}}z)\}
\\~~~~~~~	\times \mathcal{L}\{\sin(\sqrt{\lambda_{20}}z),
	\cos(\sqrt{\lambda_{20}}z)\}
\end{array}$
\\
\\
7.
%%% FIRST COMPONENT%%%%%
&  $\left\{\begin{array}{ll}
	\Phi_{1}
	=
	\frac{1}{15\lambda_{20}}
	[20({\mu_{11}}v_2+\mu_{13}v_1^2)
	-(12\mu_{14}v_1^2
	\\~~~+2{\mu_{12}}v_2)v_2
	]
	+\kappa_1^2(\gamma_{10}-2\eta_{10}v_2)
	\\	\Psi_{1}
	=\kappa_1^2\eta_{10}v_1
	+\frac{1}{15\lambda_{20}}
	[20({\mu_{11}}v_2+\mu_{13}v_1^2)
	-2{\mu_{12}}v_2^2
\\~~	-12\mu_{14}v_1^2v_2
	]	
	\\
	h_{1}=(\mu_{10}+\mu_{11}v_2+\frac{\mu_{13}}{2}v_2^2
	+\mu_{14}v_1^2)v_1	
\end{array}\right.
%%%% SECOND COMPONENT%%%%%%
\left\{\begin{array}{ll}
	\Phi_{2}
	=\kappa_1^2(\gamma_{20}-2\eta_{21}v_2)v_1
	-\frac{2}{3\lambda_{20}}
	(
	2\mu_{22}v_2
	-6\mu_{20}
	)
	\\
	\Psi_{2}
	=\kappa_1^2(3\eta_{21}v_1^2
	+\eta_{20})
	+\frac{1}{3\lambda_{20}}
	(
	{\mu_{23}}v_2^2
	+4{\mu_{22}}v_1
	)
	\\
	h_{2}=\mu_{20}v_1
	+v_2(\mu_{21}+\mu_{22}v_1+\mu_{23}v_2^2)
\end{array}\right.$
%%%% INVARIANT SPACE%%%%%%
&
$\begin{array}{ll}\mathbf{W}_4=\mathcal{L}\{\sin(\frac{\sqrt{\lambda_{20}}}{2}z),
		\cos(\frac{\sqrt{\lambda_{20}}}{2}z)\}	\\~~~~~~~~~\times\mathcal{L}\{\sin(\sqrt{\lambda_{20}}z),
		\cos(\sqrt{\lambda_{20}}z)\}
\end{array}$
\\
\\
8.
%%% FIRST COMPONENT%%%%%
&  $\left\{\begin{array}{ll}
	\Phi_{1}= \frac{\mu_{11}}{(\lambda_{21}^2+\lambda_{10})}v_2-\frac{\mu_{12}}{3\lambda_{10}}v_1^2+\kappa_1^2\gamma_{10}
\\
\Psi_{1}=-\frac{\mu_{11}}{(\lambda_{21}^2+\lambda_{10})}v_1,~~	h_{1}=(\mu_{12}v_1^2+\mu_{11}v_2~~~~~~~~~~~~
\end{array}\right.
%%%% SECOND COMPONENT%%%%%%
\left\{\begin{array}{ll}	\Psi_{2}=\kappa_1^2\eta_{20}-\frac{\mu_{22}}{4\lambda_{21}^2}v_2
,~~
	\Phi_{2}=\frac{1}{4\lambda_{10}}({\mu_{24}v_1+4\mu_{23}})
	+\mu_{10})v_1\\
	%%%% SECOND COMPONENT%%%%%%
	h_{2}=\mu_{20}+\mu_{21}v_2+\frac{1}{2}(\mu_{22}v_2^2+\mu_{23}v_1^2)
\end{array}\right.$
%%%% INVARIANT SPACE%%%%%%
&
$\begin{array}{ll}
	\mathbf{W}_4=\mathcal{L}\{\sin(\sqrt{\lambda_{10}}z),
	\cos(\sqrt{\lambda_{10}}z)\}
\\~~~~~~~~\times\mathcal{L}\{ 1,e^{-\lambda_{21}z}\}
\end{array}$
\\
\\
9.
%%% FIRST COMPONENT%%%%%
&  $\left\{\begin{array}{ll}
	\Phi_{1}= \kappa_1^2\gamma_{10}-\frac{\mu_{11}}{3\lambda_{21}^2}v_1^2
\\	\Psi_{1}=\kappa_1^2\eta_{10}	,~~ h_{1}=(\mu_{11}v_1^2+\mu_{10})v_1~~~~~~~~~~~~~~~~~~~~~
\end{array}\right.
%%%% SECOND COMPONENT%%%%%%
\left\{\begin{array}{ll}
\Phi_{2}=-\frac{1}{4\lambda_{21}^2}({4\mu_{24}v_2+\mu_{22}v_1 +4\mu_{21}})
	\\
\Psi_{2}=-\frac{\mu_{25}}{4\lambda_{21}^2}v_2+\frac{\mu_{24}}{2\lambda_{21}^2}v_1
\\
h_{2}=\mu_{20}+\mu_{21}v_2+\frac{1}{2}(\mu_{22}v_2^2 +\mu_{23}v_1^2)
\end{array}\right.$
%%%% INVARIANT SPACE%%%%%%
&
$\begin{array}{ll}
	\mathbf{W}_4=\mathcal{L}\{e^{-\lambda_{21}z},e^{\lambda_{21}z}\}
	\times\mathcal{L}\{ 1,e^{-\lambda_{21}z}\}
\end{array}$
\\
\\
10. %%% FIRST COMPONENT%%%%%
& $\left\{\begin{array}{ll}
	\Phi_{1}
	=	-\frac{1}{12\beta^2}
	(
	{\mu_{12}}v_2	-\mu_{13}v_2^2
	-2{\mu_{14}}v_1^2
	)
	+\kappa_1^2\gamma_{10}
	\\	\Psi_{1}
	=-\frac{1}{6\beta^2}
	[	(
	\mu_{13}v_2
	+2{\mu_{12}})v_1
	+6{\mu_{11}}
	]
	\\
	h_{1}=\mu_{11}v_2+[\mu_{10}+(\mu_{12}+\frac{\mu_{13}}{2}v_2)v_2+\mu_{14}v_1^2]v_1~~~
\end{array}\right.
%%%% SECOND COMPONENT%%%%%%
\left\{\begin{array}{ll}
	\Phi_{2}
	=\frac{1}{18\beta^2}
	[
	2(\mu_{23}v_1
	+3{\mu_{22}}]v_2
	+9{\mu_{20}}
	]\\
	\Psi_{2}
	=\kappa_1^2\eta_{20}-\frac{1}{18\beta^2}
	[	(
	\mu_{23}v_1
	+6{\mu_{22}})v_1
	+6{\mu_{24}}v_2^2
	]
	\\
	h_{2}=\mu_{20}v_1
	+(\mu_{21}+\mu_{22}v_1+\frac{\mu_{23}}{2}v_1^2+\mu_{24}v_2^2 )v_2
	%%%% SECOND COMPONENT%%%%%%
	
\end{array}\right.$
%%%% INVARIANT SPACE%%%%%%
&
$\begin{array}{ll}\mathbf{W}_4=\mathcal{L}\{ \sin(\sqrt{2}\beta z),
	\cos(\sqrt{2}\beta z)\}\\\qquad\,\,
	\times \mathcal{L}\{e^{\beta z},e^{-\beta z}\},
	\beta\in\mathbb{R}
\end{array}$
\\
\\
11.
%%% FIRST COMPONENT%%%%%
&  $\left\{\begin{array}{ll}
	\Phi_{1}
	=
	k_1^2\gamma_{10}+
	\frac{\mu_{14}}{3\lambda_{10}}v_1^2,~		
	\Psi_{1}
	=\frac{\lambda_{20}^2		
		4\mu_{11}-	\lambda_{10}\mu_{11}(5\lambda_{20}+\lambda_{10})
	}{\lambda_{20}(\lambda_{10}-\lambda_{20})(\lambda_{10}-4\lambda_{20})}
	\\
	h_{1}=v_1(\mu_{10}+\mu_{14}v_1^2)+\mu_{11}v_2
	%%%% SECOND COMPONENT%%%%%%
\end{array}\right.
%%%% SECOND COMPONENT%%%%%%
\left\{\begin{array}{ll}
	\Phi_{2}
	=\frac{\mu_{20}	(4\lambda_{10}^2
				+\lambda_{20}-5\lambda_{20}\lambda_{10}
		)}{{\lambda_{10}(\lambda_{10}-\lambda_{20})(4\lambda_{10}-\lambda_{20})}}	,~
	\Psi_{2}
	=k_1^2\eta_{20}+
	\frac{\mu_{24}}{3\lambda_{20}}v_2^2
\\
	h_{2}=\mu_{20}v_1
	+v_2(\mu_{21}+\mu_{24}v_2^2)
\end{array}\right.$
%%%% INVARIANT SPACE%%%%%%
&
$\begin{array}{ll}\mathbf{W}_4=\mathcal{L}\{\sin(\sqrt{\lambda_{10}}z),
	\cos(\sqrt{\lambda_{10}}z)\}
\\\qquad\,\,	\times\mathcal{L}\{\sin(\sqrt{\lambda_{20}}z),
		\cos(\sqrt{\lambda_{20}}z)\}
\end{array}$
\\
\bottomrule
\end{tabular}
%*We choose $\lambda_{20}=-\beta^2,\beta\in\mathbb{R},$ in \eqref\eqref{rd:invariant conditionn=4(a)} for this case.
\end{sidewaystable}
\subsection{Generalized separable exact solutions for the given {nonlinear two-component system} \eqref{2+1rd} }
This section provides how to construct the generalized separable exact solutions for some particular cases of the given {nonlinear two-component system} \eqref{2+1rd}  along with suitable initial and boundary conditions by using the obtained invariant linear product spaces $\mathbf{W}_4$ provided in Tables \ref{table1}-\ref{table3}.
\begin{example}\label{eg1}
	Let us consider the  cubic {nonlinear two-component system} in $(2+1)$-dimensions given in \eqref{2+1rd}
	along with the given initial  and the Dirichlet boundary conditions  \eqref{2+1:ic}-\eqref{2+1:bc}
	when the functions $f_{si}(u_1,u_2),g_{si}(u_1,u_2),$ and $h_s(u_1,u_2),s,i=1,2,$ considered as follows:
	\begin{eqnarray}
		\label{eg1: coefficients of cubic nonlinear operator}
		\begin{aligned}
			f_{s1}(u_1,u_2)=&a_{{s5}}  u_1  ^{2}+a_{{s4}}  u_{{2}}^{2}+a_{{s3}} u_{{1}} u_2		+a_{{s2}}u_{{2}} +a_{{s1}}u_{{1}} 		+a_{{s0}} ,
			\\
			f_{s2}(u_1,u_2)=& b_{{s5}}  u_{{1}}^{2}		+b_{{s4}}  u_{{2}}  ^{2}		+b_{{s3}}u_{{1}}u_{{2}} +b_{{s2}}u_{{2}}+b_{{s1}}u_{{1}}	+b_{{s0}} ,
			\\
			g_{s1}(u_1,u_2)=& p_{{s5}}  u_{{1}}  ^{2}	+p_{{s4}}  u_{{2}}  ^{2}		+p_{{s3}}u_{{1}} u_{{2}}		+p_{{s2}}u_{{2}}+p_{{s1}}u_{{1}}		+p_{{s0}} ,
			\\
			g_{s2}(u_1,u_2)=&
			q_{{s5}}  u_{{1}}^{2}		
			+q_{{s4}}  u_{{2}}^{2}	
			+q_{{s3}}u_{{1}} u_{{2}}	
			+q_{{s2}}u_{{2}}
			+q_{{s1}}u_{{1}} 	
			+q_{{s0}} ,
			\\
			h_{s}(u_1,u_2)=&
			(c_{{s9}}  u_{{1}}^{2}	+c_{{s6}}	u_{{2}} ^{2}+c_{{s5}}u_{{1}}+c_{{s1}})u_{{1}}
			+(	c_{{s2}}+c_{{s4}}  u_{{2}}
			+c_{{s3}}u_{{1}}+c_{{s7}}u_{{1}} ^{2}		\\&\quad\quad+c_{{s8}}u_{{2}}^2	) u_{{2}} 	
			+c_{{s0}} ,
		\end{aligned}
	\end{eqnarray}
	$a_{si},b_{si},p_{si},q_{si},c_{sj}\in\mathbb{R},$ $i=0,1,\dots,5,j=0,1,\dots,9,$ and $s=1,2.$
		Now, we substitute the particular form $u_s(x_1,x_2,t)=v_s(z,t),z=\kappa_1x_1+\kappa_2x_2,s=1,2,$ into the above considered cubic type  {nonlinear two-component system} \eqref{2+1rd}.
	%with $f_{si}(u_1,u_2),g_{si}(u_1,u_2),$ and $h_s(u_1,u_2),s,i=1,2,$ as in \eqref{eg1: coefficients of cubic nonlinear operator}.
	 Thus, we obtain the reduced  cubic {nonlinear two-component system} of TFRDEs in $(1+1)$-dimensions for $
	 \alpha_s\in(0,2],s=1,2,$ as follows:
	 	\begin{eqnarray}
		\begin{aligned}{\label{eg1:transformedrd}}
			\dfrac{\partial^{\alpha_s} v_s}{\partial t^{\alpha_s}}=A_s(v_1,v_2)&\equiv	\dfrac{\partial }{\partial z}\Big[\big(\kappa_1^2a_{s0}+\kappa_2^2p_{s0}
			+(\kappa_1^2a_{s1}+\kappa_2^2p_{s1})v_1
			+(\kappa_1^2a_{s2}+\kappa_2^2p_{s2})v_2
			\\&		+(\kappa_1^2a_{s3}
			+\kappa_2^2p_{s3})v_1v_2
			+(\kappa_1^2a_{s4}
				+\kappa_2^2p_{s4})v_2^2
			+(\kappa_1^2a_{s5}
	+\kappa_2^2p_{s5})v_1^2
			\big) \frac{\partial v_1}{\partial z}
			\\&+\big(\kappa_1^2b_{s0}+\kappa_2^2q_{s0}
			+(\kappa_1^2b_{s1}+\kappa_2^2q_{s1})v_1
			+(\kappa_1^2b_{s2}+\kappa_2^2q_{s2})v_2
			+(\kappa_1^2b_{s3}
			\\&+\kappa_2^2q_{s3})v_1v_2
		 +(\kappa_1^2b_{s4}  +\kappa_2^2q_{s4})v_2^2
			+(\kappa_1^2b_{s5}+\kappa_2^2q_{s5})v_1^2
			\big)\frac{\partial v_2}{\partial z}\Big]
				+c_{{s0}}		+c_{{s1}}v_{{1}}
		\\&		+c_{{s2}}v_{{2}} 	
		+c_{{s3}}v_{{1}}v_{{2}}
				+c_{{s4}}v_{{2}}^{2}	
			+c_{{s5}}v_{{1}} ^{2}	
			+c_{{s6}}	v_{{1}} v_{{2}} ^{2}	
			+c_{{s7}}v_{{1}} ^{2}	v_{{2}} 	+c_{{s8}}v_{{2}}^3		+c_{{s9}}  v_{{1}}^{3}	.
		\end{aligned}
	\end{eqnarray}
We observe that the above reduced {nonlinear two-component system} \eqref{eg1:transformedrd} admits the invariant linear product space 	
\begin{eqnarray}\label{eg1:vector space}
\mathbf{W}_4:=\mathit{W}_{1,2}\times\mathit{W}_{2,2}= \mathcal{L}\{1,e^{-\lambda_{11}z}\}
\times
\mathcal{L}\{1,z\},
\end{eqnarray}
if
\begin{eqnarray}
	\label{eg1:constraints}
\begin{array}{lllll}
			&	a_{10}=-\dfrac{\kappa_2^2p_{10}}{\kappa_1^2} +\gamma_{10},
					&	a_{20}=-\dfrac{\lambda_{11}^2\kappa_2^2 p_{20} +\mu_{21}}{\lambda_{11}^2\kappa_1^2},	
			&
	b_{12}=-\dfrac{\kappa_{2}^2q_{12}}{\kappa_{1}^2} +\eta_{12},	
			\\	& b_{10}=-\dfrac{\kappa_2^2q_{10}}{\kappa_1^2}+\eta_{10},	
			&				
			b_{22}=-\dfrac{\kappa_2^2q_{23}}{\kappa_1^2}+\eta_{21},
			&
			b_{20}=-\dfrac{\kappa_2^2 q_{20}}{\kappa_1^2}+\eta_{20},	
	%		& c_{2i}=0,i=2,4,5,\dots,9,	
		\\
			&	c_{si}=\mu_{1i},i=0,1,
			&
			b_{si}=-\dfrac{\kappa_2^2 q_{1i}}{\kappa_1^2},i=1,3,4,5,
		\\	& c_{1i}=0,
		i=2,3,\dots,9,	& \text{ and } 	a_{si}=-\dfrac{\kappa_2^2 p_{1i}}{\kappa_1^2},&i=1,2,\dots,5, s=1,2.
		\end{array}
	\end{eqnarray}
{
	Additionally, note that if $(v_1,v_2)\in\mathbf{W}_4=\mathit{W}_{1,2}\times \mathit{W}_{2,2}$ given in \eqref{eg1:vector space} with $\mathit{W}_{1,2}= \mathcal{L}\{1,e^{-\lambda_{11}z}\}
$ and $
\mathit{W}_{2,2}=	\mathcal{L}\{1,z\},$  then we get
	\begin{eqnarray*}
		\begin{aligned}
		&	A_1(v_1,v_2)=\mu_{10}+ \mu_{11}p_1+\kappa_1^2 \eta_{12} q_2^2+p_2\left(\kappa_1^2\lambda_{11}^2\gamma_{10}+\mu_{11}\right)e^{-\lambda_{11}z}\in\mathit{W}_{1,2}= \mathcal{L}\{1,e^{-\lambda_{11}z}\} \\&
		\text{and } A_2(v_1,v_2)=\mu_{20}+ \mu_{21} p_1+\kappa_1^2 \eta_{21} q_2^2\in \mathit{W}_{2,2}=	\mathcal{L}\{1,z\},
		\end{aligned}
	\end{eqnarray*}
for every   $v_1=p_1+p_2e^{-\lambda_{11}z}\in \mathit{W}_{1,2},$ and $v_2=q_1+q_2x\in\mathit{W}_{1,2},$
$p_i,q_i\in\mathbb{R},i=1,2,$ 
where $A_s(v_1,v_2), s=1,2,$  are given in \eqref{eg1:transformedrd}  along with \eqref{eg1:constraints}.	The above-discussed case is a particular case of  Table \ref{table1} of case 6.}
For this case,  the expected   generalized separable exact solution $\mathbf{V}=(v_1,v_2)$ with $v_s=v_s(z,t)$ is as follows:
	\begin{eqnarray}
		\begin{aligned}
			&		v_1(z,t)=P_1(t)+P_2(t)e^{-\lambda_{11}z},
			\text { and }
			v_2(z,t)=Q_1(t)+Q_2(t)z,
		\end{aligned}
	\end{eqnarray}
	where $ z=\kappa_1x_1+\kappa_2x_2,$  and the functions
		 $P_1(t),P_2(t), Q_1(t),$ and $Q_2(t)$ satisfy the following system of fractional-order ODEs:
	\begin{eqnarray}
		\label{eq1:reduced system of ODEs}
		\begin{aligned}
			&	\dfrac{d^{\alpha_1} P_1(t)}{d t^{\alpha_1}}=\mu_{10}+ \mu_{11}P_1(t)+\kappa_1^2 \eta_{12} Q_2^2(t),
			&	\dfrac{d^{\alpha_1} P_2(t)}{d t^{\alpha_1}}=\left(\kappa_1^2\lambda_{11}^2\gamma_{10}+\mu_{11}\right)P_2(t),
		\\
			&
			\dfrac{d^{\alpha_2} Q_1(t)}{d t^{\alpha_2}}= \mu_{20}+ \mu_{21} P_1(t)+\kappa_1^2 \eta_{21} Q_2^2(t), & \text{and }
			\dfrac{d^{\alpha_2} Q_2(t)}{d t^{\alpha_2}}= 0.\qquad\qquad\qquad
		\end{aligned}
	\end{eqnarray}
We now solve the above system of fractional-order ODEs \eqref{eq1:reduced system of ODEs} using the Laplace transformation method. We know that the Laplace transformation of Caputo fractional-order derivative is obtained in \cite{Podlubny1999} as follows:
	\begin{eqnarray}
		\begin{aligned}
			L\left( \dfrac{d^\alpha R(t)}{d t^\alpha};s\right) =s^\alpha L(R(t);s) -\sum\limits_{k=0}^{n-1}s^{\alpha-(k+1)}\dfrac{d^k R(t)}{d t^k}\Big{|}_{t=0},\alpha\in(n-1,n],n\in\mathbb{N}.
		\end{aligned}
	\end{eqnarray}
	Hence,  we apply the Laplace transformation to the second and last equations of the above system of fractional-order ODEs \eqref{eq1:reduced system of ODEs} and rearranging the terms to obtain the following:
	$$
		L(P_2(t);s)=\left\{\begin{array}{ll}
			\rho_{20}	\left( \dfrac{s^{{\alpha_1}-1}}{s^{\alpha_1}-\varrho_1}\right) ,\text{ if } {\alpha_1}\in(0,1],
			\\
		\sum\limits_{k=0}^1	\rho_{2k}	\left( 	\dfrac{s^{{\alpha_1}-k-1}}{s^{\alpha_1}-\varrho_1}\right)
			,\text{ if }{\alpha_1}\in(1,2],
		\end{array}
		\right.
		$$
		$$
 \&\, 	L(Q_2(t);s)=\left\{\begin{array}{ll}
			\dfrac{1}{s}\zeta_{20},\text{ if } {\alpha_2}\in(0,1],
			\\
			\sum\limits_{k=0}^1	\dfrac{1}{s^{k+1}}\zeta_{2i},\text{ if } {\alpha_2}\in(1,2],
		\end{array}
		\right.
	$$
	where $\varrho_1=\kappa_1^2\lambda_{11}^2\gamma_{10}+\mu_{11},$ $\rho_{20}=P_2(0),$ $\rho_{21}=\dfrac{d P_2(t)}{d t}\Big{|}_{t=0},$ $\zeta_{20}=Q_2(0)$ and $\zeta_{21}=\dfrac{d Q_2(t)}{d t}\Big{|}_{t=0}$ are arbitrary constants.
Then,  by taking inverse Laplace transformation of the above equations, we get
\begin{eqnarray}
		\begin{aligned}
		& P_2(t)=\left\{
		\begin{array}{ll}
			\rho_{20}E_{{\alpha_1},1}(\varrho_1t^{\alpha_1}),
			\text{ if } {\alpha_1}\in(0,1],
			\\
			\sum\limits_{k=0}^1\rho_{2k}t^{k+1}E_{{\alpha_1},k+1}(\varrho_1t^{\alpha_1}),
		\text{ if } {\alpha_1}\in(1,2],
		\end{array}
		\right.	
\\&	\&\	Q_2(t)=\left\{\begin{array}{ll}
			\zeta_{20},\text{ if } {\alpha_2}\in(0,1],
			\\
			\zeta_{20}+\zeta_{21}t,\text{ if } {\alpha_2}\in(1,2],
		\end{array}
		\right.
	\end{aligned}\label{p2q2}
	\end{eqnarray}
	where $E_{a_1,a_2}(\cdot)$ is the two-parameter Mittag-Leffler function, which is defined in \cite{Podlubny1999}
	as $E_{a_1,a_2}(t)=\sum\limits_{n=0}^\infty\dfrac{t^n}{\Gamma(a_1n+a_2)},$ $\mathfrak{R}({a_i})>0,i=1,2.$
	
Now, we substitute the above obtained form of $Q_2(t)$ into the first equation of the  system of fractional-order ODEs \eqref{eq1:reduced system of ODEs} and using a procedure similar to the above, we obtain $P_1(t)$ as follows:

\begin{eqnarray}
\begin{aligned}
	&
	P_1(t)=\left\{
	\begin{array}{ll}
		\rho_{10}E_{{\alpha_1},1}(\mu_{11}t^{\alpha_1}) +\varrho_2t^\alpha E_{{\alpha_1},{\alpha_1}+1}(\mu_{11}t^{\alpha_1}), \text{ if } {\alpha_s}\in(0,1],s=1,2,
		\\
		%%%%%%%%%%%%%%%%%%%%%%%%%%%%%%%%%%%%%%%%%%%%%%%%%%%%%%%%%%%%%%%%%%%%%%%%%%%%%%%%%%%%%%%%%%%%%%
		\rho_{10}E_{{\alpha_1},1}(\mu_{11}t^{\alpha_1}) +\varrho_2t^{\alpha_1} E_{{\alpha_1},{\alpha_1}+1}(\mu_{11}t^{\alpha_1})
		+2\kappa_1^2\zeta_{21}
		\eta_{12}t^{{\alpha_1}+1} 
		\Big[ \zeta_{20} E_{{\alpha_1},{\alpha_1}+2}(\mu_{11}t^{\alpha_1})	\\+\zeta_{21}t E_{{\alpha_1},{\alpha_1}+3}(\mu_{11}t^{\alpha_1}) \Big] ,
		\text{ if } {\alpha_1}\in(0,1]\, \& \,  {\alpha_2}\in(1,2],
		\\
		%%%%%%%%%%%%%%%%%%%%%%%%%%%%%%%%%%%%%%%%%%%%%%%%%%%%%%%%%%%%%%%%%%%%%%%%%%%%%%%%%%%%%%%%%%%%%%
		\rho_{10}E_{{\alpha_1},1}(\mu_{11}t^{\alpha_1})+\rho_{11}tE_{\alpha,2}(\mu_{11}t^{\alpha_1})+\varrho_2t^\alpha E_{{\alpha_1},{\alpha_1}+1}(\mu_{11}t^{\alpha_1}),
		\\
		 \text{ if } {\alpha_1}\in(1,2]\, \& \,  {\alpha_2}\in(0,1],
		\\
		%%%%%%%%%%%%%%%%%%%%%%%%%%%%%%%%%%%%%%%%%%%%%%%%%%%%%%%%%%%%%%%%%%%%%%%%%%%%%%%%%%%%%%%%%%%%%%
		\rho_{10}E_{{\alpha_1},1}(\mu_{11}t^{\alpha_1})+\rho_{11}tE_{\alpha,2}(\mu_{11}t^{\alpha_1})+2\kappa_1^2\zeta_{21}
		\eta_{12}t^{{\alpha_1}+1}
		\Big[
		\zeta_{20} E_{{\alpha_1},{\alpha_1}+2}(\mu_{11}t^{\alpha_1})
	\\	+\zeta_{21}t E_{{\alpha_1},{\alpha_1}+3}(\mu_{11}t^{\alpha_1})
		\Big]  +\varrho_2t^{\alpha_1} E_{{\alpha_1},
			{\alpha_1}+1}(\mu_{11}t^{\alpha_1}),
				\text{ if } {\alpha_s}\in(1,2],s=1,2,
	\end{array}
	\right.	
	\end{aligned}\label{p1}
\end{eqnarray}
	where $\varrho_2=\mu_{10}+\kappa_1^2\zeta_{20}^2\eta_{12}$ and $\rho_{10},\rho_{11}\in\mathbb{R}.$
	Proceeding in a similar manner, we obtain $Q_1(t)$ as follows:
\begin{eqnarray}
		\begin{aligned}
			&
		Q_1(t)=\left\{
		\begin{array}{ll}
			\zeta_{10}+\dfrac{\varrho_3t^{\alpha_2}}{\Gamma({\alpha_2}+1)}
			+\mu_{21}t^{\alpha_2}(\rho_{10}
			E_{{\alpha_1},{\alpha_2}+1}(\mu_{11}t^{\alpha_1})
\\	+\varrho_2t^{\alpha_1} E_{{\alpha_1},{\alpha_1}+{\alpha_2}+1}(\mu_{11}t^{\alpha_1})), 
						\text{ if } {\alpha_s}\in(0,1],s=1,2,\\
			%%%%%%%%%%%%%%%%%%%%%%%%%%%%%%%%%%%%%%%%%%%%%%%%%%%%%%%%%%%%%%%%%%%%%%%%%%%%%%%%%%%%%%%%%%%%%%
\dfrac{\varrho_3t^{\alpha_2}}{\Gamma({\alpha_2}+1)}+\dfrac{\varrho_4t^{{\alpha_2}+1}}{\Gamma({\alpha_2}+2)} +\dfrac{\varrho_5t^{{\alpha_2}+2}}{\Gamma({\alpha_2}+3)}
					+\Big[	\rho_{11}E_{{\alpha_1},{\alpha_2}+1}(\mu_{11}t^{\alpha_1})
					\\+\varrho_2t^{\alpha_1} E_{{\alpha_1},{\alpha_1}+{\alpha_2}+1}(\mu_{11}t^{\alpha_1})
			+2\kappa_1^2\zeta_{21}
		\eta_{12}t^{{\alpha_1}+1}\sum\limits_{i=0}^1\zeta_{2i} \\\times E_{{\alpha_1},{\alpha_1}+{\alpha_2}+i+2}(\mu_{11}t^{\alpha_1})	
			\Big] \mu_{21}t^{\alpha_2}+\zeta_{10}+\zeta_{11}t,
			\text{ if } {\alpha_1}\in(0,1]\, \& \,{\alpha_2}\in(1,2],\\
			%%%%%%%%%%%%%%%%%%%%%%%%%%%%%%%%%%%%%%%%%%%%%%%%%%%%%%%%%%%%%%%%%%%%%%%%%%%%%%%%%%%%%%%%%%
			\zeta_{10}
			+\mu_{21}t^{\alpha_2}\Big[\sum\limits_{i=0}^1\rho_{1i}
			E_{{\alpha_1},{\alpha_2}+i+1}(\mu_{11}t^{\alpha_1})
					+\varrho_2t^{\alpha_1} E_{{\alpha_1},{\alpha_1}+{\alpha_2}+1}(\mu_{11}t^{\alpha_1})\Big]
			\\	+\dfrac{\varrho_3t^{\alpha_2}}{\Gamma({\alpha_2}+1)}	, \text{ if } {\alpha_1}\in(0,1]\,\&\, {\alpha_2}\in(1,2],
			\\
			%%%%%%%%%%%%%%%%%%%%%%%%%%%%%%%%%%%%%%%%%%%%%%%%%%%%%%%%%%%%%%%%%%%%%%%%%%%%%%%%%%%%%%%%%%%%%%
	\zeta_{10}+\zeta_{11}t+	\mu_{21}t^{\alpha_2}\Big[	
		2\kappa_1^2	\zeta_{21}
		\eta_{12}t^{{\alpha_1}+1}
	\sum\limits_{i=0}^1\zeta_{2i} E_{{\alpha_1},{\alpha_1}+{\alpha_2}+i+2}(\mu_{11}t^{\alpha_1})
	\\
	+\varrho_2t^{\alpha_1}
		 E_{{\alpha_1},{\alpha_1}+{\alpha_2}+1}(\mu_{11}t^{\alpha_1})
			+\sum\limits_{i=0}^1\rho_{1i}E_{{\alpha_1},{\alpha_2}+i+1}(\mu_{11}t^{\alpha_1})
			\Big]	+\dfrac{\varrho_3t^{\alpha_2}}{\Gamma({\alpha_2}+1)}
			\\
			+\dfrac{\varrho_4t^{{\alpha_2}+1}}{\Gamma({\alpha_2}+2)} +\dfrac{\varrho_5t^{{\alpha_2}+2}}{\Gamma({\alpha_2}+3)}
			,
			\text{ if } {\alpha_s}\in(1,2],s=1,2,\\
			%%%%%%%%%%%%%%%%%%%%%%%%%%%%%%%%%%%%%%%%%%%%%%%%%%%%%%%%%%%%%%%%%%%%%%%%%%%%%%%%%%%%%%%%%%
		\end{array}
		\right.	
			\end{aligned}\label{q1}
\end{eqnarray}
	where $\varrho_3=\mu_{20}+\kappa_1^2\zeta_{20}^2\eta_{21},$
	$\varrho_4=2\kappa_1^2\zeta_{20}\zeta_{21}\eta_{21},$
	and $ \varrho_5=2\kappa_1^2\zeta_{21}^2\eta_{21}.$
	{Hence, for the given cubic {nonlinear two-component system} \eqref{2+1rd}
	along with  functions $f_{si}(u_1,u_2),g_{si}(u_1,u_2),$ and $h_s(u_1,u_2),s,i=1,2,$ as given in \eqref{eg1: coefficients of cubic nonlinear operator} and the  parameter constraints   \eqref{eg1:constraints}, we obtain the generalized separable exact solution as follows:}
	\begin{eqnarray}\label{eg1:solution-u1}
		\begin{aligned}
			&
			u_1(x_1,x_2,t)=P_1(t)+P_2(t)e^{-\lambda_{{11}}z},
%		\end{aligned}\end{eqnarray}\begin{eqnarray}\begin{aligned}\label{eg1:solution-u2}
\text{ and }			
			u_2(x_1,x_2,t)=Q_1(t)+Q_2(t)z,
		\end{aligned}
	\end{eqnarray}
where the functions $P_1(t),Q_1(t),P_2(t)$ and $Q_2(t)$ are given in \eqref{p1}, \eqref{q1}, and  \eqref{p2q2}, respectively, and $z=\kappa_1x_1+\kappa_2x_2.$
	Additionally, we observe that the obtained solution \eqref{eg1:solution-u1}
%	-\eqref{eg1:solution-u2}
	 satisfies  the given initial and the Dirichlet boundary conditions \eqref{2+1:ic}-\eqref{2+1:bc} when
	$$
	\begin{array}{lllll}
&	\delta_{11}=\rho_{10}+\rho_{11}e^{-\lambda_{{11}}z},
&\delta_{12}=\rho_{20}+\rho_{21}e^{-\lambda_{{11}}z},
&	\delta_{21}=\zeta_{10}+\zeta_{11}z,
\\&	\delta_{22}=\zeta_{20}+\zeta_{21}z,
		&\tau_{11}(t)=		P_1(t)+P_2(t),	
	&\tau_{12}(t)=P_1(t)+r_1P_2(t),		
\\	&\tau_{21}(t)= Q_1(t),\	\&			&\tau_{22}(t)=Q_1(t)+rQ_2(t),
	\end{array}$$
	where $z=\kappa_1x_1+\kappa_2x_2,$ and $r_1=e^{-\lambda_{{11}}r},r\in\mathbb{R}.$
	We would  like to point out that the above obtained exact solutions \eqref{eg1:solution-u1}
	%-\eqref{eg1:solution-u2}
	of \eqref{2+1rd} with the coefficient functions as given in \eqref{eg1: coefficients of cubic nonlinear operator} and the  parameter constraints   \eqref{eg1:constraints} hold for every $\alpha_s\in(0,2],s=1,2,$ including the integer-values of $\alpha_s=1,$ $2,s=1,2.$
	\begin{note}
	{	\label{note-eg1}
		We  wish to mention here that when
		$
		p_{10}=k_{11},
		a_{20}=k_{21},
		q_{s0}=k_{s2},$
		$
		p_{20}=-\dfrac{\kappa_1^2k_{21}}{\kappa_2^2},$
		$
\gamma_{10}=
\left( 1+\dfrac{\kappa_2^2}{\kappa_1^2}\right) k_{11},$
 $
\eta_{s0}=
\left( 1+\dfrac{\kappa_2^2}{\kappa_1^2}\right) k_{s2},
$ 	$a_{si}=b_{si}=p_{si}=q_{si}=0,i=1,2,\dots,5,$ and $c_{sj}=0,j=0,1,\dots,9,s=1,2,$	in \eqref{eg1:constraints}, then the above-discussed  two-component  system of TFRDEs in $(2+1)$-dimensions coincides with the given  physical model \eqref{reduced-zhang-model}, which was	studied in \cite{zhang2017} through the Hankel integral
transform method. For the system \eqref{reduced-zhang-model}, they derived an infinite series solution in terms Mittag-Leffler functions. Also, we observe that the obtained exact solution in generalized separable form \eqref{eg1:solution-u1}
%-\eqref{eg1:solution-u2}
is  valid for the physical model \eqref{reduced-zhang-model} with initial and boundary conditions \eqref{2+1:ic}-\eqref{2+1:bc} under the above-mentioned parametric constraints.}
\end{note}
\end{example}
\begin{example}
	\label{eg2}
	The linear product space
	\begin{eqnarray}\label{eg2:vector space}
		\mathbf{W}_4= \mathcal{L}\{ \sin(\theta z) , \cos( {\theta} z)\}
		\times
		\mathcal{L}\{\sin( \sqrt{\lambda_{20}}z) , \cos( \sqrt{\lambda_{20}}z) \},
	\end{eqnarray}
$\theta= \sqrt{\frac{5\lambda_{20}}{2}},\lambda_{20}>0,z=\kappa_1x_1+\kappa_2x_2$	is invariant under the  given {nonlinear two-component system} \eqref{2+1rd} with $f_{si}(u_1,u_2),g_{si}(u_1,u_2),$ and $h_s(u_1,u_2),i,s=1,2,$ as given in \eqref{eg1: coefficients of cubic nonlinear operator} when the constants read as follows:
	\begin{eqnarray}
	\small	\begin{array}{lll}
			\label{eg2:constraints}
			a_{10}=\gamma_{10}-\dfrac{\kappa_2^2p_{10}}{\kappa_1^2}, 		&
			a_{12}=\dfrac{2\mu_{12}-3\kappa_2^2\lambda_{20} p_{12}}{3\kappa_1^2\lambda_{20}},
			&a_{14}=-\dfrac{3\kappa_2^2\lambda_{20} p_{14}+\mu_{13}}{3\kappa_1^2\lambda_{20}},
			\\
			a_{15}=\dfrac{2\mu_{14}-15\kappa_2^2\lambda_{20} p_{15}}{15\kappa_1^2\lambda_{20}},
			&
			b_{10}=\dfrac{\mu_{11}-\kappa_2^2\lambda_{20} q_{10}}{\kappa_1^2\lambda_{20}},
			&
			b_{11}=\dfrac{2\mu_{12}-3\kappa_2^2\lambda_{20} q_{11}}{3\kappa_1^2\lambda_{20}},
			\\
			b_{13}=-\dfrac{3\kappa_2^2\lambda_{20} q_{13}-2\mu_{13}}{3\kappa_1^2\lambda_{20}},
			&
			a_{1i}=-\dfrac{\kappa_2^2 p_{1i}}{\kappa_1^2},i=1,3,
			&
			b_{1i}=-\dfrac{\kappa_2^2 q_{1i}}{\kappa_1^2},i=2,4,5,\\
			c_{16}=\dfrac{\mu_{13}}{2}, c_{27}=\dfrac{\mu_{23}}{2},
			&
			a_{20}=-\dfrac{5\kappa_2^2\lambda_{20} p_{20}-2\mu_{20}}{5\kappa_1^2\lambda_{20}},
			&
			a_{22}=-\dfrac{3\kappa_2^2\lambda_{20} p_{22}-2\mu_{22}}{3\kappa_1^2\lambda_{20}},
			\\	
			a_{23}=-\dfrac{9\kappa_2^2\lambda_{20} p_{23}-\mu_{23}}{9\kappa_1^2\lambda_{20}},
			&
			b_{20}=-\dfrac{\kappa_2^2\lambda_{20} q_{20}-\kappa_1^2\lambda_{20} \eta_{20}}{\kappa_1^2\lambda_{20}},
			&
			b_{21}=-\dfrac{3\kappa_2^2\lambda_{20} q_{21}+2\mu_{22}}{3\kappa_1^2\lambda_{20}},
			\\	b_{24}=-\dfrac{3\kappa_2^2 \lambda_{20} q_{24}-\mu_{24}}{3\kappa_1^2\lambda_{20}},
			&	b_{25}=-\dfrac{18\kappa_2^2\lambda_{20} q_{25}-\mu_{23}}{18\kappa_1^2\lambda_{20}},& a_{2i}=-\dfrac{\kappa_2^2 p_{2i}}{\kappa_1^2},i=1,4,5,
%	\end{array}\end{eqnarray*}\begin{eqnarray*}\begin{array}{lll}
\\
		 		b_{2i}=-\dfrac{\kappa_2^2 q_{2i}}{\kappa_1^2},i=2,3,&
			c_{si}=\mu_{si-1},i=1,2,3,s=1,2,	&c_{16}=\mu_{13},	c_{19}=\mu_{14},
			\\
			c_{27}=\mu_{23},	c_{28}=\mu_{24},& c_{1i}=0,i=0,4,5,7,8, & \text{ and } c_{2j}=0,j=0,4,5,6,9.
		\end{array}
	\end{eqnarray}
	Substituting the particular form $(u_1(x_1,x_2,t),u_2(x_1,x_2,t))=(v_1(z,t),v_2(z,t)),z=\kappa_1x_1+\kappa_2x_2,$ into the {{nonlinear two-component system} \eqref{2+1rd} with the coefficient functions $f_{si}(u_1,u_2), $ $g_{si}(u_1,u_2),$ and $h_s(u_1,u_2),s,i=1,2,$ as given in \eqref{eg1: coefficients of cubic nonlinear operator} under the above-given  parameter constraints,} %\eqref{eg2:constraints},
 we obtain the following  reduced two-component system in $(1+1)$-dimensions as:
	\begin{eqnarray}
	\small	\begin{aligned}{\label{eg2:transformedrd}}
			\dfrac{\partial^{\alpha_1} v_1}{\partial t^{\alpha_1}}=A_1(v_1,v_2)\equiv&	\dfrac{\partial }{\partial z}\left[ \dfrac{1}{15\lambda_{20}} ({15\kappa_1^2\lambda_{20}\gamma_{10} +10\mu_{12}v_2+5\mu_{13}v_2^2+2\mu_{14}v_1^2}
			)\frac{\partial v_1}{\partial z}		\right.	\\&
			\left.+\dfrac{1}{3\lambda_{20}}(3c_{12} -2\mu_{12}v_1
					+2\mu_{13}v_2v_1
			)\frac{\partial v_2}{\partial z}\right]
			+\mu_{{10}}v_{{1}}
			+\mu_{{11}}v_{{2}} 	
			+\mu_{{12}}v_{{1}}v_{{2}}
		\\&	+\dfrac{1}{2}\mu_{13}v_{{1}} v_{{2}} ^{2}	
			+	\mu_{{14}}  v_{{1}}^{3}	
			,
			\\
			\dfrac{\partial^{\alpha_2} v_2}{\partial t^{\alpha_2}}=A_2(v_1,v_2)\equiv&	\dfrac{\partial }{\partial z}\left[ \dfrac{1}{45\lambda_{20}}(18\mu_{20} +5\mu_{23}v_1v_2
			+30\mu_{22}v_2
			) \frac{\partial v_1}{\partial z}		+\dfrac{1}{18\lambda_{20}}(18\kappa_1^2\lambda_{20}\eta_{20}	\right.		\\ & 	\left. -12\mu_{22}v_1
					+6\mu_{24}v_2^2-\mu_{23}v_1^2
			)\frac{\partial v_2}{\partial z} \right]
			+\mu_{{20}}v_{{1}}
			+\mu_{{21}}v_{{2}} 	
		\\&	+\mu_{{22}}v_{{1}}v_{{2}}
			+\dfrac{1}{2}\mu_{23}v_{{1}} ^{2}	 v_{{2}}
			+	\mu_{{24}}  v_{{2}}^{3}, \alpha\in(0,2].
		\end{aligned}
	\end{eqnarray}
{
Note that if $(v_1,v_2)\in\mathbf{W}_4=\mathit{W}_{1,2}\times \mathit{W}_{2,2}$ given in \eqref{eg2:vector space} with $\mathit{W}_{1,2}= \mathcal{L}\{sin(\theta z) ,\cos( {\theta} z) \},$  and  $\mathit{W}_{2,2}= \mathcal{L}\{sin( \sqrt{\lambda_{20}}z) ,\cos( \sqrt{\lambda_{20}}z) \},$  then we obtain
		\begin{eqnarray*}
			\begin{aligned}
				&	A_1(v_1,v_2)=  p_1(\mu_{11}-\dfrac{5}{2}\kappa_1^2\lambda_{20}\gamma_{10})\sin(\theta z) +p_2(\mu_{11}-\dfrac{5}{2}\kappa_1^2\lambda_{20}\gamma_{10})\cos( {\theta} z)  \in\mathit{W}_{1,2}
				%= \mathcal{L}\{sin(\theta z) ,\cos( {\theta} z) \}, 
				\text{ and }
				\\ &
				A_2(v_1,v_2)=  q_1(\mu_{21}-\kappa_1^2\lambda_{20}\eta_{20})\sin( \sqrt{\lambda_{20}}z) +q_2(\mu_{21}-\kappa_1^2\lambda_{20}\eta_{20})\cos( \sqrt{\lambda_{20}}z) \in \mathit{W}_{2,2}%=\mathcal{L}\{sin( \sqrt{\lambda_{20}}z) ,\cos( \sqrt{\lambda_{20}}z) \}
				,
			\end{aligned}
		\end{eqnarray*}
	whenever  $v_1=p_1\sin(\theta z) +p_2\cos( {\theta} z)$ and $v_2
	=q_1\sin( \sqrt{\lambda_{20}}z) +q_2\cos( \sqrt{\lambda_{20}}z) \},$	where $A_s(v_1,v_2), s=1,2,$ are given in \eqref{eg2:transformedrd}, $p_1,p_2,q_1,q_2\in\mathbb{R},$ $\theta= \sqrt{\frac{5\lambda_{20}}{2}},$ and  $\lambda_{20}>0.$
		This case is discussed in case 3 of Table \ref{table3}.}
	Thus, following a similar procedure as discussed in Example \ref{eg1}, we obtain the generalized separable exact solution for the given {{nonlinear two-component system} \eqref{2+1rd} along  with the coefficient functions as given in \eqref{eg1: coefficients of cubic nonlinear operator}  and $ \mu_{s3}=\mu_{s4}=0,s=1,2,$ in the given parameter constraints \eqref{eg2:constraints}}  as follows:
	\begin{eqnarray}
		\begin{aligned}
			\label{eg2:solutionu1}
			&u_1(x_1,x_2,t)=v_1(z,t)=\left\{\begin{array}{ll}
				\Big[ \rho_{10}\sin\big( \sqrt{\frac{5\lambda_{20}}{2}}z\big) +\rho_{20} \cos\big( \sqrt{\frac{5\lambda_{20}}{2}}z\big)\Big] E_{{\alpha_1},1}(\varrho_1t^{\alpha_1}) ,
				\\		 \text{ if }  {\alpha_1}\in(0,1]\,\&\, {\alpha_2}\in(0,2],
				\\
				\big[ \rho_{10}\sin\big( \sqrt{\frac{5\lambda_{20}}{2}}z\big) +\rho_{20} \cos\big( \sqrt{\frac{5\lambda_{20}}{2}}z\big)\big] E_{{\alpha_1},1}(\varrho_1t^{\alpha_1})
			\\	+ 
				\big[ \rho_{11}\sin\big( \sqrt{\frac{5\lambda_{20}}{2}}z\big) 
								+\rho_{21} \cos\big( \sqrt{\frac{5\lambda_{20}}{2}}z\big)\big] tE_{{\alpha_1},2}(\varrho_1t^{\alpha_1}),
				\\ \text{ if } {\alpha_1}\in(1,2]\,\&\, {\alpha_2}\in(0,2],
			\end{array}\right.
		\end{aligned}	
\\
		\begin{aligned}
				&
		%	\label{eg2:solutionu2}
			u_2(x_1,x_2,t)=v_2(z,t)=\left\{\begin{array}{ll}
				\big[ \zeta_{10}\sin\big( \sqrt{\lambda_{20}}z\big) + \zeta_{20} \cos\big( \sqrt{\lambda_{20}}z\big)\big] E_{{\alpha_2},1}(\varrho_2t^{\alpha_2}) ,
						\\	 \text{ if }  {\alpha_1}\in(0,2]\, \& \,{\alpha_2}\in(0,1],
				\\
				\big[  \zeta_{10}\sin\big( \sqrt{\lambda_{20}}z\big) + \zeta_{20} \cos\big(\sqrt{\lambda_{20}}z\big)\big] E_{{\alpha_2},1}(\varrho_2t^{\alpha_2})\\
				+ \big[  \zeta_{11}\sin\big( \sqrt{\lambda_{20}}z\big)
				\\
				+ \zeta_{21} \cos\big( \sqrt{\lambda_{20}}z\big)\big] tE_{{\alpha_2},2}(\varrho_2t^{\alpha_2 }), 
				\\
				\text{ if }  {\alpha_1}\in(0,2]\, \& \, {\alpha_2}\in(1,2],
			\end{array}\right.
		\end{aligned}
	\end{eqnarray}
	where $z=\kappa_1x_1+\kappa_2x_2,$ $\varrho_1=-\dfrac{5}{2}\kappa_1^2\lambda_{20}\gamma_{10}+ \mu_{11},$ $ \varrho_2=-\kappa_1^2\lambda_{20}\eta_{20}+\mu_{21},$ and $\rho_{i0},\rho_{i1},\zeta_{i0},\zeta_{i1}\in\mathbb{R},$ $i=1,2. $
	We observe that the above exact solution \eqref{eg2:solutionu1}
	%-\eqref{eg2:solutionu2} 
	agrees with the given initial and the Dirichlet boundary condition \eqref{2+1:ic}-\eqref{2+1:bc} when
		$$
		\begin{array}{ll}
		&	\delta_{11}(x_1,x_2)=\rho_{10}\sin\big( \sqrt{\frac{5\lambda_{20}}{2}}z\big) +\rho_{20} \cos\big( \sqrt{\frac{5\lambda_{20}}{2}}z\big), 
			\\&	\delta_{21}(x_1,x_2)=\zeta_{10}\sin\big( \sqrt{\lambda_{20}}z\big) +\zeta_{20} \cos\big( \sqrt{\lambda_{20}}z\big),
			\\&	\delta_{12}(x_1,x_2)=\rho_{11}\sin\big( \sqrt{\frac{5\lambda_{20}}{2}}z\big) +\rho_{21} \cos\big( \sqrt{\frac{5\lambda_{20}}{2}}z\big), 
			\\
			&	\delta_{22}(x_1,x_2)=\zeta_{11}\sin\big( \sqrt{\lambda_{20}}z\big) +\zeta_{21} \cos\big( \sqrt{\lambda_{20}}z\big),
		\end{array}
$$
	$$
	\begin{aligned}
		%%%%%%%%%%%%%%%%%%%%%%%%%%%%%%%%%%%%%%%%%%%%%%%%%%%%%%%%%%%%%BC_1%%%%%%%%%%%%%%%%%%%%%%%%%%%%%%%%
		\tau_{11}(t)= &
		\left\{\begin{array}{ll}
		 \rho_{20} E_{{\alpha_1},1}(\varrho_1t^{\alpha_1}) ,
					\text{ if }  {\alpha_1}\in(0,1]\,\&\, {\alpha_2}\in(0,2],
			\\
			 \rho_{20} E_{{\alpha_1},1}(\varrho_1t^{\alpha_1})
			+ \rho_{21} tE_{{\alpha_1},2}(\varrho_1t^{\alpha_1})
		 ,
			\text{ if } {\alpha_1}\in(1,2]\,\&\, {\alpha_2}\in(0,2],
		\end{array}\right.
		%%%%%%%%%%%%%%%%%%%%%%%%%%%%%%BC@@2%%%%%%%%%%%%%%%%%%%%%%%%%%%%%%%%%%%%%%%%%%%%%%%%%%%%%%%%
		\\	\tau_{21}(t)=&
		\left\{\begin{array}{ll}
			\zeta_{20} E_{{\alpha_2},1}(\varrho_2t^{\alpha_1}) ,
			\text{ if }  {\alpha_2}\in(0,1]\,\&\, {\alpha_1}\in(0,2],
			\\
			\zeta_{20} E_{{\alpha_2},1}(\varrho_2t^{\alpha_1})
			+ \zeta_{21} tE_{{\alpha_2},2}(\varrho_2t^{\alpha_1})
			,
			\text{ if } {\alpha_2}\in(1,2]\,\&\, {\alpha_1}\in(0,2], 	\end{array}\right.
		\\
		\text{and }				\tau_{i2}(t)= &
		\left\{\begin{array}{ll}
			r_{i0} E_{{\alpha_i},1}(\varrho_it^{\alpha_i}) ,
			\text{ if }  {\alpha_i}\in(0,1]\,\&\, {\alpha_j}\in(0,2], i\neq j,i,j=1,2,
			\\
			r_{i0} E_{{\alpha_i},1}(\varrho_it^{\alpha_1})
			+ r_{i1} tE_{{\alpha_i},2}(\varrho_it^{\alpha_i})
		,
			\text{ if } {\alpha_i}\in(1,2]\,\&\, {\alpha_j}\in(0,2],  i\neq j,i,j=1,2,
		\end{array}\right.
		%%%%%%%%%%%%%%%%%%%%%%%%%%%%%%BC@@2%%%%%%%%%%%%%%%%%%%%%%%%%%%%%%%%%%%%%%%%%%%%%%%%%%%%%%%%		&	\tau_{22}(t)=&		\left\{\begin{array}{ll}			r_{20} E_{{\alpha_1},1}(\varrho_1t^{\alpha_1}) , 			\text{ if }  {\alpha_2}\in(0,1],\&\, {\alpha_1}\in(0,2], 			\\		r_{20} E_{{\alpha_1},1}(\varrho_1t^{\alpha_1})			+r_{21} tE_{{\alpha_1},2}(\varrho_1t^{\alpha_1})			,		\\					\text{ if } {\alpha_2}\in(1,2],\&\, {\alpha_1}\in(0,2], 	\end{array}\right.
				\end{aligned}$$
	%%%%%%%%%%%%%%%%%%%%%%%%%%%%%%%%%%%%%%%%%%%%%%%%%%%%%%%%%%%%%%%%%%%%BC@@@3%%%5
	where $z=\kappa_1x_1+\kappa_2x_2,r_{1i}=\rho_{1i}\sin\big( \sqrt{\frac{5}{2}\lambda_{20}}r\big) +\rho_{2i} \cos\big( \sqrt{\frac{5}{2}\lambda_{20}}r\big),$ and  $r_{2i}=\zeta_{1i}\sin\big( \sqrt{\lambda_{20}}r\big) +\zeta_{2i} \cos\big( \sqrt{\lambda_{20}}r\big),i=0,1
	.$
It is important to note that the  obtained exact solutions \eqref{eg2:solutionu1}
%-\eqref{eg2:solutionu2}
 of the  {{nonlinear two-component system}  \eqref{2+1rd} along  with the coefficient functions given in \eqref{eg1: coefficients of cubic nonlinear operator} and the given parameter constraints \eqref{eg2:constraints}} hold for every $\alpha_s\in(0,2],s=1,2,$  including the integer values of  $\alpha_s=1,2,$ $s=1,2.$
Additionally,  we have presented the two and three dimensional plots of the obtained exact solution \eqref{eg2:solutionu1}
%-\eqref{eg2:solutionu2}
 of the {nonlinear two-component system} of \eqref{2+1rd} in the Figures \ref{Fig1} and \ref{Fig2} when the parameters $\rho_{10}=400,\rho_{20}=\zeta_{20}=\zeta_{11}=500,\zeta_{10}=-200,\rho_{11}=110,\rho_{21}=30,\zeta_{21}=-300,\mu_{11}=14,\mu_{21}=-7,\gamma_{10}=-\eta_{20}=\kappa_2=2,\lambda_{20}=3,$ and $\kappa_1=-1$ for chosen values of $x_1$ and $x_2. $
\begin{figure}\centering
	\begin{subfigure}[b]{0.435\textwidth}\centering
	\includegraphics[width=\textwidth]{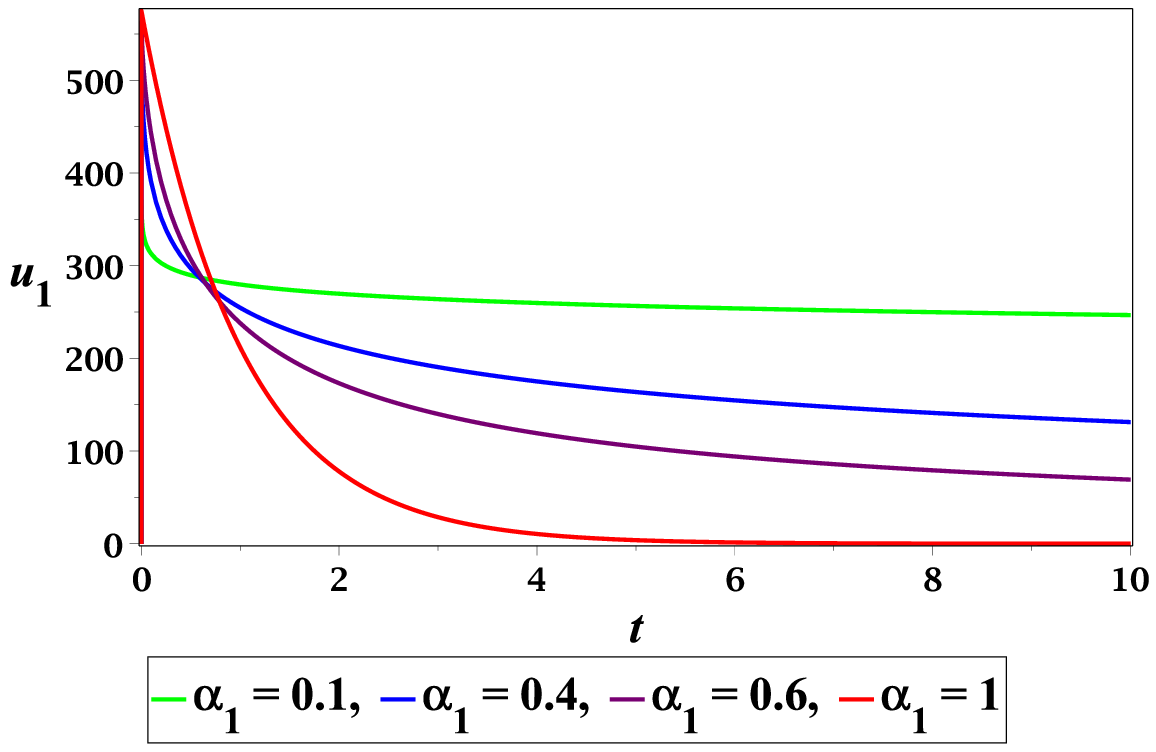}
	\caption{$x_1=3$ and $x_2=4$}
\end{subfigure}	
\begin{subfigure}[b]{0.45\textwidth}\centering
	\includegraphics[width=\textwidth]{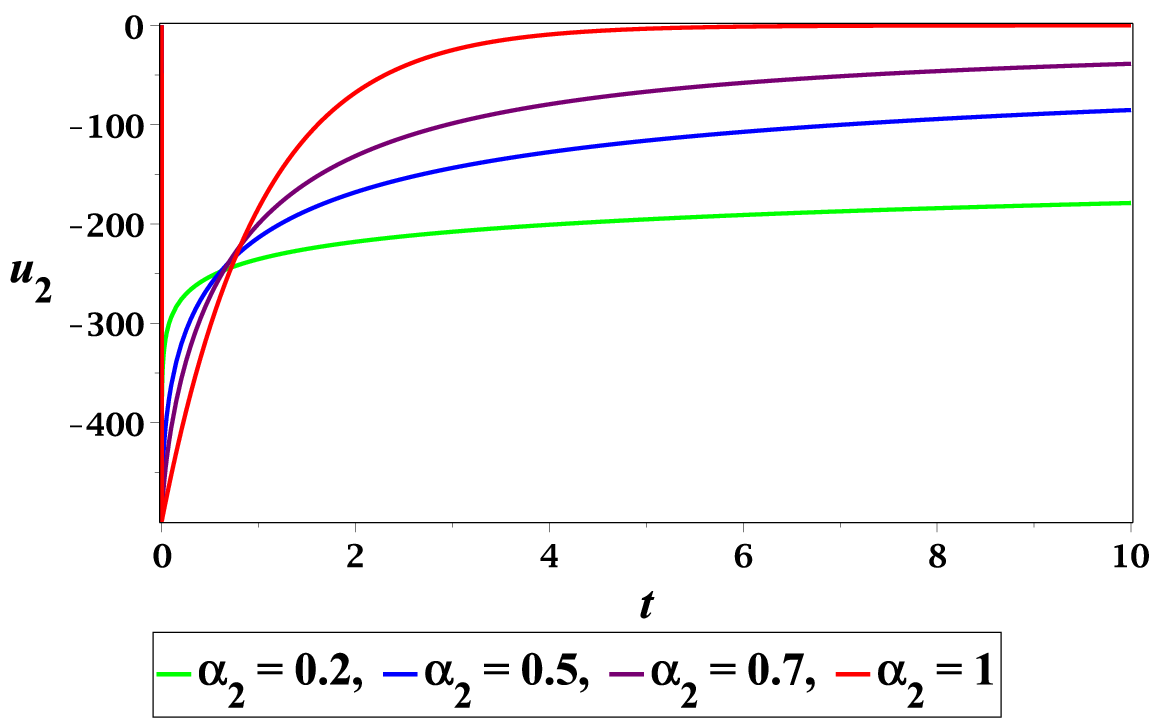}
	\caption{$x_1=3$ and $x_2=4$}
\end{subfigure}	
\begin{subfigure}[b]{0.44\textwidth}\centering
	\includegraphics[width=\textwidth]{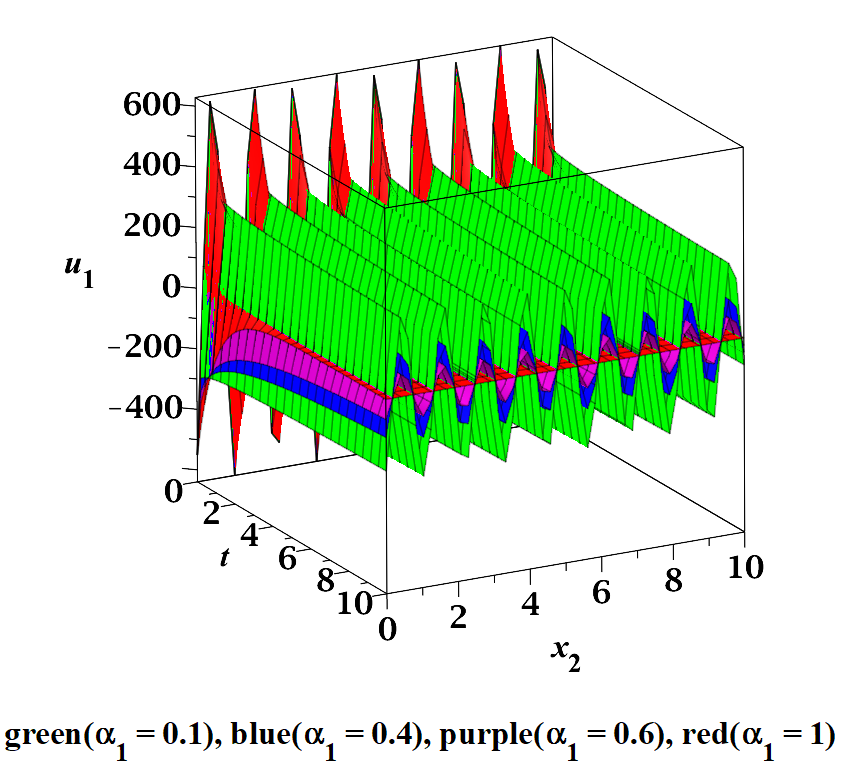}
	\caption{$x_1=3$}
\end{subfigure}	
\begin{subfigure}[b]{0.47\textwidth}\centering
	\includegraphics[width=\textwidth]{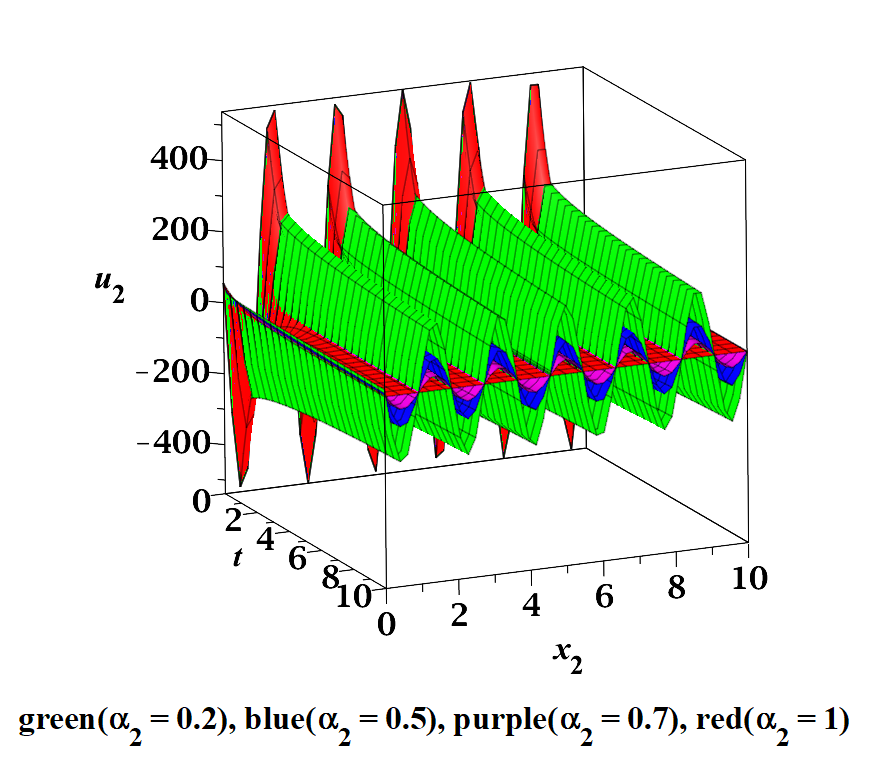}
	\caption{$x_1=3$}
\end{subfigure}
\caption{2D and 3D plots of the exact solution \eqref{eg2:solutionu1} of the  cubic type of {nonlinear two-component system}  \eqref{2+1rd}  when $\alpha_1,\alpha_2\in(0,1].$}
\label{Fig1}
\end{figure}
\begin{figure}\centering
\begin{subfigure}[b]{0.42\textwidth}
	\includegraphics[width=\textwidth]{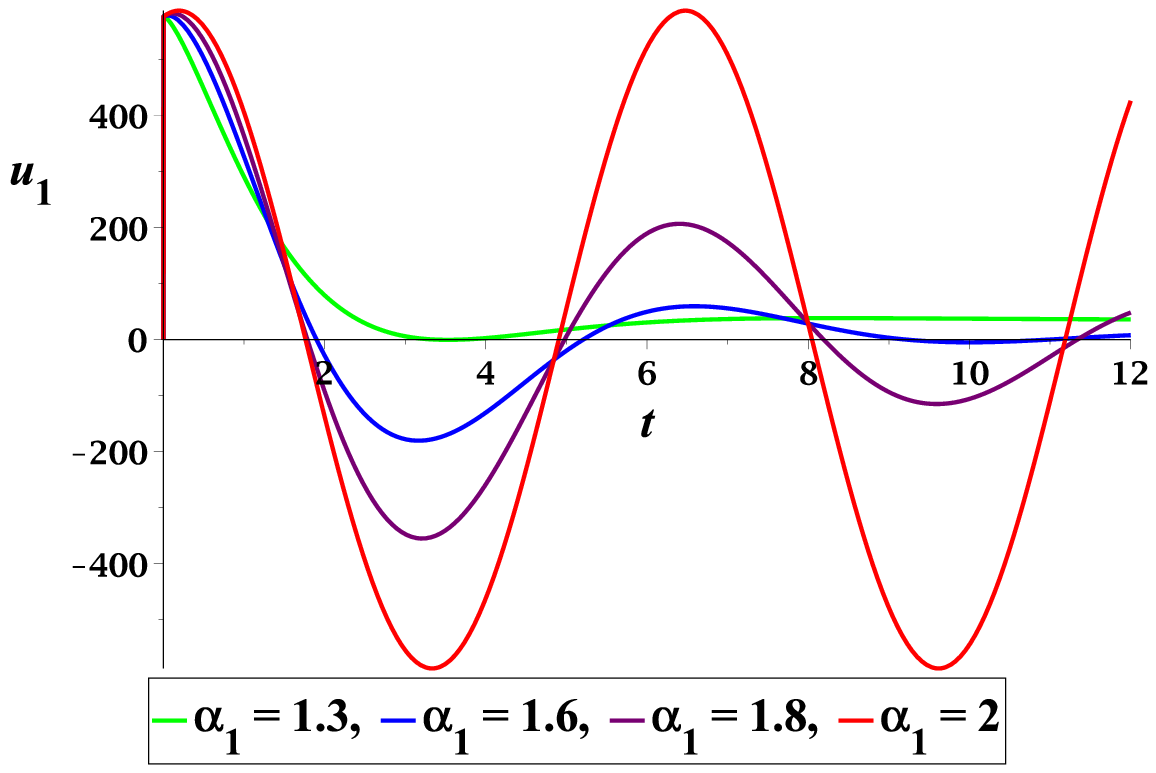}
	\caption{$x_1=3$ and $x_2=4$}
\end{subfigure}	
\begin{subfigure}[b]{0.42\textwidth}\centering
	\includegraphics[width=\textwidth]{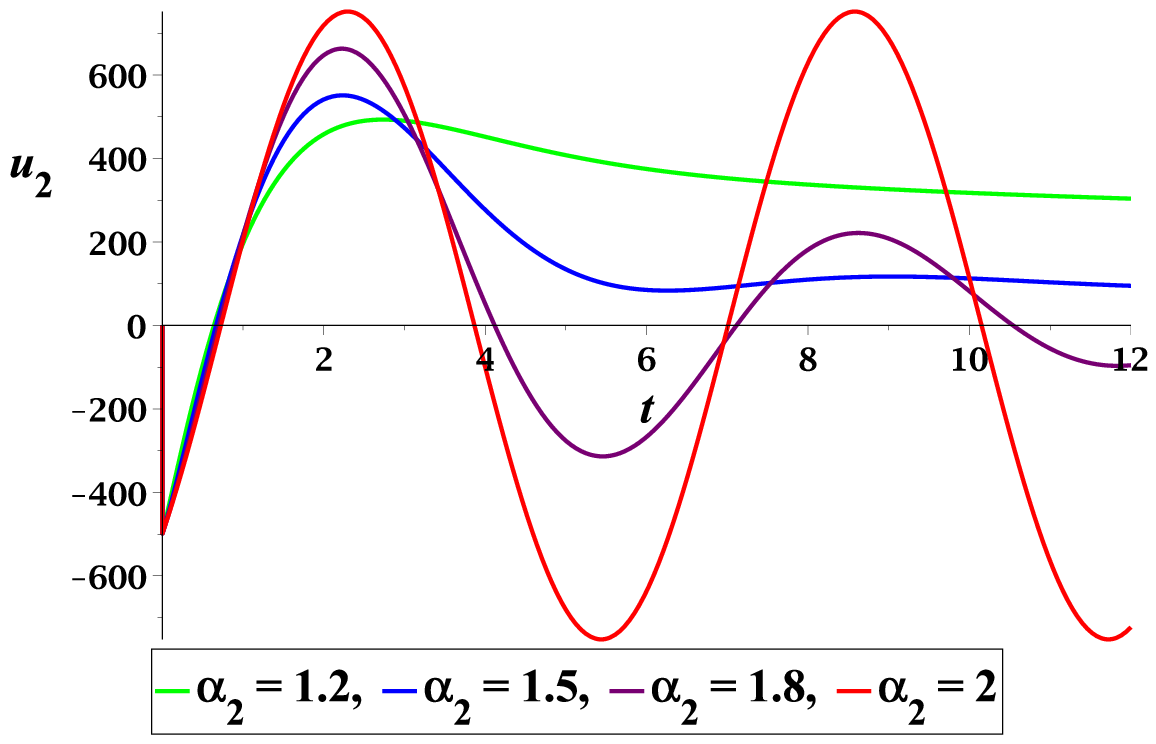}
	\caption{$x_1=3$ and $x_2=4$}
\end{subfigure}	
\begin{subfigure}[b]{0.42\textwidth}\centering
	\includegraphics[width=\textwidth]{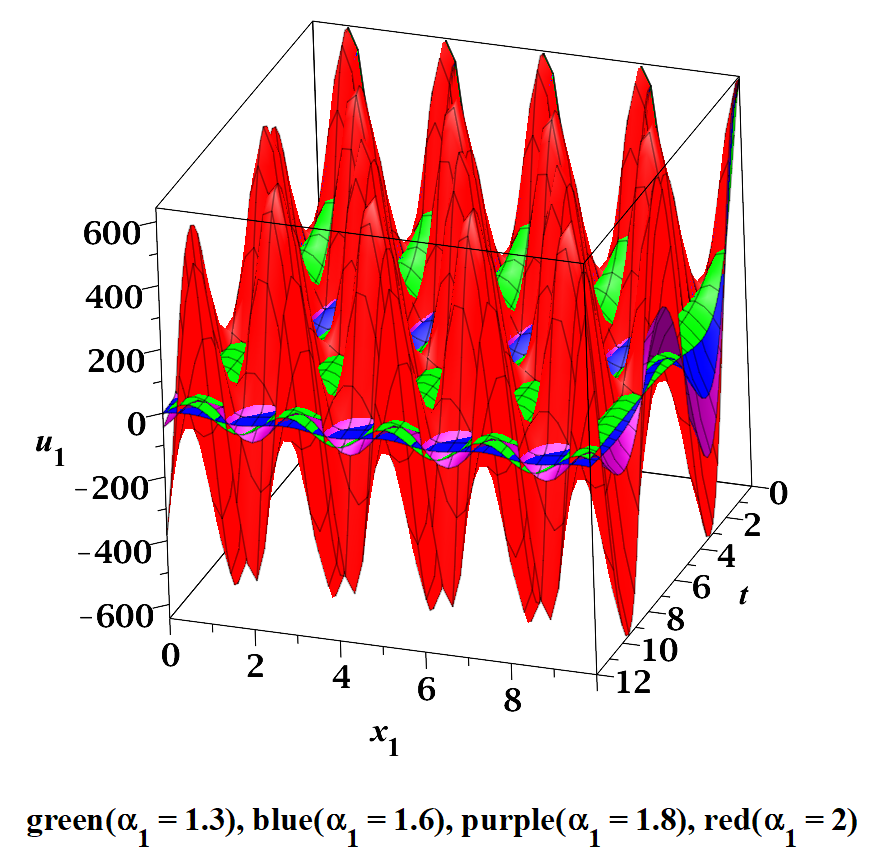}
	\caption{$x_2=4$}
\end{subfigure}	
\begin{subfigure}[b]{0.45\textwidth}\centering
	\includegraphics[width=\textwidth]{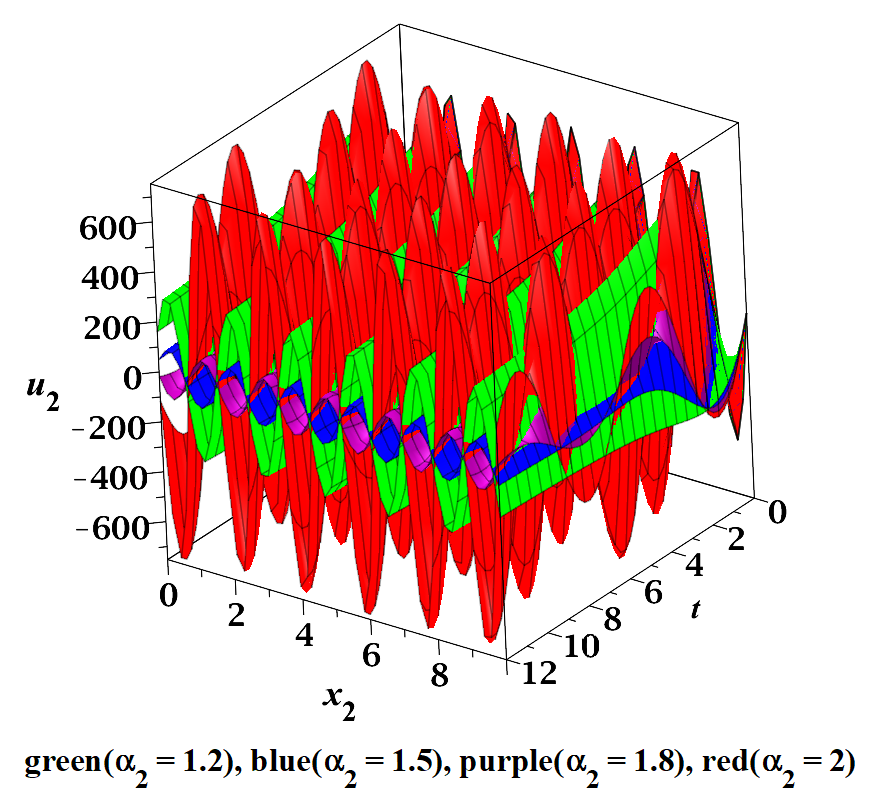}
	\caption{$x_1=3$}
\end{subfigure}
\caption{2D and 3D plots of the exact solution \eqref{eg2:solutionu1} of the cubic type of {nonlinear two-component system}   \eqref{2+1rd}  when $\alpha_1,\alpha_2\in(1,2].$}
\label{Fig2}
\end{figure}
\begin{note}
	From the two and three-dimensional graphs given in Figure \ref{Fig1}, we observe that the obtained exact solution \eqref{eg1:solution-u1} shows  the sub-diffusion behaviour of  the particles corresponding  to $u_1$ and $u_2$ when the values of $\alpha_1$ and $\alpha_2$ lie between 0 and 1, respectively, as the time varies. Also, we wish to point out that the change in the rate of diffusion increases gradually when the value of $\alpha_s,s=1,2,$ increases from 0 to 1. Additionally,  from the two and three-dimensional graphs in Figure \ref{Fig2}, we observe  the transition  from the diffusive nature to the wave behaviour of the particles corresponding  to $u_1$ and $u_2$  when the values of $\alpha_1$ and $\alpha_2$ increase from 1 to 2. It is interesting to see that the wave nature of the particles corresponding  to $u_1,$ and $u_2$ is visible clearly when $\alpha_1=\alpha_2=2.$
\end{note}

\end{example}
\begin{example}
	\label{eg3}
	Now, we consider $\alpha_s=\alpha,s=1,2,$ in the following  cubic {nonlinear two-component system} of TFRDEs in $(2+1)$-dimensions  of the form
	\begin{eqnarray}
		\begin{aligned}
			\label{eg3:2+1rd}
			\dfrac{\partial^{\alpha} u_s}{\partial t^{\alpha}}=&	
			\dfrac{\partial }{\partial x_1}
			\left[
			(a_{{s5}}  u_{{1}}  ^{2}
			+(a_{{s4}}  u_{{2}} 		
			+a_{{s3}}u_{{1}} 	+a_{{s2}})u_{{2}}+a_{{s1}}u_{{1}}		+a_{{s0}} )
			\dfrac{\partial u_1}{\partial x_1}
			+	(b_{{s5}}  u_{{1}}  ^{2}	
			\right.	\\
		&\left.	+(	b_{{s4}}  u_{{2}}
				+b_{{s3}}u_{{1}}
			+b_{{s2}})u_{{2}}
			+b_{{s1}}u_{{1}}	
			+b_{{s0}} )
			\dfrac{\partial u_2}{\partial x_1}\right]
			+\dfrac{\partial }{\partial x_2}\left[ (p_{{s5}}  u_{{1}}  ^{2}	
			+p_{{s1}}u_{{1}}	
		+p_{{s0}}	\right.	\\
		&\left.	+(p_{{s4}}  u_{{2}} 		
			+p_{{s3}}u_{{1}} 		
			+p_{{s2}})u_{{2}}
		 )
			\frac{\partial u_1}{\partial x_2}
				+(q_{{s5}}  u_{{1}}  ^{2}
			+
		(	q_{{s4}}  u_{{2}}
			+q_{{s3}}u_{{1}}	+q_{{s2}})u_{{2}}+q_{{s1}}u_{{1}}			\right.	\\
			&\left.+q_{{s0}} )\frac{\partial u_2}{\partial x_2}\right]
		+	c_{{s0}	}+c_{{s1}}u_{{1}}
					+(c_{{s2}} 	
			 		+c_{{s3}}u_{{1}}	+	c_{{s4}}  u_{{2}} )u_{{2}}	+	c_{{s5}}  u_{{1}} ^{2}	
		\\
		&  + u_{{1}}(	c_{{s6}} u_{{2}} ^{2}	+	c_{{s7}}  u_{{1}} 	u_{{2}})
		 	+	c_{{s8}}  u_{{2}}^{3}	+	c_{{s9}}  u_{{1}}^{3}	,
		\end{aligned}
	\end{eqnarray}
	for $	\alpha\in(0,2],s=1,2.$
	Note that the given {nonlinear two-component system} \eqref{eg3:2+1rd} admits the  invariant linear product space
	\begin{eqnarray}
	\label{eg3:vector space}
	\mathbf{W}_4:= \mathit{W}_{1,2}\times \mathit{W}_{2,2}=\mathcal{L}\{ \sin(\sqrt{2}\beta z), \cos(\sqrt{2} \beta z)\}
	\times
	\mathcal{L}\{e^{-\beta z}, e^{\beta z}\},\beta\in\mathbb{R},z=\kappa_1x_1+\kappa_2x_2,
	\end{eqnarray}
	when the constants are obtained as follows:
\begin{eqnarray}
		\begin{array}{lllll}
		a_{{10}}=-\dfrac{\kappa_2^2p_{10}}{\kappa_1^2}	+\gamma_{10},
		&
		a_{12}=\dfrac{\mu_{12}-3\kappa_2^2\beta^2p_{12}}{3\kappa_1^2\beta^2},
		&
		a_{14}=\dfrac{\mu_{13}-12\kappa_2^2\beta^2p_{14}}{12\kappa_1^2\beta^2},
	\\
		a_{15}=\dfrac{\mu_{14}-6\kappa_2^2\beta^2p_{15}}{6\kappa_1^2\beta^2}, 
		&				b_{{10}}=\dfrac{\mu_{11}}{\kappa_1^2\beta_1^2}-\dfrac{\kappa_2^2q_{10}}{\kappa_1^2},
		&
		a_{{1i}}=-\dfrac{\kappa_2^2p_{1i}}{\kappa_1^2},i=1,3,
			\\
		b_{11}=-\dfrac{3\kappa_2^2\beta^2q_{12}+\mu_{12}}{3\kappa_1^2\beta^2},
	&	b_{13}=-\dfrac{6\kappa_2^2\beta^2q_{13}+\mu_{13}}{6\kappa_1^2\beta^2},
&
	a_{{20}}=\dfrac{\mu_{20}}{2\kappa_1^2\beta^2}-\dfrac{\kappa_2^2p_{20}}{\kappa_1^2},	\\b_{{1i}}=-\dfrac{\kappa_2^2q_{1i}}{\kappa_1^2},i=2,4,5,
	&
	a_{22}=-\dfrac{3\kappa_2^2\beta^2p_{22} +\mu_{22}}{3\kappa_1^2\beta^2},	
	&
	a_{23}=-\dfrac{9\kappa_2^2\beta^2p_{23} -\mu_{23}}{9\kappa_1^2\beta^2},	\label{eg3:constraints}
\\
		b_{{20}}=-\dfrac{\kappa_2^2q_{20}}{\kappa_1^2}	+\eta_{20},
	&	c_{16}=\dfrac{\mu_{13}}{2},c_{27}=\dfrac{\mu_{23}}{2},
		& c_{1i}=0,i=0,4,5,7,8,
		\\	a_{{2i}}=-\dfrac{\kappa_2^2p_{2i}}{\kappa_1^2},i=1,4,5,
&
	c_{19}=\mu_{14},c_{28}=\mu_{24}
		&
		%%%%%%%%%%%%%%%%%%%%%%%%%
			b_{21}=-\dfrac{3\kappa_2^2\beta^2q_{21}+\mu_{22}} {3\kappa_1^2\beta^2},
			\\	b_{24}=-\dfrac{3\kappa_2^2\beta^2q_{24}+\mu_{24}}{3\kappa_1^2\beta^2},&
			b_{25}=-\dfrac{18\kappa_2^2\beta^2q_{25}+\mu_{23}}{18\kappa_1^2\beta^2},
&
		b_{{2i}}=-\dfrac{-\kappa_2^2q_{2i}}{\kappa_1^2},i=2,3,	\\
		c_{2i}=0,i=0,4,5,6,9, & \text{and }c_{si}=\mu_{si-1},&i=1,2,3,s=1,2,
	\end{array}
\end{eqnarray}
Thus, substituting  the particular form $(u_1(x_1,x_2,t),u_2(x_1,x_2,t))=(v_1(z,t),v_2(z,t)),$ $z=\kappa_1x_1+\kappa_2x_2,$ into to the above {nonlinear two-component system} \eqref{eg3:2+1rd} with the  parametric restrictions \eqref{eg3:constraints}, we obtain  the following  reduced {nonlinear two-component system} of TFRDEs in $(1+1)$-dimensions:
	\begin{eqnarray}
		\begin{aligned}{\label{eg3:transformedrd}}
			\dfrac{\partial^{\alpha} v_1}{\partial t^{\alpha}}=A_1(v_1,v_2)\equiv&	\dfrac{\partial }{\partial z}\Big[ \dfrac{1}{12\beta^2} (12\kappa_1^2\beta^2\gamma_{10} +2\mu_{14}v_1^2+\mu_{13}v_2^2+4\mu_{12}v_2
				)\frac{\partial v_1}{\partial z}
				\\&-  \dfrac{1}{6\beta^2}(6\mu_{11} +2\mu_{12}v_1
			 				+\mu_{13}v_2v_1
			)\frac{\partial v_2}{\partial z}\Big]
			+\mu_{10}v_{{1}}
			+\mu_{11}v_{{2}} 	
			+\mu_{12}v_{{1}}v_{{2}}
		\\&	+\dfrac{1}{2}\mu_{13}v_{{1}} v_{{2}} ^{2}	
			+	\mu_{14}  v_{{1}}^{3}	
			,
			%\end{aligned}
			\\%\begin{aligned}{\label{eg3:transformedrdb}}
			\dfrac{\partial^{\alpha} v_2}{\partial t^{\alpha}}=A_2(v_1,v_2)\equiv&	\dfrac{\partial }{\partial z}\left[ \dfrac{1}{18\beta^2}(9\mu_{20} +2\mu_{23} v_1v_2
			+6\mu_{22} v_2
			) \frac{\partial v_1}{\partial z}		\right.		\\ & 	\left.
			-\dfrac{1}{18\beta^2}(-18\kappa_1^2\beta^2\eta_{20} +6\mu_{22} v_1	+6\mu_{24} v_2^2
						+\mu_{23}v_1^2
			)\frac{\partial v_2}{\partial z} \right]
		\\&	+\mu_{20}v_{{1}}
			+\mu_{21}v_{{2}} 	
			+\mu_{22}v_{{1}}v_{{2}}
			+\dfrac{1}{2}\mu_{23}v_{{1}} ^{2}	 v_{{2}}
			+\mu_{24} v_{{2}}^{3},\alpha\in(0,2].
		\end{aligned}
	\end{eqnarray}
{Note that if $(v_1,v_2)\in \mathbf{W}_4=\mathit{W}_{1,2}\times \mathit{W}_{2,2}$ with $\mathit{W}_{1,2}=\mathcal{L}\{\sin( \sqrt{2}\beta z) ,\cos( \sqrt{2}\beta z) \}$ and $\mathit{W}_{2,2}=\mathcal{L}\{e^{\beta z},e^{-\beta z}\},$ as given in \eqref{eg3:vector space}, then we have
	\begin{eqnarray*}
		\begin{aligned}
			&	A_1(v_1,v_2)=  p_1(\mu_{10}-2\beta^2\kappa_1^2\gamma_{10})\sin(\sqrt{2}\beta z) +p_2(\mu_{10}-2\beta^2\kappa_1^2\gamma_{10})\cos( \sqrt{2}\beta z)  \in  \mathit{W}_{1,2},\text{ and }
			\\ &
			A_2(v_1,v_2)=  q_1(\mu_{21}-\beta^2\kappa_1^2\eta_{20})e^{\beta z} +q_2(\mu_{21}-\beta^2\kappa_1^2\eta_{20})e^{-\beta z}\in \mathit{W}_{2,2},
		\end{aligned}
	\end{eqnarray*}
for every $ v_1=p_1\sin( \sqrt{2}\beta z) +p_2\cos( \sqrt{2}\beta z)\in \mathit{W}_{1,2}, $ and $v_2
=q_1e^{\beta z}+q_2e^{-\beta z} \in \mathit{W}_{2,2},$ 	where $A_s(v_1,v_2), s=1,2,$  are given in \eqref{eg3:transformedrd}, and $p_1,p_2,q_1,q_2,\beta\in\mathbb{R}.$ 
Thus, the linear product space $\mathbf{W}_4$ given in \eqref{eg3:vector space} is invariant under the  above {nonlinear two-component system} \eqref{eg3:transformedrd}, which was  listed in  case 10 of Table \ref{table3}.}
	Thus, applying a  procedure similar to the one  discussed in Example \ref{eg1}, we obtain the generalized separable exact solution for the above-considered {nonlinear two-component system} \eqref{eg3:2+1rd} along with $\mu_{s3}=\mu_{s4}=0,s=1,2,$ as follows:
	\begin{eqnarray}
	\small	\begin{aligned}
			\label{eg3:solutionu1}
			&u_1(x_1,x_2,t)=v_1(z,t)=\left\{\begin{array}{ll}
				\big[ \rho_{10}\sin( \sqrt{2}\beta z) +\rho_{20} \cos( \sqrt{2}\beta z)\big] E_{\alpha,1}(\varrho_1t^\alpha) , \text{ if }  \alpha\in(0,1],
				\\
				\big[ \rho_{10}\sin( \sqrt{2}\beta z) +\rho_{20} \cos( \sqrt{2}\beta z)\big] E_{\alpha,1}(\varrho_1t^\alpha)+ tE_{\alpha,2}(\varrho_1t^\alpha)\\
				\times \big[ \rho_{11}\sin( \sqrt{2}\beta z) +\rho_{21} \cos( \sqrt{2}\beta z)\big] , \text{ if } \alpha\in(1,2],
			\end{array}\right.
			\\
			&
		%	\label{eg3:solutionu2}
			u_2(x_1,x_2,t)=v_2(z,t)=\left\{\begin{array}{ll}
				\big(
				\zeta_{10}e^{-\beta z} + \zeta_{20} e^{\beta z}\big)E_{\alpha,1}(\varrho_2t^\alpha) , \text{ if }  \alpha\in(0,1],
				\\
				\big(\zeta_{10}e^{-\beta z} + \zeta_{20} e^{\beta z}\big) E_{\alpha,1}(\varrho_2t^\alpha)+t\big(
				\zeta_{11}e^{-\beta z}
				+ \zeta_{21} e^{\beta z}\big) E_{\alpha,2}(\varrho_2t^\alpha),
				\\ \text{ if } \alpha\in(1,2],
			\end{array}\right.
		\end{aligned}
	\end{eqnarray}
	where $\varrho_1=\mu_{10}-2\beta^2\kappa_1^2\gamma_{10},\varrho_2=\mu_{21}-\beta^2\kappa_1^2\eta_{20},$ and $\rho_{i0},\rho_{i1},\zeta_{i0},\zeta_{i1}\in\mathbb{R},$ $i=1,2. $
	We note that the above exact solution \eqref{eg3:solutionu1} agrees with  the given initial and the Dirichlet boundary condition \eqref{2+1:ic}-\eqref{2+1:bc}  when
	$\delta_{11}(x_1,x_2)=\rho_{10}\sin( \sqrt{2}\beta z)  +\rho_{20} \cos( \sqrt{2}\beta z) ,
	$
	$	\delta_{21}(x_1,x_2)=\zeta_{10}e^{-\beta z} +\zeta_{20} e^{\beta z},
$ $	\delta_{12}(x_1,x_2)=\rho_{11}\sin( \sqrt{2}\beta z)  +\rho_{21} \cos( \sqrt{2}\beta z) ,
	$ $	\delta_{22}(x_1,x_2)=\zeta_{11}e^{-\beta z} +\zeta_{21} e^{\beta z},$
	$$
\small	\begin{aligned}
		%%%%%%%%%%%%%%%%%%%%%%%%%%%%%%%%%%%%%%%%%%%%%%%%%%%%%%%%%%%%%BC_1%%%%%%%%%%%%%%%%%%%%%%%%%%%%%%%%
	&	\tau_{11}(t)=
		\left\{\begin{array}{ll}
			\rho_{20} E_{{\alpha},1}(\varrho_1t^{\alpha}) ,
			\text{ if }  {\alpha}\in(0,1],
			\\
			\rho_{20} E_{{\alpha},1}(\varrho_1t^{\alpha})
			+ \rho_{21} tE_{{\alpha},2}(\varrho_1t^{\alpha})
			,
		\\	\text{ if } {\alpha}\in(1,2],
		\end{array}\right.
		&	\tau_{12}(t)=
	\left\{\begin{array}{ll}
		r_{10} E_{{\alpha},1}(\varrho_1t^{\alpha}) ,
		\text{ if }  {\alpha}\in(0,1],
		\\
		r_{10} E_{{\alpha},1}(\varrho_1t^{\alpha})
		+ r_{11} tE_{{\alpha},2}(\varrho_1t^{\alpha})
		,\qquad\qquad
	\\	\text{ if } {\alpha}\in(1,2],
	\end{array}\right.\\
&	\tau_{21}(t)=
	\left\{\begin{array}{ll}
		(	\zeta_{10}+	\zeta_{20}) E_{{\alpha},1}(\varrho_1t^{\alpha}) ,
		\text{ if }  {\alpha}\in(0,1],
		\\
		(\zeta_{10}+	\zeta_{20}) E_{{\alpha},1}(\varrho_1t^{\alpha})
		+ (\zeta_{11}+\zeta_{21}) \\
		\times tE_{{\alpha},2}(\varrho_1t^{\alpha})
		,
		\text{ if } {\alpha}\in(1,2],	\end{array}\right.
		%%%%%%%%%%%%%%%%%%%%%%%%%%%%%%BC@@2%%%%%%%%%%%%%%%%%%%%%%%%%%%%%%%%%%%%%%%%%%%%%%%%%%%%%%%%
			%%%%%%%%%%%%%%%%%%%%%%%%%%%%%%BC@@2%%%%%%%%%%%%%%%%%%%%%%%%%%%%%%%%%%%%%%%%%%%%%%%%%%%%%%%%
	&\&\,\tau_{22}(t)=
		\left\{\begin{array}{ll}
			r_{20} E_{{\alpha},1}(\varrho_1t^{\alpha}) ,
			\text{ if }  {\alpha}\in(0,1],
			\\
			r_{20} E_{{\alpha},1}(\varrho_1t^{\alpha})
			+r_{21} tE_{{\alpha},2}(\varrho_1t^{\alpha})
			,
		\\
		\text{ if } {\alpha}\in(1,2],	\end{array}\right.\qquad
	\end{aligned}$$
	%%%%%%%%%%%%%%%%%%%%%%%%%%%%%%%%%%%%%%%%%%%%%%%%%%%%%%%%%%%%%%%%%%%%BC@@@3%%%5
	where $z=\kappa_1x_1+\kappa_2x_2,r_{1i}=\rho_{1i}\sin( \sqrt{2}\beta r) +\rho_{2i} ( \sqrt{2}\beta r),$ and  $r_{2i}=\zeta_{1i}e^{-\beta r} +\zeta_{2i} e^{\beta r},i=0,1
	.$
	We note that the obtained exact solution \eqref{eg3:solutionu1} is valid for all values of  $\alpha\in(0,2],$ including the   integer values $\alpha=1$ and $\alpha=2.$
\end{example}
\section{Discussions and concluding remarks}\label{Discussions-conclusions}
In this section, we wish to point out the importance of the fractional-order reaction-diffusion systems that arise in various areas of science and engineering disciplines. Additionally, we will present the    concluding remarks of the developed method and the obtained results of the discussed systems in the paper.
\subsection{Discussions}
It is well-known that the study of system of diffusion-reaction equations has  gained considerable attention due to the wide range of applications \cite{cherniha2017,murray2002,turing1952,lenzi2016,povstenko2013,zhang2017,yang2019,qu2009,dat1,dat2}. It has been a topic of interest ever since Turing \cite{turing1952} proposed the system of diffusion-reaction equations that can be used to describe the chemical basis of morphogenesis. In contrast to the fractional-order case,  the class of system of reaction-diffusion equations of the integer-order case has been discussed extensively in the literature \cite{qu2009,cherniha2017}.
It is well-known that  the anomalous diffusion processes are described more accurately using the fractional diffusion equation of the form \eqref{diffusionequation}.
The power-law memory property of the fractional derivatives helps in preserving the generic nature of nonlinear time-dependency of the anomalous diffusion processes \cite{lenzi2016,hanygad2022}, that is, $\langle x^2(t)\rangle\sim t^\alpha,\alpha\in(0,2].$  Many researchers have studied the  system of TFRDEs to describe various physical processes. For example, to analyze the heat conduction in an infinite one-dimensional composite
medium,  Povestenko \cite{povstenko2013} has formulated  the following system of time-fractional diffusion equations:
  \begin{equation}\label{povstenko-diffusionequation}
  	\dfrac{\partial ^{\alpha_s} u_s}{\partial t^{\alpha_s}}=C_s\dfrac{\partial ^2 u}{\partial x^2},{\alpha_s}\in(0,2],s=1,2,
  \end{equation}
  where $C_s$ represent the coefficients of heat conduction of the quantity $u_s=u_s(x,t),x\in\mathbb{R},t>0,$ and $ \dfrac{\partial ^{\alpha_s} }{\partial t^{\alpha_s}}(\cdot)$ is the fractional time derivative of ${\alpha_s}$-th order in the sense of the Caputo \eqref{caputo}, $s=1,2.$ Povstenko \cite{povstenko2013} used the Laplace transformation on the system \eqref{povstenko-diffusionequation}  to find solutions in terms of the Mittag-Leffler function and Mainardi function.
    Additionally, Lenzi \cite{lenzi2016} has discussed the above system \eqref{povstenko-diffusionequation} to investigate the diffusion and transport across a membrane in a heterogeneous medium. Also, we note that in \cite{dat1,dat2}, the authors have studied the bifurcations and complex spatio-temporal solutions for one-dimensional fractional-order reaction-diffusion system.

Also, we observe that some particular cases of the discussed  two-component  system of TFRDEs \eqref{2+1rd}  in $(2+1)$-dimensions  have been used widely to investigate  many  phenomena in real-world complex systems \cite{zhang2017,yang2019,choudary2019a,mo2022}.
Recently, Yang et al.\cite{yang2019}   studied a particular case of the given  {nonlinear two-component system} of TFRDEs \eqref{2+1rd} in $(2+1)$-dimensions  in order  to describe the anomalous diffusion process of the sodium and potassium ions in neuronal phenomena having cross-diffusion.  This particular  case of the {nonlinear two-component system} of TFRDEs \eqref{2+1rd} in $(2+1)$-dimensions  is also known as the nonlinear system of time-fractional Nernst-Planck equations \cite{yang2019}, which can be read as follows:
\begin{eqnarray}
	\label{yang-model}
	\begin{aligned}	\dfrac{\partial^\alpha u_s }{\partial t^\alpha}	=&	\sum\limits_{i=1}^{2}\dfrac{\partial }{\partial x_i}\left( (c_1u_s+c_2)\frac{\partial u_1}{\partial x_i}+(c_1u_s+c_2)\frac{\partial u_2}{\partial x_i}\right)
	-I_s,\alpha\in(0,1],s=1,2,
\end{aligned}
\end{eqnarray}
 where $\dfrac{\partial^\alpha  }{\partial t^\alpha}(\cdot)$ represents the Caputo fractional  derivative given in \eqref{caputo}, $u_s=u_s(x_1,x_2,t),s=1,2,$ represent the concentration of sodium and potassium ions with constant initial concentration and potential differences, and $I_s,s=1,2,$ denote the current densities in the physical process.
%Note that in the discussed system \eqref{2+1rd}, if the orders of the Caputo fractional derivative  is  $\alpha_s=\alpha\in(0,1],$ the diffusion coefficients are $f_{si}(u_1,u_2)=g_{si}(u_1,u_2)=c_1u_s+c_2,c_i\in\mathbb{R},s,i=1,2,$ and the reaction terms $h_s(u_1,u_2)=I_s,s=1,2,$
Additionally, they  have used  the Jacobi special collocation method to  find approximate solutions for the above-system \eqref{yang-model}.
 In addition, a Legendre wavelet collocation method \cite{mo2022} was used to study a particular case of the given    two-component  system of TFRDEs \eqref{2+1rd} in $(2+1)$-dimensions  when $f_{si}(u_1,u_2)=g_{si}(u_1,u_2)=-1,s,i=1,2$ and  $\alpha_1=\alpha_2=1$ along with initial and Dirichlet boundary conditions.

  However,  the analytical solutions of {nonlinear multi-component system} of TFRDEs   have been discussed only in a few  studies \cite{choudary2019a}, particularly in higher dimensions.
  In our paper, we developed the invariant subspace method combined with variable transformation to solve the initial and boundary value problems of the generalized  {nonlinear two-component system} of TFRDEs  \eqref{2+1rd} in $(2+1)$-dimensions, for the first time.

\subsection{Concluding remarks}
In this article, we have  systematically developed and applied the invariant subspace method along with the particular form $(u_1(x_1,x_2,t),u_2(x_1,x_2,t))=(v_1(z,t),v_2(z,t)), z=\kappa_1x_1+\kappa_2x_2$ for finding the  exact solutions of the   {nonlinear two-component system} of TFPDEs given in \eqref{coupledsystem} in $(2+1)$-dimensions.
Under the above  particular form \eqref{transformation}, the {nonlinear two-component system} of TFPDEs \eqref{coupledsystem} in $(2+1)$-dimensions converted into  the   {nonlinear two-component system} of TFPDEs  \eqref{transformedcoupledsystem} in $(1+1)$-dimensions.
Additionally, we wish to point out that the obtained
finite-dimensional linear product space $\mathbf{W}_n$ given in \eqref{invariantproductspaces} is an invariant linear product space   of the considered {nonlinear two-component system} \eqref{coupledsystem} as well as the  reduced {nonlinear two-component system} \eqref{transformedcoupledsystem}.
Thus, the obtained generalized separable exact solutions satisfied both the $(2+1)$ and $(1+1)$-dimensional {nonlinear two-component systems}  \eqref{coupledsystem} and \eqref{transformedcoupledsystem} using the obtained invariant linear product spaces $\mathbf{W}_n.$
Also, we illustrated the significance and efficacy of the invariant subspace method combined with variable transformation by deriving the generalized separable exact solutions for initial and boundary value problems of the generalized  {nonlinear two-component system} of TFRDEs   \eqref{2+1rd} in $(2+1)$-dimensions. We have constructed the different types of invariant linear product spaces $\mathbf{W}_n$ for given the generalized {nonlinear two-component system} of TFRDEs   \eqref{2+1rd}.
Moreover, we  derived the generalized separable exact solutions in terms of the exponential, trigonometric, polynomial, Euler-Gamma, and  Mittag-Leffler functions for the {nonlinear two-component system} of TFRDEs  \eqref{2+1rd} in $(2+1)$-dimensions.
% In addition, We have  plotted two and three dimensional plots of the  exact solutions he considered  two-component  nonlinear coupled system of TFRDEs in $(2+1)$-dimensions, which corresponds to different  fractional-order derivatives $\alpha_s,\alpha_s\in(0,2],s=1,2.$

This study shows that the developed method is a very effective and powerful analytic tool to derive exact solutions for the different {nonlinear two-component systems} of TFPDEs, which will be beneficial to analyze and understand  the quantitative and qualitative properties. In the future, we plan to systematically investigate  how to extend the invariant subspace method to derive  exact solutions for the nonlinear systems of time and space fractional PDEs, which has enormous applications in science and engineering.
 %{\textbf{{Data accessibility: }}}Data sharing does not apply to this article as no data sets were generated or analyzed during the current study.	\\
%	{\textbf{{Declaration of AI use:}} }We have not used AI-assisted technologies in creating this article.	\\
%		{\color{jobcolor}\textbf{Author contributions:}} 	\textbf{P. Prakash: }Conceptualization;  Formal analysis;  Investigation; Methodology;   Software; Supervision; Validation; Visualization;Writing–original draft; Writing–review \& editing.
%	\textbf{K.S. Priyendhu:} Conceptualization;  Formal analysis;  Investigation; Methodology;   Software;  Validation; Visualization; Writing–original draft; Writing–review \& editing.
% \textbf{M. Lakshmanan:} Conceptualization;   Investigation; Methodology; Supervision;  Validation; Visualization; Writing–original draft; Writing–review \& editing.\\
%All authors gave final approval for publication and agreed to be held accountable for the work performedtherein.\\
		{{\section*{Acknowledgments}} }
		%The authors wish to thank the Editor and anonymous referees for their constructive suggestions for improving the manuscript.
		The author (K.S.P.) would like to thank International Mathematical Union (IMU), Germany,  for providing financial support in the form of IMU Breakout Graduate fellowship-2023 (IMU-BGF-2023-06). Another author (M.L.) is supported by a Department of Science and Technology, India-SERB National Science Chair position (NSC/2020/000029).
	%	\\
	%	\\
%		{{\textbf{Conflict of interest}:} }	 We declare we have no competing interests.

\end{document}